\documentclass[12pt]{article}

\usepackage{epsf,epsfig,epstopdf}
\usepackage{graphics}

\usepackage{url}
\usepackage{amsmath,amssymb}
\usepackage{cite}

\def\slash#1{#1 \hskip-0.45em /}
\def\Slash#1{#1 \hskip-0.59em /}

\usepackage{a4wide}
\numberwithin{equation}{section}

\DeclareMathOperator{\tr}{tr}

\def\ket#1{\mathinner{|{#1}\rangle}}
\def\braket#1{\mathinner{\langle{#1}\rangle}}

\begin{document}

\allowdisplaybreaks
\thispagestyle{empty}

\begin{flushright}
{\small

TTK-10-45\\
IPPP/10/61\\ 
DCPT/10/122\\
FR-PHENO-2010-026\\
SFB/CPP-10-65\\
arXiv:1007.5414 [hep-ph] \\[0.2cm]
September 28, 2010}
\end{flushright}

\vspace{\baselineskip}

\begin{center}
\vspace{0.1\baselineskip}
\textbf{\Large\boldmath
Threshold resummation for pair production of\\[0.2cm] 
 coloured heavy (s)particles at hadron colliders
}
\\
\vspace{3\baselineskip}
{\sc M.~Beneke$^a$, P.~Falgari$^b$, C.~Schwinn$^c$}\\
\vspace{0.7cm}
{\sl ${}^a$Institute f\"ur Theoretische Teilchenphysik und 
Kosmologie,\\
D--52056 Aachen, Germany\\
\vspace{0.3cm}
${}^b$IPPP, Department of Physics, University of Durham, \\
Durham DH1 3LE, England\\
\vspace{0.3cm}
${}^c$ Albert-Ludwigs Universit\"at Freiburg, 
Physikalisches Institut, \\
D-79104 Freiburg, Germany }

\vspace*{1.2cm}
\textbf{Abstract}\\
\vspace{1\baselineskip}
\parbox{0.9\textwidth}{ We derive a factorization formula for the
  production of pairs of heavy coloured particles in hadronic
  collisions near the production threshold that establishes 
  factorization of soft and Coulomb effects.  This forms the basis for
  a combined resummation of Coulomb and soft corrections, including the
  non-trivial interference of the two effects. We develop a resummation
  formalism valid at NNLL accuracy using the momentum-space
  approach to soft gluon resummation. We present numerical results
  for the NLL resummed squark-antisquark production cross section at the LHC
  and Tevatron, including also the contribution of squark-antisquark
  bound states below threshold. The total correction on top of the 
  next-to-leading order approximation is found to be sizeable, and
  amounts to (4--20)\% in the squark mass region 200 GeV -- 3 TeV at 
  the 14 TeV LHC. The scale dependence of the total cross section is also
  reduced.  }

\end{center}

\newpage
\setcounter{page}{1}

\section{Introduction}

In perturbative calculations of partonic cross sections at hadron
colliders there often arise terms that are kinematically
enhanced in certain regions in phase space.
In the Drell-Yan process, for instance,
Sudakov logarithms of the form $[\alpha_s \ln^2(1-z)]^n$ arise, 
where $z = Q^2/\hat s$, and $Q^2$ and $\hat s$ represent the invariant 
mass squared of the lepton pair and the partonic centre-of-mass (cms) 
energy, respectively. In the threshold region 
$z \rightarrow 1$, these terms are large, and spoil the convergence of the
perturbative expansion in the QCD coupling $\alpha_s$. If the dominant 
contribution to the hadronic cross sections originates from the 
partonic threshold region, which is certainly the case when 
$Q^2$ approaches the cms energy $s$, these logarithms 
need to be resummed to all orders in perturbation theory to attain 
a reliable theoretical description. This was accomplished 
in \cite{Sterman:1986aj,Catani:1989ne} for the inclusive Drell-Yan 
cross section by solving evolution equations in Mellin space. 
In production processes of pairs of heavy coloured particles such as
top quarks or coloured particles in extensions of the standard model,
e.g.  squarks and gluinos in supersymmetric extensions, the partonic
cross section contains terms of the form $(\alpha_s \log^2\beta\,)^n$
(``threshold logarithms''), where $\beta=(1-4M^2/\hat s)^{1/2}$ is the heavy
particle velocity, and $(\alpha_s/\beta)^n$ (``Coulomb
singularities''), which are also enhanced near the partonic threshold
$\hat s\approx 4 M^2$.  Resummation of threshold logarithms for
heavy-particle and di-jet production has also been implemented in
Mellin
space~\cite{Laenen:1991af,Berger:1996ad,Catani:1996yz,Kidonakis:1996aq,Kidonakis:1997gm,Kidonakis:1998nf,Bonciani:2003nt}
up to now, and has been used for improved predictions of the top-pair
production cross section at hadron
colliders~\cite{Catani:1996dj,Bonciani:1998vc,Kidonakis:2001nj,Cacciari:2003fi,Moch:2008qy,Cacciari:2008zb,Kidonakis:2008mu}.
Recently the resummation of threshold logarithms for production
processes of supersymmetric coloured
particles~\cite{Kulesza:2008jb,Kulesza:2009kq,Langenfeld:2009eg,Beenakker:2009ha}
and colour octet scalars~\cite{Idilbi:2009cc,Idilbi:2010rs} has been studied as
well.  The resummation of multiple exchanges of Coulomb gluons and
bound-state effects has been studied for the total top-antitop cross
section~\cite{Fadin:1990wx,Catani:1996dj} and the invariant
mass distribution of top quarks and
gluinos~\cite{Hagiwara:2008df,Kiyo:2008bv,Hagiwara:2009hq}.

The theoretical basis for resummation is a factorization of the 
partonic cross section $\hat \sigma$ in the partonic threshold region 
into hard and soft contributions of the schematic form
\begin{equation}
\label{eq:factor-thresh}
\hat \sigma =H\otimes S
\end{equation}
with a hard function $H$ and a soft function $S$ both of which are 
matrices in colour space. If the invariant mass $Q^2=(p_1+p_2)^2$ of the
heavy-particle pair $H(p_1)H'(p_2)$ is held fixed, the ``partonic
threshold'' is then defined more generally as the kinematical region
where  $Q^2$ is close to the partonic centre-of-mass energy $\hat
s$. Arguments from perturbative QCD~\cite{Contopanagos:1996nh}, the
properties of Wilson lines~\cite{Korchemsky:1992xv,Korchemsky:1993uz},
or effective
theories~\cite{Manohar:2003vb,Becher:2006nr,Becher:2006mr,Becher:2007ty}
can then be used to demonstrate exponentiation of the enhanced
contributions by solving evolution equations for the functions $S$ and
$H$.  Traditionally this resummation is performed in Mellin space
which requires a numerical transformation of the resummed
result back to momentum space. A resummation method directly in
momentum space was proposed in~\cite{Becher:2006nr} and has been
applied subsequently e.g. to Drell-Yan or Higgs
production~\cite{Becher:2007ty,Ahrens:2008qu,Ahrens:2008nc} and
recently to the top-quark invariant mass
distribution~\cite{Ahrens:2010zv}.

In this paper we consider the production of a pair of heavy 
coloured particles $H, H'$ with masses $m_H$ and $m_{H^\prime}$, 
respectively, in the collision of hadrons $N_1$ and $N_2$,
\begin{equation}\label{eq:heavy-pair}
N_1(P_1) N_2(P_2) \rightarrow H(p_1) H'(p_2) + X
\end{equation}
and concentrate on the resummation of the total cross section. In this case, 
after integration over the invariant mass of $H H'$, the enhanced logarithms 
appear when 
\begin{equation}
\label{eq:threshold}
\hat s \approx (m_H+m_{H^\prime})^2.
\end{equation}
In this limit, the factorization of soft gluons is complicated by the
fact that the non-relativistic energy of the heavy particles is of the
same order as the momenta of the soft gluons, in contrast to the
assumptions made in the derivation of the usual factorization
formula~\eqref{eq:factor-thresh}. We show that in this kinematical
regime the partonic cross sections factorizes into three
contributions,
\begin{equation}
\label{eq:factor-sketch}
  \hat\sigma=H\otimes  W\otimes J \, ,
\end{equation}
where $H$ is determined by hard fluctuations, $W$ by soft fluctuations
and $J$ accounts for the propagation of the heavy-particle pair
including Coulomb-gluon exchange. The short-hand $W\otimes J$ includes
a convolution which accounts for the energy loss of the heavy
particles due to soft emissions (the precise form of the factorization
formula is given in~\eqref{eq:fact-final} below).

The main points of the 
present approach are as follows:

\begin{itemize}
\item 
Our approach is largely model-independent and highlights the universal 
features of soft-gluon and Coulomb resummation: the precise nature of 
the heavy particles and the physics model enter the factorization
formula~\eqref{eq:factor-sketch} only in the hard function $H$; the
soft function and the Coulomb function depend only on the colour 
and (in the latter case) the spin quantum
numbers of the heavy-particle pair. The hard function can be obtained 
directly from the  $H H'$ production amplitude, expanded 
near threshold, without any need to perform a full cross section 
calculation.
\item Eq.~(\ref{eq:factor-sketch}) generalizes previous results of the
  form~\eqref{eq:factor-thresh} by factorizing the Coulomb effects in addition
  to those from soft gluons.  Multiple exchange of Coulomb gluons associated
  to corrections $\sim \alpha_s^n/\beta^m$ can be resummed in the function $J$
  using methods familiar from non-relativistic QCD~(NRQCD).  The presence of
  the Coulomb function $J$ leads to a more complicated colour structure of the
  soft functions $W$ compared to previous treatments based on a factorization
  of the form~\eqref{eq:factor-thresh}. The factorization
  formula~\eqref{eq:factor-sketch} allows a combined resummation of soft and
  Coulomb gluons and justifies earlier
  treatments~\cite{Bonciani:1998vc,Hagiwara:2008df,Kiyo:2008bv,Kulesza:2009kq}
  where the factorization of Coulomb from soft gluons was put in as an
  assumption.
\item The kinematical structure of the soft function $W$ simplifies in
  the threshold region~\eqref{eq:threshold} compared to the more
  general situation considered in~\cite{Kidonakis:1997gm}. This has
  allowed us to construct a basis in colour space that diagonalizes
  the soft functions relevant for hadron collider pair production
  processes of heavy particles in arbitrary representations of the
  colour gauge group to all orders of perturbation
  theory~\cite{Beneke:2009rj}.  The diagonal colour bases for
  resummation of threshold logarithms for all production processes of
  pairs of coloured supersymmetric particles at hadron colliders, i.e.
  squark-antisquark, squark-squark, squark-gluino and gluino-gluino
  production have been provided in~\cite{Beneke:2009rj}, extending
  explicit one-loop results for soft anomalous dimensions at
  threshold~\cite{Kidonakis:1997gm,Kulesza:2008jb,Kulesza:2009kq,Beenakker:2009ha}. This
  result greatly simplifies threshold resummations at NNLL accuracy
  and has allowed us to extract the two-loop soft anomalous dimension for
  arbitrary colour representations~\cite{Beneke:2009rj} from results
  of~\cite{Becher:2009kw,Kidonakis:2009ev,Korchemsky:1991zp}, in
  agreement with an independent two-loop study for top-pair
  production~\cite{Czakon:2009zw}. The more complicated colour
  structures in the two-loop soft anomalous dimension for massive
  particles with generic
  kinematics~\cite{Ferroglia:2009ep,Ferroglia:2009ii,Mitov:2010xw}
  have been shown not to contribute to the NNLO total production cross
  section of a heavy-particle pair produced in an $S$-wave at
  threshold~\cite{Beneke:2009ye}.
\item We use the method 
of~\cite{Becher:2006nr,Becher:2006mr} to
perform the resummation of threshold logarithms $\log \beta $ 
directly in momentum space solving evolution equations for the 
functions $H$ and $W$. The
resummation of the total cross section in the threshold region has
been performed previously in Mellin
space~\cite{Bonciani:1998vc,Bonciani:2003nt}. Our derivation gives a
field theoretical definition of the quantities appearing in that
approach and the resummation in momentum space allows an analytical
treatment, since no numerical inverse Mellin transform is necessary.
The relation between the Mellin-space and momentum-space formalisms has been
discussed in~\cite{Becher:2006nr,Becher:2006mr,Becher:2007ty,Beneke:2009rj}.
\item The formalism includes the case of heavy particles with sizeable 
decay widths. To first approximation
finite-width effects enter only in the Coulomb function and can be
included by a shift of the non-relativistic energy $E\to E+i\Gamma$ as 
familiar from top-quark pair production at electron-positron colliders. 
A systematic description of finite width effects beyond leading order 
can be achieved in the effective theory 
approach~\cite{Beneke:2003xh,Beneke:2004km,Beneke:2004xd,Beneke:2007zg}.
\end{itemize}

In this paper we focus on establishing the factorization of soft and
Coulomb gluons and demonstrate the new features arising in a combined
resummation by presenting numerical results for the example of
squark-antisquark production. While we set up a method suitable for
resummation at NNLL accuracy, in our initial application we restrict
ourselves to NLL accuracy since the required colour separated hard
production coefficients for squark-antisquark production are currently
unknown. We also do not include finite squark decay widths which are
small in typical supersymmetry (SUSY) scenarios and would introduce a
dependence on the SUSY parameters but, as mentioned
above, could easily be included in our approach.  Note that, since in
the present work we are concerned with the total cross section, there
are no sizable finite-width corrections as long as $\Gamma\ll M$,
unlike the case of the invariant mass distribution near threshold.

The paper is organized as follows. In section~\ref{sec:factorize} we
derive the factorization formula~\eqref{eq:factor-sketch} using
effective field theory methods. We compare to previous treatments of 
combined soft and Coulomb corrections and extend the diagrammatic argument
of~\cite{Beneke:2009ye} on the absence of corrections from subleading
couplings of soft gluons to non-relativistic particles at NNLL to all 
orders in the strong coupling using the effective theory approach. In
section~\ref{sec:resum} we discuss the process-independent ingredients
in the factorization formula, the soft function $W$ and the Coulomb
function $J$, collect their explicit results and perform the
summation of threshold logarithms using evolution equations in
momentum space.  We also give a simple prescription for how to obtain
the hard-function $H$, the only process-dependent ingredient in the
resummation formalism, from a standard fixed-order calculation.  In
section~\ref{sec:squarks} we perform the combined
soft-Coulomb resummation at NLL accuracy for squark-antisquark
production and present numerical results for the Tevatron and the
LHC.\footnote{Preliminary results have appeared in 
\cite{Beneke:2009nr,Beneke:2010gm}.}  
A number of additional results concerning parton densities, 
the Coulomb potential in various colour representations and 
the resummed and expanded cross section are collected in the 
appendix.

\section{Factorization for coloured 
heavy-particle pair pro\-duction 
near the partonic threshold}
\label{sec:factorize}

In this section we derive a factorization formula of the
form~\eqref{eq:factor-sketch} for the process~\eqref{eq:heavy-pair}
near the partonic threshold. The heavy particles are assumed to 
transform under representations $R$ and $R'$ of the colour gauge 
group $SU(3)_C$. The inclusive production cross section   
is described by the factorization formula
\begin{equation}
\label{eq:parton-model}
\sigma =
\sum_{p,p'} 
 \int d x_1 d x_2\,f_{p/N_1}(x_1,\mu) f_{p'/N_2}(x_2,\mu)\,
\hat\sigma_{pp'}(x_1x_2s, \mu), 
\end{equation}
where $\hat \sigma_{pp'}$ are the hard-scattering cross sections for 
the partonic subprocesses
\begin{equation}
\label{eq:subprocess}
p(k_1) p'(k_2)\rightarrow H(p_1)H'(p_2)+X \\
\end{equation}
and $pp'\in\{qq, q\bar q, \bar q\bar q, gg, gq,g\bar q\}$. 
As usual the nucleon 
masses are neglected and parton momenta 
$k_{1,2}$ are related to the incoming nucleon momenta by
\begin{equation}
  k_1=x_1 P_1\;,\quad k_2=x_2 P_2.
\end{equation}
The partonic cms energy is given by
\begin{equation}
  \hat s \equiv(k_1+k_2)^2=x_1x_2 s.
\end{equation}
At hadron colliders $\hat s$ is not a fixed quantity but one
may argue that the steep rise of the parton distribution functions with
decreasing $x$ leads to an enhanced contribution to the total production cross
section from the \emph{partonic} threshold region
\begin{equation}
  z\equiv  \frac{4 M^2}{\hat s}\sim 1,
\end{equation}
$M=(m_H+m_{H^\prime})/2$ denoting the average heavy-particle mass.  According
to a recent study of threshold effects in the Drell-Yan
process~\cite{Becher:2007ty} this dynamical enhancement of the partonic
threshold region remains effective for values $4 M^2/s = z x_1x_2
\gtrsim 0.2$. 
Therefore we expect the resummation of threshold logarithms to be 
relevant for particles with masses $M\gtrsim
3$ TeV ($1.6, 2.2 $ TeV) at the LHC at $14$ TeV ($7,10$ TeV) and
$M\gtrsim 450$ GeV at the Tevatron which are at the border of the
estimated reach for coloured SUSY particles at the LHC~\cite{Baer:2009dn,Baer:2010tk} and the exclusion limits obtained at the Tevatron~\cite{Abazov:2007ww,Aaltonen:2008rv}. However, experience with the NLO
corrections to top-quark production suggests that also for smaller masses there
is a sizable contribution to the total cross section from those terms
in the partonic cross section that are enhanced for $z\to 1$ so that
resummation can serve as an estimate of higher-order effects. In
addition, expansions of resummed cross sections can also be used to
predict the threshold-enhanced terms in higher-order perturbative
calculations~\cite{Kidonakis:2001nj,Beneke:2009ye,Ahrens:2009uz}.

The heavy particles produced near threshold are  
non-relativistic. In terms of dimensionless variables the partonic 
threshold is characterized by the smallness of 
\begin{equation}
\beta=\sqrt{1-\frac{4M^2}{\hat s}}=\sqrt{1-z},
\end{equation}
which for equal masses also corresponds to the non-relativistic velocity of
the heavy particles in the partonic cms. The aim of this paper is to perform a
resummation of threshold logarithms $\log\beta$ and Coulomb corrections
$1/\beta$.  In order to discuss the systematics of the combined resummations
of the two corrections we count both $\alpha_s \ln\beta$ and $\alpha_s/\beta$
as quantities of order one and introduce a parametric representation of the
expansion of the cross section of the form~\cite{Beneke:2009rj}
\begin{eqnarray}
\label{eq:syst}
\hat{\sigma}_{p p'} &\propto& \,\hat\sigma_{p p'}^{(0)}\, 
\sum_{k=0} \left(\frac{\alpha_s}{\beta}\right)^k \,
\exp\Big[\underbrace{\ln\beta\,g_0(\alpha_s\ln\beta)}_{\mbox{(LL)}}+ 
\underbrace{g_1(\alpha_s\ln\beta)}_{\mbox{(NLL)}}+
\underbrace{\alpha_s g_2(\alpha_s\ln\beta)}_{\mbox{(NNLL)}}+\ldots\Big]
\nonumber\\[0.2cm]
&& \,\times
\left\{1\,\mbox{(LL,NLL)}; \alpha_s,\beta \,\mbox{(NNLL)}; 
\alpha_s^2,\alpha_s\beta,\beta^2 \,\mbox{(NNNLL)};
\ldots\right\}.
\end{eqnarray}
With this counting, the resummed
cross section at LL accuracy includes all terms of order
$1/\beta^{\,k}\times\alpha_s^{n+k}\ln^{2n}\beta$ relative to the Born cross
section near threshold.  Next-to-leading summation includes in
addition all terms of order $\alpha_s \ln\beta; \,\alpha_s^2
\left\{1/\beta\times\ln\beta, \ln^3 \beta \right\};\ldots$, while 
furthermore all terms $\alpha_s ; \,\alpha_s^2
\left\{1/\beta,\ln^{2,1} \beta\right\}; \ldots$ are included
in NNLL approximation.\footnote{The NNLL terms odd in $\beta$ at order
$\alpha_s$ and $\alpha_s^2$, 
$\alpha_s\,\beta\ln^{2,1}\beta;\,\alpha_s^2\,\beta\ln^{4,3}\beta$, 
vanish due to rotational invariance for the total cross section 
when the heavy-particle pair is dominantly produced in an $S$-wave state
at tree level~\cite{Beneke:2009rj,Beneke:2009ye}. See also below 
in section~\ref{sec:comments}.}  

In the threshold region, the cross section of the 
process~\eqref{eq:subprocess} receives contributions from hard, potential, 
soft and two collinear momentum regions defined by
\begin{equation}
\label{eq:regions}
\begin{aligned}
&\\[-0.5cm]
\text{hard } (h)&:  \quad k\sim M\\
  \text{potential } (p) &: \quad k_0\sim M \lambda,\,\,
  \vec{k}\sim M \sqrt \lambda \\
  \text{soft } (s)&:  \quad k_0\sim \vec{k} \sim  M \lambda \\
  \text{$n$-collinear } (c) &:   \quad k_- \sim M,\, k_+\sim M\lambda,\,
  k_\perp \sim M\sqrt\lambda \\
  \text{$\bar n$-collinear } (\bar c) &: \quad k_+ \sim M,
  \, k_-\sim M\lambda,\, k_\perp \sim M\sqrt\lambda\,, \\[-0.5cm]
&
\end{aligned}
\end{equation}
assuming equal heavy-particle masses for simplicity. 
Here we introduced the generic power counting parameter 
$\lambda \sim 1-z = \beta^2 \ll 1$.
For the collinear momentum we employ the light-cone
decomposition
\begin{equation}
  p^\mu=\frac{p_-}{2}\,n^\mu+\frac{p_+}{2}\,\bar n^\mu+p_\perp^\mu
\end{equation}
with two light-like vectors $n$ and $\bar n$ satisfying $n\cdot \bar
n=2$.  Due to the threshold kinematics $(k_1+k_2)^2\sim 4M^2$, no
collinear modes can appear in the final state so that the particles
denoted by $X$ in~\eqref{eq:heavy-pair} are entirely given by soft
modes.

In order to achieve the separation of the effects corresponding to the
different regions~\eqref{eq:regions} it is useful to describe the
heavy-pair production process in the framework of effective field
theories. The relevant effective Lagrangian contains elements of
non-relativistic QCD (NRQCD), describing the interaction of potential
and soft modes, and soft-collinear effective theory
(SCET)~\cite{Bauer:2000yr,Bauer:2001yt,Beneke:2002ph,Beneke:2002ni} 
describing the
interaction of soft and collinear modes.  Hard modes with virtuality
of order $M^2$ are integrated out and are not part of the effective
Lagrangian.  This effective theory is similar to that used
in~\cite{Beneke:2004xd,Beneke:2007zg,Actis:2008rb} to describe
$W$-pair production near threshold in $e^+ e^-$ collisions and the
construction of the effective Lagrangian proceeds along similar lines.
As in the discussion of the Drell-Yan process in~\cite{Becher:2007ty}
the following considerations will be adequate to establish
factorization of the partonic cross section; the discussion of the
hadronic cross section requires to include additional modes with a
scaling given by powers of
$\Lambda_{\text{QCD}}/M$~\cite{Becher:2006mr}. We also note the
existence of an additional region with scaling $k\sim M \sqrt\lambda$,
usually referred to as ``soft'' in the NRQCD literature. It
is related to potential forces between the heavy particles, and will
not be relevant for the following discussion of the leading Coulomb
corrections.
 
In the EFT framework, the factorization formula is derived by a
series of steps involving first a matching calculation from QCD (or
another ``full theory'' such as the MSSM) to the EFT followed by field
redefinitions that decouple
soft gluons from collinear~\cite{Bauer:2001yt,Bauer:2002nz} and
potential~\cite{Beneke:2010gi} fields. These will be described
in the remainder of this section.\footnote{The corresponding discussion of
factorization of electromagnetic effects and resummation for the
$W$-pair production cross section at $e^+e^-$ colliders has been given
in~\cite{Falgari:2009zz}.}  We also show that, up to NNLL accuracy,
corrections from subleading interactions to the effective Lagrangians
not included in the derivation of the factorization formula 
either vanish or can be straightforwardly incorporated. 

\subsection{Matching to SCET and NRQCD}
\label{sec:matching}

The first step in the derivation of the factorization formula consists of 
representing the scattering amplitude for the process~\eqref{eq:heavy-pair}
in terms of effective theory matrix elements of production 
operators $\mathcal{O}^{(\ell)}_{\{a;\alpha\}}$ 
corresponding to the partonic sub-processes $p(k_1)p'(k_2)\to H(p_1)H'(p_2)X$, 
multiplied by coefficient functions that contain the hard  
part of the amplitude. After this step the partonic scattering 
amplitude is given by 
\begin{equation}
\label{eq:matrix}
\mathcal{A}(pp'\to HH'X) = \sum_{\ell}
\;C^{(\ell)}_{\{a;\alpha\}}(\mu)\;
\langle H H'X |\mathcal{O}^{(\ell)}_{\{a;\alpha\}}(\mu) |
p p'\rangle_{\rm EFT} \,.
\end{equation} 
The subscript ``EFT'' means that the matrix element is evaluated with 
the effective Lagrangian discussed below. 
We now explain the definition of the operators and the calculation 
of the coefficient functions.

The hard production subprocess corresponds to the $2\to 2$ process 
$p(k_1)p'(k_2)\to H(p_1)H'(p_2)$. Near the partonic threshold,
kinematics forbids additional massive or energetic particles in the 
production operator, while additional soft fields connected directly 
to the hard process imply additional highly off-shell propagators and
therefore lead to operators power-suppressed in $\lambda\sim 1-z$ 
relative to the leading four-particle operators. The required
four-particle operators take the form 
\begin{equation}
\label{eq:opdef}
\mathcal{O}^{(0)}_{\{a;\alpha\}}(\mu) = 
\Big[\phi_{c;a_1\alpha_1}\phi_{\bar c;a_2\alpha_2} 
\psi^\dagger_{a_3\alpha_3}\psi^{\prime \,\dagger}_{a_4\alpha_4}\Big](\mu). 
\end{equation}
The form of the production operators is thus similar to
those describing the production of non-relativistic
$W$-pairs in $e^+ e^-$ collisions discussed 
in~\cite{Beneke:2004xd,Beneke:2007zg}. Several remarks on 
the notation are in order:

(1) The fields $\psi^\dagger$ and $\psi^{\prime \,\dagger}$ are
non-relativistic fields which create the heavy particles $H$ and $H'$ 
with momenta
\begin{equation}
\label{eq:nr-mom}
  p^\mu_{1}=m_H w^\mu+\tilde p_{1}^\mu\quad,\quad  
  p^\mu_{2}=m_{H'} w^\mu+\tilde p^\mu_{2}
\end{equation}
with the cms velocity $w^\mu$ ($w^2=1$) and small 
residual momenta $\tilde p_{1/2}$ in the potential region
(\ref{eq:regions}). In the partonic cms frame
$w=(1,\vec 0\,)$  and $\tilde p_1=-\tilde p_2\equiv\tilde p$ 
for the spatial components of $\tilde{p}_{1,2}$. The 
fields $\phi_c$ ($\phi_{\bar c}$) are collinear (anticollinear) 
fields that destroy the initial state partons $p$ and $p'$
with momenta
\begin{equation}
k_1^\mu  \approx M n^\mu \; ,\quad k_2^\mu\approx M \bar n^\mu,
\end{equation}
respectively.\footnote{We do not have to consider operators
  where the $\phi_c$ destroy particles different from $p$ and
  $p'$. These would arise from a splitting of a collinear gluon
  into a collinear and a soft quark or of a collinear quark into a
  soft quark and a collinear gluon, but these interactions are present
  only in the power-suppressed part of the SCET 
  Lagrangian~\cite{Beneke:2002ph}.}  We provide more
details on the definition of these fields and their effective
Lagrangians in the following subsection.

(2) Greek indices $\alpha_i$ denote the spin (or Lorentz group
representation) index of the field. Where convenient we use the
multi-index notation $\{\alpha\} =
\alpha_1\alpha_2\alpha_3\alpha_4$. Repeated multi-indices are summed
by summing over all four components. A similar convention is used for
colour indices, which we denote by Latin letters $\{a\}=a_1\dots
a_4$. We employ creation fields for the heavy final state particles
and destruction fields for the initial state. If, for example, $H'$ is
the antiparticle of $H$, then this convention implies that $\psi'$
transforms in the complex conjugate SU(3) representation of
$\psi$. Similarly, an initial-state antiquark transforms in the
antifundamental representation. In effective-theory calculations it
is conventional to decompose operators into a complete basis in spin
and colour, making operators and coefficient functions
``scalars''. Here we prefer to work with operators and coefficient
functions that carry open spin and colour indices, see
(\ref{eq:opdef}). This turns out to be convenient for the spin
indices, since the soft gluon contributions are spin-independent, as
well as the Coulomb contributions needed at NLL accuracy.  Thus, it is
more direct to perform the spin summations on the square of the
amplitude as in standard unpolarized cross section calculations
instead of decomposing the amplitude.  At NNLL accuracy the 
spin-dependence of the potential corrections has to be taken into
account in order to obtain all terms of the order
$\alpha_s^2\ln\beta$. For this purpose a spin decomposition of 
the potential function is introduced in section~\ref{sec:factorization}.
As concerns colour, given the representations $r, r^\prime$ and $R,
R^\prime$ of the initial and final state particles, 
we can introduce a set of independent colour matrices
$c^{(i)}$ that form an orthonormal basis 
of invariant tensors in the representation $r\otimes r'\otimes
\bar R\otimes\bar R'$.
The colour structure of the Wilson coefficients can be decomposed in
this basis according to
\begin{equation}
\label{eq:colour-wilson}
C^{(\ell)}_{\{a;\alpha\}}=
\sum_i c^{(i)}_{\{a\}} C^{(\ell, i)}_{\{\alpha\}}
\end{equation}
This allows us to write
\begin{equation}
C^{(\ell)}_{\{a;\alpha\}}(\mu)\,
\mathcal{O}^{(\ell)}_{\{a;\alpha\}}(\mu) 
= \sum_i C^{(\ell,i)}_{\{\alpha\}}(\mu)\,
\Big[c^{(i)}_{\{a\}}\mathcal{O}^{(\ell)}_{\{a;\alpha\}}(\mu)\Big]
\equiv \sum_i C^{(\ell,i)}_{\{\alpha\}}(\mu)\,
\mathcal{O}^{(\ell,i)}_{\{\alpha\}}(\mu)
\end{equation}
in (\ref{eq:matrix}).  For example, in a $3+\bar 3\to 3+\bar 3$
scattering process (such as $q\bar q\to t\bar t$ and 
squark-antisquark production), a convenient colour basis 
is~\cite{Kidonakis:1997gm}
\begin{equation}
\label{eq:basis_33}
\begin{aligned}
c^{(1)}_{\{a\}} &= \frac{1}{N_c} \delta_{a_2a_1} \delta_{a_3a_4}  
\\
c^{(2)}_{\{a\}} &=\frac{2}{\sqrt{D_A}} \,T^b_{a_2a_1} T^b_{a_3a_4} \, ,
\end{aligned}
\end{equation}
with $D_A=N_c^2-1$, expressing colour conservation and the fact that the
initial and final states can be in a colour-singlet or a colour-octet state.
Suitable colour bases for the different colour scattering processes have been
constructed in~\cite{Beneke:2009rj} where also more details on the properties
of the basis tensors can be found.

(3)  In (\ref{eq:matrix}) the argument $\mu$ denotes the
factorization scale dependence of the operators and hard functions. We
drop this in the following, but note that exploiting the invariance 
of the physical amplitude under changes of the factorization scale
will be a key element in deriving the resummation formula. The 
superscript $\ell$ stands for higher order terms in the expansion 
of the amplitude, which correspond to operators of the form 
(\ref{eq:opdef}) with derivatives acting on the fields, 
as discussed below.

(4) Fields without space-time arguments are at $x=0$. In general, 
the SCET representation of a scattering amplitude involves a
convolution of the Wilson coefficient with collinear fields evaluated 
at different positions along the light ray~\cite{Beneke:2002ph}, 
instead of a product with a local bilinear operator 
$\phi_{c;a_1\alpha_1}\phi_{\bar c;a_2\alpha_2}$
as in (\ref{eq:matrix}). However, since here there is only one collinear 
field of a given type in the operator, translation invariance 
may be used to rewrite the operator in a local
form~\cite{Beneke:2002ph,Becher:2003kh}. Applying the translation 
operators on the initial state turns the convolution integrals 
with the coefficient function into the Fourier-transform of the 
position-space coefficient functions to momentum-space coefficient 
functions. These then depend on the large momentum components of 
the partons, and are the natural objects to work with.  The hard 
functions $C^{(\ell)}_{\{a;\alpha\}}$ in (\ref{eq:matrix}) refer 
to the momentum-space quantities.

Despite the somewhat complicated formalism and notation, there is 
a straightforward general prescription for the computation of the matching 
coefficients $C^{(\ell)}_{\{a;\alpha\}}$ in the 
expression~\eqref{eq:matrix}. One simply calculates the 
renormalized scattering amplitude for the partonic process 
$p(k_1)p^\prime(k_2)\to H(p_1)H^\prime(p_2)$ without any averages 
or sums over colour and spin states, and expands it in 
$\tilde p_{1,2}$ defined by~\eqref{eq:nr-mom} around the 
production threshold $(k_1+k_2)^2=(m_{H}+m_{H'})^2$.
Beyond tree level, the matching coefficients can be extracted directly
by expanding the loop integrand under the assumption that all loop 
momenta are hard \cite{Beneke:1997zp} and by performing the loop 
integration of the expanded integrand 
in dimensional regularization. As a special case,
the leading term in the kinematical expansion is found by calculating the full 
amplitude directly at threshold. In this case the loop corrections to 
the effective theory matrix element evaluated directly at threshold 
vanish since they are given by scaleless integrals. Therefore only
tree diagrams have to be considered on the effective theory side
of the matching calculation and the hard function is determined 
from 
\begin{equation}
\mathcal{A}(pp'\to HH^\prime)_{|\hat s=4 M^2} = \sum_i
\;C^{(0)}_{\{a;\alpha\}}\;
\langle H H'|\mathcal{O}^{(0)}_{\{a;\alpha\}} |
p p'\rangle_{\rm EFT \,\, tree} \,.
\label{eq:match}
\end{equation} 
Since the matrix element on the right-hand side is simply a product of
spinors (polarization vectors) and the operator renormalization 
factor, the $\overline{\rm MS}$ scheme coefficient function 
$C^{(0)}_{\{a;\alpha\}}$ equals the scattering amplitude at 
threshold with the spinors (polarization
vectors) stripped off, and all $1/\epsilon$ poles set to zero.
The Born-level matching required for squark-antisquark production is 
discussed explicitly in section~\ref{sec:squarks}.

The leading term in the kinematical expansion describes the production
of $HH^\prime$ in an $S$-wave state. The next term is related to 
the terms linear in $\tilde p_{1,2}$, more precisely their components 
orthogonal to $w^\mu$, denoted by the symbol $\top$, 
since the other component is of higher order in the potential 
region. The $P$-wave production operators corresponding to this term 
are of the form 
(\ref{eq:opdef}) with the replacement
\begin{equation}
\label{eq:pwave}
\psi^\dagger_{a_3\alpha_3}\psi^{\prime \,\dagger}_{a_4\alpha_4}
\to  \psi^\dagger_{a_3\alpha_3}
\left(-\frac{i}{2} \overleftrightarrow D^\mu_\top\right)
\psi^{\prime \,\dagger}_{a_4\alpha_4} 
\equiv 
-\frac{i}{2} \left(\psi^\dagger_{a_3\alpha_3}
[\overrightarrow D^\mu_\top \psi^{\prime \,\dagger}]_{a_4\alpha_4} 
- [\psi^\dagger\overleftarrow D^\mu_\top]_{a_3\alpha_3}
\psi^{\prime \,\dagger}_{a_4\alpha_4}
\right)
\end{equation}
in the non-relativistic part of the operator, with the covariant
derivative $D^\mu$ in the appropriate colour representation. The
series in $\ell$ in (\ref{eq:matrix}) accounts for the threshold
expansion in powers of $\lambda$. In the following, we will mostly
consider only the leading term $\ell=0$. In particular, the
factorization formula we derive is valid only for the case that the
Born cross section is dominated by $S$-wave production.

\subsection{Effective Lagrangians}

\subsubsection{Non-relativistic}

The propagation and interactions of the  non-relativistic fields $\psi$ and
$\psi'$ are described by the potential NRQCD Lagrangian
\cite{Pineda:1997bj,Beneke:1998jj,Brambilla:1999xf,Beneke:1999zr}.
That is, we assume that the modes with momentum $k\sim
M\sqrt{\lambda}$ and potential gluons have already been integrated
out. At leading order, in the partonic cms frame, the effective 
Lagrangian, including the decay widths of the heavy particles, is given by
\begin{equation}
\begin{aligned}
\label{eq:NRQCD}
\mathcal{L}_{\mbox{\tiny PNRQCD}} = \,\,&
\psi^\dagger \left(i D_s^0+\frac{\vec{\partial}^2}{2 m_H}
 +\frac{i\Gamma_{H}}{2}\right) \psi
+\psi^{\prime\,\dagger} \left(i D_s^0+\frac{\vec{\partial}^2}{2 m_{H'}}
 +\frac{i\Gamma_{H^\prime}}{2}\right) \psi'
\\
&+\int d^3 \vec{r}\, 
\left[\psi^{\dagger}{\bf T}^{(R)a} \psi\right]\!(\vec r\,)
\left(\frac{\alpha_s}{r}\right)
\left [\psi^{\prime\,\dagger}{\bf T}^{(R')a}\psi^\prime\right]\!(0)
\,.
\end{aligned}
\end{equation}
Here ${\bf T}^{(R)}$ are the $SU(3)$ generators in the appropriate
representation of the heavy particle. In this convention $\psi'$
transforms in the antifundamental representation, if $H'$ is an
antiparticle in the fundamental representation. This deviates from
the standard NRQCD notation for quark-antiquark production that
employs a field $\chi$ that creates an antiquark and transforms in the
fundamental representation. Our notation is convenient since it
accommodates the case where the heavy particles produced are not
particle-antiparticle pairs. This convention leads to a different sign
of the potential term compared to the conventional formulation. 

The only relevant interactions in~\eqref{eq:NRQCD} are the exchange
of Coulomb (potential) gluons, rewritten as an instantaneous, but
spatially non-local operator in the second line, and an interaction
with the zero component (that is, $w\cdot A_s$, in a general frame)
of the soft gluon field in the soft covariant derivative
$iD_s^0\psi(x)=(i\partial^0 +g_s {\bf T}^{(R)a}A^{a0}(x_0))\psi(x)$, and the
analogous expression for $\psi'$ with the generator in the
representation $R'$.  Using the canonically normalized NRQCD
Lagrangian~\eqref{eq:NRQCD}, one has to take into account a 
normalization factor
$\sqrt{2E_H}\equiv(2 m_H (1+\vec{\tilde p}^{\,2}/m_H^2))^{1/2}=\sqrt{2m_H}+\mathcal{O}(\lambda^2)$ in the definition of the
 state $\ket{H}$ (similarly for  $\ket{H'}$) on the NRQCD side of the matching
 relation.\footnote{For $\Gamma_H=0$. 
See \cite{Beneke:2004km} for the general case.}

The NRQCD Lagrangian~\eqref{eq:NRQCD} is invariant under soft gauge transformations 
in the appropriate representation:
\begin{equation}
\label{eq:soft-nrqcd}
  \psi(x)\to U^{(R)}_s(x_0)\psi(x)\quad,\qquad
  \psi'(x)\to U^{(R')}_s(x_0)\psi'(x).
\end{equation} 
The soft gauge transformation depends only on $x_0$ since soft fields 
have to be multipole expanded when multiplied with potential fields 
in order to maintain a uniform 
power counting~\cite{Labelle:1996en,Grinstein:1997gv,Beneke:1999zr}.
The invariance of the Coulomb potential for arbitrary representations 
follows from the property
\begin{equation}
\label{eq:trafo-generator}
  U^{(R)\dagger} {\bf T}^{(R)a}U^{(R)}= U_{ab}^{(8)}{\bf T}^{(R)b}
\end{equation}
where $U^{(R)}$ is a $SU(3)$ transformation in the representation $R$ 
and ``8'' denotes the adjoint. 
Therefore under a soft gauge transformation:
\begin{equation}
  \left[\psi^{\dagger}{\bf T}^{(R)a} \psi\right]\!(x+\vec r\,)\Rightarrow
  \left[\psi^{\dagger}{\bf T}^{(R)b} \psi\right]\!(x+\vec r\,) 
U_{s,ab}^{(8)} (x_0)
\end{equation}
and analogously for the $\psi'$ fields. The gauge invariance of the
Coulomb potential term follows then from the fact that the
transformations in the adjoint representation are real so that
$U^{(8)}_{ab}U_{ac}^{(8)}=(U^{(8)\dagger} U^{(8)})_{bc}=\delta_{bc}$.
For this is essential that only the time component enters the soft
gauge transformation since the $\psi$ fields and the $\psi'$ fields in
the Coulomb potential are defined at different points in space but at the
same time.

\subsubsection{Soft-collinear}

The propagation and interactions of quark and gluon collinear modes with 
large momentum proportional to $n^\mu$ is described by the SCET 
Lagrangian \cite{Bauer:2000yr}. At leading order, in the 
position-space formalism of SCET \cite{Beneke:2002ph,Beneke:2002ni} 
it is given by
\begin{eqnarray}
{\cal L}_c &=&\bar{\xi}_c \left(in \cdot D +i\Slash{D}_{\perp c}\frac{1}{i
    \bar n \cdot D_c}
i\Slash{D}_{\perp c}\right)\frac{\slash{\bar n}}{2}\,\xi_c 
 -\frac{1}{2}\,\mbox{tr}\left(F_c^{\mu\nu}
  F^c_{\mu\nu}\right) \,.
\label{lagrangian}
\end{eqnarray}
Here $\xi_c$ denotes the $n$-collinear quark field, which satisfies
$\slash n\xi_c=0$ and the projection identity $(\slash n\slash{\bar
  n}/4)\xi_c=\xi_c$.  The covariant derivatives are defined as $iD_c =
i\partial + g_s A_c$ with $A_c$ the matrix-valued gluon field in the
fundamental representation. The covariant derivative $D$ without subscript, 
however, contains both the
collinear and soft gluon field. The quantity $F_{\mu\nu}^c$ is the
field-strength tensor built from the collinear gauge field in the
usual way, except that $n_- \cdot D$ rather than $n_- \cdot D_c$
appears \cite{Beneke:2002ni}.  The collinear and soft fields are
evaluated at $x$, but in products of soft and collinear fields the
soft fields are evaluated at $x_+^\mu=(\bar n\cdot x/2)\,n^\mu \equiv
x_- n^\mu$, according to the multipole expansion.  In this notation
$x_-$ is a scalar, while $x_+^\mu$ is a vector.  The $\bar
n$-collinear fields are described by an identical Lagrangian with the
roles of $n$ and $\bar n$ interchanged. The corresponding quark field
satisfies $\slash {\bar n}\xi_{\bar c}=0$ and $(\slash {\bar n}\slash
n/4)\xi_{\bar c}=\xi_{\bar c}$. The two collinear sectors couple only
via soft gluon interactions.

It is convenient to express the collinear part of the 
operators (\ref{eq:opdef}) in 
terms of fields that are invariant under the collinear gauge
transformation as defined in \cite{Beneke:2002ni}. Let 
$W_{c}$ be the collinear Wilson line in the $\bar n$ direction, 
\begin{equation}
W_{c}(x) =
\mbox{P} \exp \left[i g_s \int_{-\infty}^0 d t 
\, \bar n \cdot A_c(x+\bar n t) \right].
\end{equation}
Then $\phi_c$ in (\ref{eq:opdef}) is given by the 
combinations
\begin{equation}
\label{eq:quark-scet}
W^\dagger_{c}  \xi_{c}(x)\;,\quad 
\bar{\xi}_{c}W_{c}(x)
\end{equation}
for the quark and antiquark initial state, respectively. 
Note that our conventions imply that the conjugate fields $\phi_c^\dagger$ are 
$\xi_c^\dagger W_c=\bar\xi_c\gamma^0W_c$ for the quark and $W_c^\dagger\gamma^0 \xi_c$ for the antiquark initial state.
For the gluon initial state we use the following definition that is 
 invariant under the collinear gauge transformations~\cite{Hill:2002vw}:
\begin{equation}
\label{eq:gluon-scet}
\mathcal{A}_c^{\perp \mu}(x)= g_s^{-1} (W^\dagger_{c}
[i D_c^{\mu_\perp}W_{c}])(x)=
\int_{-\infty}^0 ds \, \bar n_\nu 
\left(W^\dagger_{c} F_c^{\nu\mu_\perp} W_c\right)(x+s\bar n) \, .
\end{equation}
The anticollinear fields are defined 
similarly with $\bar n$ replaced by $n$. 
These collinear operators transform under the soft 
gauge symmetry of the SCET Lagrangian \cite{Beneke:2002ni} according to
\begin{equation}
\label{eq:soft-scet}
\begin{aligned}
  (W^\dagger_{c}  \xi_{c})(x)&\to U^{(3)}_s(x_-)(W^\dagger_{c}  \xi_{c})(x)\\
  \mathcal{A}^a_c(x) T^a & 
  \to U^{(3)}_s(x_-) \mathcal{A}^a_c(x) T^a U^{(3)\dagger}_s(x_-)
  =U^{(8)}_{ab}(x_-)\mathcal{A}^b_c(x) T^a,
\end{aligned}
\end{equation}
where $U_s^{(R)}(x_-)$ is a soft gauge transformation in the representation
$R$ at the point $x_- n^\mu$.

\subsection{Decoupling of soft gluons}
\label{sec:decouple}

Since the collinear fields $\phi_c$, anticollinear fields $\phi_{\bar
  c}$ and the potential fields $\psi^\dagger$ in the production
operators~\eqref{eq:opdef} interact with each other only via exchange
of soft gluons, an essential step in deriving a factorization of the
hadronic cross section 
is the decoupling of soft gluons from the collinear and
potential degrees of freedom.  In SCET it is well known that the soft
gluons can be decoupled from the collinear fields at leading power by
performing field redefinitions involving Wilson
lines~\cite{Bauer:2001yt}. Here we will show that an analogous
transformation also decouples the soft gluons from the potential
fields at leading order in the non-relativistic expansion.  As a
result, the production operators factorize into products of
non-interacting collinear, anticollinear, potential and soft
contributions.

In order to decouple the soft gluons from the collinear and
anticollinear fields, the required redefinitions are familiar from
the derivation of factorization formulas in
SCET~\cite{Bauer:2001yt,Bauer:2002nz}
\begin{align}
\xi_{c}(x) &= S^{(3)}_{n}(x_-) \xi^{(0)}_{c}(x)\nonumber\\
\bar \xi_{c}(x) &=  \bar\xi^{(0)}_{c}(x)S^{(3)\dagger}_{n}(x_-)
= \bar \xi^{(0)}_{c}(x)S^{(\bar 3)T}_{n}(x_-)\nonumber\\
A_c^{a,\mu}(x)&=S^{(8)}_{n,ab}(x_-)A_c^{(0)b,\mu} 
\end{align}
with the soft Wilson lines $S^{(R)}_n$ for a particle in the representation 
$R$ of $SU(3)$ given by 
\begin{equation}
S^{(R)}_{n}(x) = \mbox{P} \exp 
\left[i g_s \int_{-\infty}^0 d t \,n \cdot  A^c_s(x+n t){\bf T}^{(R)c}\right]
\end{equation}
and similarly for particles collinear to $\bar{n}$.
Consistent with our treatment of incoming antiparticles as particles
in the complex conjugate representations, we have rewritten the
 transformation of the conjugate collinear spinor as a transformation 
in the antifundamental representation
with the generators $ {\bf T}^{(\bar 3)}=-T^T$.

These transformations induce analogous transformations on the collinear fields $\phi_c\in\{W_c^\dagger\xi_c, \bar\xi_c W_c, \mathcal{A}_c^\perp\}$ entering the production operators:
\begin{equation}
  \phi_{c;a\alpha}(x)=S^{(r)}_{n,ab}(x_-)\phi_{c;b\alpha}^{(0)}(x)
\label{eq:decouple-gluon}
\end{equation}
as follows from the transformation
\begin{equation}
  W_c(x)\to S^{(3)}_{n}(x_-)  W^{(0)}_c(x)S^{(3)\dagger}_{n}(x_-)
\end{equation}
of the collinear Wilson line. For the incoming parton in the 
anticollinear direction we exchange
$n$ by $\bar n$, $x_-$ by $x_+$ in the argument of the Wilson lines, 
and the representation $r$ by $r'$. In the expressions
with the superscript $(0)$ all collinear fields are replaced by the
decoupled fields $A_c^{(0)}$ and $\xi_c^{(0)}$ and the soft gluon
fields are set to zero.

Turning to the non-relativistic sector, we first note that from 
the diagrammatic perspective the coupling between the 
non-relativistic and soft fields is non-trivial, since the
non-relativistic energy of the heavy fields and the energy of soft
gluons is of the same order $M\lambda\sim M\beta^2$. Thus, soft 
gluons can attach to the non-relativistic lines, to and in between the 
Coulomb ladder rungs, leaving potential lines in the potential 
region. However, a redefinition of the non-relativistic fields 
by a time-like Wilson line decouples the soft gluons also from the potential 
fields at leading order in the non-relativistic expansion. 
Namely, the interaction with soft gluons in the PNRQCD Lagrangian
given in~\eqref{eq:NRQCD} can be eliminated through
\begin{align}
\label{eq:Sw}
\psi_a(x) &= S^{(R)}_{w,a b}(x_0) \psi^{(0)}_b(x)\\
\psi^\dagger_a(x) &= {\psi}^{(0)\dagger}_b(x)S^{(R)\dagger}_{w,ba}(x_0)\, ,
\label{eq:decouple-nr}
\end{align}
where the soft Wilson lines $S^{(R)}_w$ are defined as\footnote{ Note
  that in contrast to~\eqref{eq:decouple-gluon} we do not use a field
  redefinition of the non-relativistic annihilation fields $\psi$ with
  Wilson lines extending from $-\infty$ to $x_0$ but (\ref{eq:Sw}),
  since the former 
  definition would obscure the decoupling of the soft gluons in
  the PNRQCD Lagrangian. As discussed
  in~\cite{Arnesen:2005nk}, the two different forms of the redefinition 
  are equivalent if one takes an appropriate
  phase for external non-relativistic in-states into account. Since we
  encounter only external outgoing non-relativistic particles, this
  subtlety is irrelevant in the present context.  }
\begin{align}
S^{(R)}_{w}(x) &=
\overline{\mbox{P}}\exp \left[-i g_s \int_0^{\infty} d t \; 
w\cdot A^c_s(x+w t)  \,{\bf T}^{(R)c} \right]\\
S^{(R)\dagger}_{w}(x) &=
\mbox{P}
 \exp \left[i g_s \int_{0}^\infty d t\; w \cdot A^c_s(x+w t)\,
   {\bf T}^{(R)c}\right]
\end{align}
and analogously for the primed field transforming under the
representation $R'$.  This transformation eliminates the interaction
contained in the soft covariant derivative $D_s^0$ in
(\ref{eq:NRQCD}), since 
\begin{equation}
S^{(R)\dagger}_{w} iD_s^0 S^{(R)}_{w} = i\partial^0,
\end{equation}
but at first 
sight introduces the soft Wilson lines into the Coulomb potential 
interaction. However, since the
identity~\eqref{eq:trafo-generator} also holds if the group elements
$U$ are the Wilson lines $S_w$~\cite{Beneke:2009rj}, and since the $S_w$ in
the transformations~\eqref{eq:Sw}, \eqref{eq:decouple-nr} only depends 
on the time
coordinate, the soft gluon Wilson lines drop out from the Coulomb
potential term expressed in terms of the redefined fields. 
In other words, in terms of the redefined fields 
the \mbox{PNRQCD} Lagrangian takes exactly the same form as (\ref{eq:NRQCD}) 
except that $D_s^0 \to \partial^0$, so that the Lagrangian is 
independent of the soft-gluon field.
At the next order in the velocity expansion, the \mbox{PNRQCD}
Lagrangian contains an interaction involving the gauge invariant
chromoelectric operator $\vec x\cdot \vec
E_s(t,0)$\cite{Voloshin:1978hc,Pineda:1997bj} that is not removed by
the above transformation so the decoupling of soft and potential modes
is only valid at leading order in the non-relativistic
expansion.\footnote{In order to achieve factorization of soft gluons
  in a kinematical regime where relativistic corrections become
  important one can describe the heavy fields by two copies of heavy
  quark effective theory with different velocities $w_1$ and
  $w_2$, see also~\cite{Ahrens:2010zv}. Performing separate field redefinitions
  for $H$ and $H'$ with Wilson lines defined in terms of the two
  velocities decouples the soft gluons up to corrections of the order
  $1/m_{H}, 1/m_{H^\prime}$.}  We demonstrate in 
  section~\ref{sec:comments}
that the corrections to the total cross section from a single
insertion of this operator vanish to all orders.

Assembling the different contributions shows that the
operators~\eqref{eq:opdef} for $S$-wave production
($\ell=0$) factorize into an expression of the form
\begin{equation}
\label{eq:opdef-dec}
\mathcal{O}^{(0,i)}_{\{a;\alpha\}}(x,\mu) = \mathcal{S}^{(i)}_{\{j\}}(x)
\Big[\phi^{(0)}_{c;j_1\alpha_1}\phi^{(0)}_{\bar c;j_2\alpha_2} 
\psi^{(0)\dagger}_{j_3\alpha_3}\psi^{\prime (0)\,\dagger}_{j_4\alpha_4}
\Big](x,\mu) \,  ,
\end{equation}
 with the universal soft contribution
\begin{equation}
\label{eq:soft-amp}
\mathcal{S}^{(i)}_{\{j\}}(x)= c^{(i)}_{\{a\}} S_{n,a_1j_1}(x_-)
S_{\bar n,a_2j_2}(x_+) S_{w,j_3a_3}^\dagger(x_0) S^\dagger_{w,j_4a_4}(x_0) \, .
\end{equation}
Here and in the following we suppress the representation labels on the
soft Wilson lines, and take it as understood that the Wilson lines
always are in the representations appropriate for the respective
particles.  The decoupled fields $\psi^{(0)\dagger}$, $\phi^{(0)}_c$
and $\phi^{(0)}_{\bar c}$ in~\eqref{eq:opdef-dec}
do not interact with the soft Wilson lines and with each other at
leading power in SCET and at leading order in the non-relativistic
expansion.

Near the partonic threshold only soft radiation can occur in the
final state so the state $\ket{X}$ is free of collinear radiation.
Therefore the Fock space is a direct product of potential, soft, collinear
and anticollinear contributions,
\begin{equation}
\ket{pp'}=\ket{p}_{c}\otimes\ket{p'}_{\bar c}\otimes\ket{0}_s\otimes \ket{0}_p
\,,\qquad
\ket{HH'X}=\ket{0}_{c}\otimes\ket{0}_{\bar c}\otimes 
\ket{X}_{s}\otimes \ket{HH'}_p\,,
\end{equation}
and we obtain the scattering amplitude in the factorized form:
\begin{align}\label{eq:matrix_element}
\mathcal{A}(pp'\to HH'X) = &\sum_i
\;C^{(0,i)}_{\{\alpha\}}(\mu)\;
\braket{ X|\mathcal{S}^{(i)}_{\{j\}}(0)| 0}\,
 \langle HH' |\psi^{(0)\dagger}_{j_3\alpha_3}\psi^{\prime (0)\,\dagger}_{j_4\alpha_4}
|0\rangle \nonumber\\
&\times \braket{0|\phi^{(0)}_{c;j_1\alpha_1}|p}
\braket{0|\phi^{(0)}_{\bar c;j_2\alpha_2} |p'}\, .
\end{align}
As remarked above, soft-potential and soft-collinear factorization is 
only valid at leading order in the non-relativistic and collinear 
expansion.  Similarly for a
$P$-wave production operator~\eqref{eq:pwave} the
field redefinitions~\eqref{eq:decouple-nr} do not eliminate the
soft gluons in the covariant
derivative.

\subsection{Factorization formula for the cross section near threshold}
\label{sec:factorization}

We now show that the total partonic cross section for the production
of a heavy-particle pair near the production threshold factorizes into
hard, potential and soft contributions as sketched
in~\eqref{eq:factor-sketch}.  To this end, we consider the cross
section for the production of a heavy-particle pair near threshold in an
$S$-wave from the partonic process~\eqref{eq:subprocess} 
for on-shell, massless initial state partons $p$ and $p'$ with
zero transverse momentum. 
 In this case only the hard, soft, collinear
and anticollinear modes~\eqref{eq:regions} contribute to the cross
section since no smaller scale related to $\Lambda_{\text QCD}$,
parton masses or off-shellness is introduced and the modes included in
the effective theory introduced in section~\ref{sec:matching} are
sufficient. We can therefore use the expression for the 
amplitude~\eqref{eq:matrix_element}
 in terms of decoupled collinear and potential fields and soft Wilson
 lines to compute the cross section
as usual  by squaring the scattering amplitude, averaging over initial
state and summing over final state polarizations and colours, and
integrating over the final-state phase space:
\begin{equation} \label{eq:pp-cross}
\sigma_{pp'}(s)
= \frac{1}{2 s} \int d \Phi_1 d \Phi_{2} d\Phi_X  \sum_{\mbox{\tiny
    pol}} 
\sum_{\mbox{\tiny colour}} |\mathcal{A}(pp'\to HH'X)|^2
(2 \pi)^4 \delta^{(4)}(p_1+p_2+P_X-P)
\end{equation}
Here $P$ is the total incoming momentum and $P_X$ is the total momentum of the
hadronic state $X$.  We suppress the normalization factors for the
initial-state averages. We do not aim here at a proof of the
factorization of the hadronic cross section, that would also require
the treatment of modes related to the QCD scale, as in the case of
deep inelastic scattering for $x\to 1$ in~\cite{Becher:2006mr}. 
Rather, we are concerned with the dimensionally regularized 
partonic cross section (\ref{eq:pp-cross}) and its short-distance 
counterpart with the initial-state infrared divergences subtracted.

The factorized form~\eqref{eq:matrix_element} of the scattering
amplitude holds if the expectation value of the EFT operator is
evaluated with the leading order PNRQCD and SCET Lagrangians. This is
a priori appropriate for NLL accuracy in the counting~\eqref{eq:syst}.
Since according to (\ref{eq:syst}) the NNLL approximation includes 
${\cal O}(\beta)$ corrections, subleading soft interactions in the
effective Lagrangian that are not removed by the decoupling
transformations could contribute to the cross section starting from 
NNLL. At fixed-order NNLO accuracy it was shown
in~\cite{Beneke:2009ye} that these corrections do not, in
fact, contribute to the total cross section and the
only corrections not generated by the leading order effective
Lagrangian are the one-loop hard function and NLO Coulomb and
non-Coulomb potentials. These considerations will be elaborated on 
in section~\ref{sec:comments} and extended 
to all orders in the coupling expansion. 
Therefore the derivation of the simple factorization of
the form~\eqref{eq:factor-sketch} given in the following holds at NNLL
accuracy for production processes dominated by $S$-wave production. 

As for the hadronic cross
section~\eqref{eq:parton-model}, standard QCD factorization implies
that the cross section for the parton initial states factorizes
according to
\begin{equation}
\label{eq:pp-factor}
\sigma_{pp'}(s) =
\sum_{\tilde p,\tilde p'} 
 \int d x_1 d x_2\,f_{p/\tilde p}(x_1,\mu) f_{p'/\tilde p'}(x_2,\mu)\,
\hat\sigma_{\tilde p\tilde p'}(x_1x_2s, \mu), 
\end{equation}
where the parton distribution functions of parton $\tilde p$ in parton
$p$, $f_{p/\tilde p}(x,\mu)$, contain the collinear
modes. We will consider this cross section near the production
threshold, $s\approx 4M^2$, and show that the short-distance cross
section $\hat\sigma_{\tilde p\tilde p'}$ itself factorizes into hard,
soft and potential contributions as sketched
in~\eqref{eq:factor-sketch}.  At threshold, the PDFs
in~\eqref{eq:pp-factor} necessarily appear in the limit $x\to 1$, where
the flavour off-diagonal PDFs are suppressed, so it is sufficient to
consider the case $p=\tilde p$, $p'=\tilde p'$.  This is consistent
with the fact that, at leading power, we only have to consider EFT
matrix elements
in~\eqref{eq:matrix} where the collinear (anticollinear) field
$\phi_c$ ($\phi_{\bar c}$) annihilates the state $\ket{p}$
($\ket{p'}$). According to QCD
factorization, $\hat\sigma_{\tilde p \tilde p'}$ is independent of the
initial states $p$, $p'$ 
so it is identical to the hard-scattering cross section
to be used in the hadronic cross
section~\eqref{eq:parton-model}. Therefore the factorized form of the
short-distance cross section~\eqref{eq:factor-sketch} derived in this
way can be used in the hadronic cross section in the \emph{partonic}
threshold region where $x_1x_2 s\sim 4M^2$. As discussed above, this
region is expected to give a sizable contribution to the total cross
section for heavy-particle masses $4M^2\sim 0.2 s$. For values of
$x_{1,2}$ outside the partonic threshold region, the factorized
formula is not necessarily a good approximation to the cross section 
and the resummed cross section should be matched to the fixed-order 
expansion as discussed further in section~\ref{sec:match}.

We now proceed with the derivation of the factorization formula which
follows similar steps as that for the Drell-Yan process
in~\cite{Becher:2007ty}. The new ingredients here are the presence of
the potential fields and additional Wilson lines for the heavy
coloured particles, resulting in a  more involved colour and spin structure.
After inserting the factorized matrix
element~\eqref{eq:matrix_element} into the standard expression for the
cross section~\eqref{eq:pp-cross}, we use an integral representation
of the $\delta$ function,
\begin{equation}
(2 \pi)^4 \delta^{(4)}(p_1+p_2+P_X-P)=
\int d^4 z \,e^{-i z \cdot(p_1+p_2+P_X-P)}.
\end{equation}
Rewriting the exponential in terms of the momentum operator acting on
the external states, we can translate the collinear, potential and
soft operators in the conjugate matrix element to $z$, e.g.
$e^{iz\cdot
  k_1}\braket{p(k_1)|\phi^{(0)\dagger}_{c}(0)|0}=\braket{p(k_1)|e^{iz\cdot
    \hat P}\phi^{(0)\dagger}_{c}(0)e^{-iz\cdot \hat
    P}|0}=\braket{p(k_1)|\phi^{(0)\dagger}_{c}(z)|0}$.  
We then obtain for the cross section:
\begin{eqnarray}\label{eq:cross}
\sigma_{pp'}(s) &\!\!=\!\!& 
\frac{1}{2s} \sum_{\mbox{\tiny pol}}\sum_{\mbox{\tiny colour}}\sum_{i,i'}
C_{\{\alpha\}}^{(0,i)}C_{\{\beta\}}^{(0,i')*}\!
\int d^4 z  \,e^{- 2 i M w\cdot z}
 J_{k_3k_4j_3j_4}^{\beta_3\beta_4\alpha_3\alpha_4 }(z)
\nonumber\\[0.2cm]
&&\times\,
\braket{p|\phi^{(0)\dagger}_{c;k_1\beta_1}(z)|0}
\braket{p'|\phi^{(0)\dagger}_{\bar c;k_2\beta_2}(z)|0}
\braket{0|
\phi^{(0)}_{c;j_1\alpha_1}(0)|p}\braket{0|\phi^{(0)}_{\bar c;j_2\alpha_2}(0) |p'} 
\nonumber\\[0.2cm]
&& \times 
\int d\Phi_X \braket{0|\mathcal{S}^{(i')*}_{\{k\}}(z) |X}
\braket{ X| \mathcal{S}^{(i)}_{\{j\}}(0)|0} \, .
\end{eqnarray}
Here we introduced the definition of the \emph{potential function} 
in terms of the PNRQCD matrix elements: 
\begin{equation}
J_{\{a\}}^{\{\alpha\}}(z)
=\int d\Phi_1 d\Phi_{2}
\sum_{\mbox{\tiny pol},\mbox{\tiny colour}}
\braket{0|[\psi_{a_2\alpha_2}^{'(0)} \psi_{a_1\alpha_1}^{(0)} ](z) |HH' }
\braket{HH'| [\psi_{a_3\alpha_3}^{(0)\dagger} 
\psi_{a_4\alpha_4}^{'(0)\dagger}](0)|0}.
\end{equation}
The factor $e^{-2i M w\cdot z}=e^{-i (m_H+m_{H'}) w\cdot z}$ arises 
because, by definition, the non-relativistic fields depend
only on the residual momentum so that
 $\psi(z) = e^{i z \cdot\hat P}\psi(0)e^{-iz\cdot \hat P}=
e^{-i z\cdot \tilde p}\psi(0)$.
We can introduce a colour basis for the potential function $J$ by 
introducing projectors $P^{R_{\alpha}}$ onto the
irreducible representations $R_{\alpha}$ appearing in the
decomposition  of the final-state representations,
$R\otimes R'=\sum_{R_\alpha} R_{\alpha}$:
\begin{equation}
\label{eq:potential_colour}
J^{\{\alpha\}}_{\{a\}}
=\sum_{R_{\alpha}} P^{R_{\alpha}}_{\{a\}} J^{\{\alpha\}}_{R_{\alpha}} \, .
\end{equation}
A systematic procedure for the construction of the projectors for
arbitrary representations out of Clebsch-Gordan coefficients for the
coupling $R\otimes R'\to R_\alpha$ is reviewed
in~\cite{Beneke:2009rj}
where
the projectors needed for all production processes of pairs of
coloured SUSY particles are collected and further properties of the projectors
can be found.

Since the leading order PNRQCD Lagrangian is spin independent, the
spin structure of the leading order potential function (resumming
multiple Coulomb exchange) is trivial:
\begin{equation}
\label{eq:spin-trivial}
 J_{R_\alpha}^{\{\alpha\}}(x)=
\delta_{\alpha_1\alpha_3}\delta_{\alpha_2\alpha_4}\; J_{R_\alpha}(x).
\end{equation}
However, as mentioned above, spin-dependent non-Coulomb potentials become
relevant starting from NNLL accuracy in the combined counting in
$\alpha_s\ln\beta$ and $\alpha_s/\beta$ defined
in~\eqref{eq:syst}. Therefore we introduce a decomposition of the
potential function into different spin states:
\begin{equation}
J_{R_\alpha}^{\{\alpha\}}(x)=\sum_{S=|s-s'|}^{s+s'}
\Pi_S^{\{\alpha\}}\; J^{S}_{R_\alpha}(x).
\end{equation}
Here $s$ ($s'$) is the spin of the heavy particle $P$ ($P'$). The projectors
$\Pi_S$ project on the $H H^\prime$ state with total spin $S$. 
They satisfy the completeness relation
\begin{equation}
\sum_{S=|s-s'|}^{s+s'}\Pi^S_{\{\alpha\}}=
\delta_{\alpha_1\alpha_3}\delta_{\alpha_2\alpha_4}\,.
  \end{equation}
As an example, for two spin
$\frac{1}{2}$ particles, the projectors on singlet and triplet states
read in the cms frame:
\begin{equation}
\label{eq:projectors-spin}
\begin{aligned}
  \Pi_{0}^{\{\alpha\}}&=\frac{1}{2}\delta_{\alpha_2\alpha_1}
  \delta_{\alpha_3\alpha_4},\\
\Pi_{1}^{\{\alpha\}}&=
\frac{1}{2}\sigma^i_{\alpha_2\alpha_1}\sigma^i_{\alpha_3\alpha_4}.
\end{aligned}
\end{equation}
The explicit expression for the function $J$ is given 
in section~\ref{sec:coulomb}.

In order to bring~\eqref{eq:cross} into the form of the standard
factorization formula~\eqref{eq:pp-factor} we have to identify the
collinear matrix elements with the PDFs which  requires to combine the
products of the matrix elements of the collinear fields into a single
expectation value (the same discussion applies for the anticollinear
fields).  This can be achieved by recalling that due to the threshold
kinematics collinear radiation from the initial state is inhibited. We
can therefore formally sum over the states of the collinear final
state Fock space $\ket{C}$ since the only contribution comes from
the vacuum. After subsequent use of the completeness relation we can
simplify the product of collinear matrix elements to
\begin{align}
\braket{p|\phi^{(0)\dagger}_{c}(z)|0}
\braket{0|\phi^{(0)}_{c}(0) |p}
&\Rightarrow
\sum_C 
\int d\Phi_C \braket{p|\phi^{(0)\dagger}_{c}(z)|C}
\braket{C|\phi^{(0)}_{c}(0) |p}\nonumber\\
&= \braket{p|\phi^{(0)\dagger}_{c}(z)
\phi^{(0)}_{c}(0) |p} \, .
\end{align}
The same result for the collinear matrix element is obtained in the
standard collinear factorization away from threshold where the final state
contains collinear radiation and the sum over the collinear Fock space
is present from the beginning. We therefore define
\begin{equation}
\label{eq:pdf-general}
\braket{p(P_1)|\phi^{(0)\dagger}_{c;k_1\beta_1}(z)
    \phi^{(0)}_{c;j_1\alpha_1}(0)|p(P_1)} |_{\text{avg.}} =
  \delta_{k_1j_1} \int_0^1 \!\frac{dx_1}{x_1} 
  \,e^{i x_1 (z\cdot P_1)}\, N^{p}_{\alpha_1\beta_1}(x_1 P_1)
  \,f_{p/\tilde p}(x_1,\mu), 
 \end{equation}
where $\tilde p$ is the parton annihilated
by the field $\phi$ and the average refers to colour and
polarization. We show in appendix~\ref{app:pdfs} that 
the so defined functions $f_{p/\tilde p}(x_1,\mu)$ coincide 
with the standard definition of the quark, antiquark and gluon 
parton distribution functions. Since the momentum of the external state is 
$P_1^\mu = \bar n\cdot P_1\,n^\mu/2$, the matrix element depends on 
$z$ only through  $z_+=n\cdot z\propto z\cdot P_1$ since 
$z\cdot P_1 = (n\cdot z) (\bar n\cdot P_1)/2$, which has been used 
to represent it as a one-dimensional Fourier transform. This also 
allows us to replace $z^\mu$ by $(n\cdot z) \,\bar n^\mu/2$ on the 
left-hand side.
For all parton species, the prefactor is given by the helicity average of
the polarization wave functions $\chi_\alpha$ of the SCET fields
\begin{equation}
\label{eq:spin-sum}
  N_{\alpha\beta}^p(x_1 P_1)=\frac{1}{n_s D_r}
  \sum_\lambda \chi_\alpha^\lambda(x_1 P_1)\chi_\beta^{\lambda*}(x_1 P_1)\, ,
\end{equation}
where the sum extends over the physical polarization states. 
$D_r$ denotes the dimension of the SU(3) colour representation of the
initial state parton $p$ and the number of spin-degrees of freedom is
$n_s=2$ for all relevant cases. Note that the polarization wave
functions are evaluated for a particle with momentum $x_1 P_1$, 
which corresponds to the standard rule 
for calculating parton scattering cross sections. At the same 
time the factors $1/x_1$ (and $1/x_2$ for the other parton) 
from the definition 
(\ref{eq:pdf-general}) combine with the prefactor $1/(2 s)$ in 
(\ref{eq:cross}) to yield the standard flux factor $1/(2\hat s)$ 
for partonic scattering.

The soft matrix elements in the cross section~\eqref{eq:cross} can be
collected in the \emph{soft function}
\begin{equation}
\hat W^{R_\alpha}_{ii'}(z,\mu)=
\sum_X \int d\Phi_X \braket{0|\mathcal{S}^{(i')*}_{\{j_1j_2k_1k_2\}}(z) |X}
P^{R_\alpha}_{\{k\}}\braket{ X|\mathcal{S}^{(i)}_{\{j_1j_2k_3k_4\}}(0)|0},
\end{equation}
which is a matrix in the colour basis given by the $c^{(i)}_{\{a\}}$.
Since the collinear matrix elements are diagonal in colour and due to
the colour decomposition of the potential
function~\eqref{eq:potential_colour}, the initial-state colour indices
$j_1$, $j_2$ are summed over and the final state indices are projected
on the irreducible representations $R_\alpha$.  Since we are concerned
with the total cross section, the soft matrix elements can be combined
in a correlation function by summing over the hadronic final state and
using the completeness relation 
\begin{equation}
\sum_X \int d \Phi_X |X\rangle\;\langle X| = I 
\end{equation}
of the soft Hilbert space.
The soft function can then be written as 
\begin{equation}
\label{eq:project-soft-coulomb}
\hat W^{R_\alpha}_{ii'}(z,\mu)=  P^{R_\alpha}_{\{k\}}
c^{(i)}_{\{a\}} \hat W^{\{k\}}_{\{ab\}}(z,\mu)c^{(i')*}_{\{b\}}
\end{equation}
with the correlation function of soft Wilson lines
\begin{equation}
\label{eq:soft_general}
\hat W^{\{k\}}_{\{a,b\}}(z,\mu)=
\langle 0|\overline{\mbox{T}}[ 
S_{w,b_4 k_2} S_{w,b_3 k_1} S^\dagger_{\bar{n},jb_2} S^\dagger_{n, ib_1}](z)
\mbox{T}[S_{n,a_1i} S_{\bar{n},a_2j} S^\dagger_{w,k_3 a_3} 
S^\dagger_{w,k_4 a_4}](0)|0\rangle .
\end{equation}
In combining the soft matrix elements into a correlation function we
have introduced time- and anti-time-ordering symbols as discussed
in~\cite{Becher:2007ty}. In evaluating this correlation function, a
soft gluon propagator connecting fields in the time-ordered  and
anti-time-ordered products is given by a cut propagator, therefore this
prescription reproduces the usual rules for the real soft corrections.
A colour basis that diagonalizes
the soft matrices $W^{R_\alpha}_{ii'}$ to all orders in perturbation theory
has been constructed 
in~\cite{Beneke:2009rj} as reviewed in section~\ref{sec:colour-soft}
below. 

Inserting these definitions, the cross section becomes a convolution
of the PDFs with the potential and soft functions:
\begin{align}
\sigma_{p p' } &=
 \int d x_1 d x_2 \,f_{p/\tilde p}(x_1,\mu) f_{p'/\tilde p'}(x_2,\mu)\;
\sum_{i,i'}\sum_{S=|s-s'|}^{s+s'}H_{ii'}^{S}(\mu)
\nonumber\\
&\phantom{=}\times \int \frac{d^4 q}{(2 \pi)^4}\;
J_{R_\alpha}^{S}(q)\;
 \int d^4 z \,e^{i (x_1 P_{1}+x_2 P_{2} -2M w -q) \cdot z} \,
 \hat W^{R_\alpha}_{ii'}(z,\mu)\,,
\label{eq:factorization1}
\end{align}
where we have introduced the Fourier transform of the potential function
\begin{equation}
\label{eq:fourier-potential}
J_{R_\alpha}^{S}(q)
= \int d^4 z\, e^{iq\cdot z}\,
J_{R_\alpha}^{S}(z)
\end{equation}
and the spin and colour dependent \emph{hard functions} defined in 
terms of the
short-distance coefficients averaged over the initial state spins and
projected on the spin state of the heavy-particle system:
\begin{equation}
\label{eq:hard-colour}
\begin{aligned}
H_{ii'}^{S}(\mu)&=  \frac{1}{2 \hat s}
\left[ N^{p}_{\alpha_1\beta_1}(x_1 P_1) N^{p'}_{\alpha_2\beta_2}(x_2 P_2)
C_{\{\alpha\}}^{(0,i)}(\mu) \Pi_S^{\beta_3\beta_4\alpha_3\alpha_4}
 C_{\{\beta\}}^{(0,i')*}(\mu)\right]
\\
&=\frac{1}{2 \hat s}\frac{1}{4D_r D_{r'}}\sum_{\lambda_1\lambda_2}
(C_{\{\alpha\}}^{(0,i)}(\mu)\chi_{\alpha_1}^{\lambda_1}
\chi_{\alpha_2}^{\lambda_2})\Pi_S^{\beta_3\beta_4\alpha_3\alpha_4}
(C_{\{\beta\}}^{(0,i')}(\mu)\chi^{\lambda_1}_{\beta_1}
\chi_{\beta_2}^{\lambda_2})^\ast\,.
\end{aligned}
\end{equation}
In the second line we have inserted the prefactors~\eqref{eq:spin-sum}
(omitting the momentum argument of the polarization wave 
functions).\footnote{In (\ref{eq:hard-colour}) we could 
replace $\hat s$ by $4M^2$ to make the hard function
energy-independent. However, in the numerical evaluation it 
is trivial to keep the kinematic factor $1/(2\hat s)$ and this is 
the convention we adopt in Section~\ref{sec:squarks}.}
Summing over the spin states and using the completeness relation of
the spin projectors we obtain the spin-averaged hard function that 
is sufficient for NLL resummations:
\begin{equation}\label{eq:hard-spin-sum}
\begin{aligned}
  H_{ii'}(\mu)= \sum_{S=|s-s'|}^{s+s'}H_{ii'}^{S}(\mu)=& \frac{1}{2
    \hat s}
  \left[ N^{p}_{\alpha_1\beta_1}(x_1 P_1) N^{p'}_{\alpha_2\beta_2}(x_2 P_2)
    C_{\alpha_1\alpha_2\gamma\delta}^{(0,i)}(\mu)
    C_{\beta_1\beta_2\gamma\delta}^{(0,i')*}(\mu)\right]
\end{aligned}
\end{equation}
 A simple prescription for the computation of
the hard functions directly from the scattering amplitudes will be
given in section~\ref{sec:hard}.

Comparing~\eqref{eq:factorization1} to the factorization 
formula~\eqref{eq:pp-factor} we obtain the factorized expression for 
the short-distance cross section by stripping off the integrals 
with the parton distributions functions. The remaining expression 
can be further simplified in the
threshold region.  For notational simplicity, we will perform these
manipulations in the partonic cms frame where $w=(1,\vec 0\,)$. We first
shift the $q$ integral by introducing the new integration variable
$q'=(x_1 P_{1}+x_2 P_{2}-2M w-q)= (\sqrt{\hat s}-2M)w-q$ and obtain
\begin{equation}
\label{eq:factorization}
\hat \sigma_{p p' }(\hat s,\mu) =
\sum_{i,i'}\sum_{S=|s-s'|}^{s+s'}H_{ii'}^{S}(\mu)
\int \frac{d^4 q'}{(2 \pi)^4}\;
J_{R_\alpha}^{S}(E w-q')\;
 \int d^4 z \,e^{i q' \cdot z} \,
 \hat W^{R_\alpha}_{ii'}(z,\mu)\,.
\end{equation}
Here we defined the partonic centre-of-mass energy measured from
threshold, $E=\sqrt{\hat s}-2 M=M(1-z)+\mathcal{O}(\lambda^2)$. The
function $\hat W$ contains soft fields that, by definition, vary
significantly only on distances $z \sim 1/\lambda$.  Hence, only soft
momenta ($q' \sim \lambda$) contribute to the $q'$ integration.  On
the other hand the function $J$, which is defined in terms of
potential fields alone, is a function of $q=(\sqrt{\hat s}-2M) w-q'$.
In the partonic centre of mass frame we have by assumption $x_1 P_{1}
+x_2 P_{2}=\sqrt{\hat s}w= p_1+p_2+k_s=2M w +k_s^\prime$ with soft momenta
$k_s$, $k_s^{\prime}$, therefore $q$ is also soft.  Since potential fields with
scalings $(k_0,\vec k\,) \sim (\lambda,\sqrt{\lambda})$ can depend on
soft momenta only through their small time-like components, $J$ is a
function of $q_0=\sqrt{\hat s}-2M-q_0'$ alone. This enables us to
perform the $\vec{q}$ integration in~\eqref{eq:factorization} which
sets $\vec z=0$ in the argument of the soft function.  We then obtain
the final expression for
the factorized short-distance cross sections:
\begin{equation}
\label{eq:fact-final}
\hat \sigma_{pp'}(\hat s,\mu) = 
\sum_{S=|s-s'|}^{s+s'}
\sum_{i,i'}H_{ii'}^S(\mu)
\int d \omega\;
\sum_{R_\alpha}\,J_{R_\alpha}^S(E-\frac{\omega}{2})\,
W^{R_\alpha}_{ii'}(\omega,\mu).
\end{equation}
Here we defined the Fourier transform of the soft function
\begin{equation}
\label{eq:soft-fourier}
W^{R_\alpha}_{ii'}(\omega,\mu)=
\int \frac{d z_0}{4 \pi}
e^{i \omega z_0/2} \,\hat W^{R_\alpha}_{ii'}(z_0,\vec 0,\mu)\,.
\end{equation}
The formula~\eqref{eq:pp-factor}, with the partonic cross
section~\eqref{eq:fact-final}, establishes the factorization into 
collinear (the parton distribution functions), 
potential (the function $J$) and  soft (the soft function
$W$) contributions for heavy particles produced in an $S$-wave, and 
constitutes the main theoretical result of this work. 
Some further comments on the structure and validity range of the 
factorization formula as well as a comparison to previous results are 
given in section~\ref{sec:comments} below. The simplification of 
the colour sum over $i,i^\prime$ due to the existence of a diagonal 
basis~\cite{Beneke:2009rj} is reviewed in section~\ref{sec:colour-soft}.

\subsection{Some comments on the factorization formula}
\label{sec:comments}

\subsubsection{Gauge invariance}
Let us briefly comment on the gauge invariance of the ingredients in
the factorization formula. The hard function is defined in terms of
on-shell fixed-order scattering amplitudes projected on a given colour
and spin state. The gauge invariance
of the hard function then follows from the gauge invariance of the
on-shell scattering amplitude and the linear independence of the
elements of the colour and spin bases. The effective-theory Lagrangian
is invariant under the collinear gauge transformations of
SCET~\cite{Beneke:2002ni} and the soft gauge
transformations~\eqref{eq:soft-scet}
and~\eqref{eq:soft-nrqcd}. Collinear gauge invariance is built into
the formalism since the operators are constructed from the invariant
fields $\phi_c$.  The invariance of the components of the soft
function~\eqref{eq:soft-fourier} under soft gauge transformations
follows since the elements of the colour basis are invariant tensors
satisfying
\begin{equation}
\label{eq:gauge-basis}
c^{(i)}_{\{a\}}= U^{(R)\dagger}_{a_3b_3}  U^{(R')\dagger}_{a_4b_4} 
c^{(i)}_{\{b\}}  \, U^{(r)}_{b_1a_1} U^{(r')}_{b_2a_2}.
\end{equation}
Note that for this it is essential that only the soft
function with vanishing spatial argument, $W^{R_\alpha}_{ii'}(x_0,\vec 0,\mu)$,
enters the final factorization formula since for the soft function at 
arbitrary $x$ the gauge transformations of the collinear,
anticollinear and potential fields appear with different arguments
$x_-$, $x_+$
and $x_0$. Finally the collinear matrix elements and the potential function
are defined in terms of the decoupled fields that transform trivially under
soft gauge transformations.

\subsubsection{Subleading corrections to the cross section}
\label{sec:subleading}

In our derivation of the factorization formula we relied on field
redefinitions that decouple the soft gluons from the leading SCET
and PNRQCD Lagrangians.  The subleading effective Lagrangians include
higher-order potentials, but also interactions of the soft gluons to
the potential fields via the $\vec{x}\cdot \vec{E}$ interaction
mentioned above and analogous couplings to the collinear fields. These
terms cannot be eliminated using the decoupling transformations.
However, being sub-leading, these interactions can be treated as
perturbations in $\beta$ and, since the soft gluons decouple from the
leading effective Lagrangian, the cross section can be written to all
orders in the non-relativistic and SCET expansions
in the schematic form
\begin{equation} 
 \hat\sigma=\sum_aH^a  \,[W^a\otimes J^a] \, ,
  \label{eq:factform}
\end{equation}
with higher-order hard, soft and potential functions labeled by the
index $a$.  At NNLL accuracy, several effects could be relevant that
are not included in the leading term $a=0$ in~\eqref{eq:factform}
considered so far in this work.  We now discuss these effects and show
that they either do not contribute at NNLL or are incorporated in a
straightforward way without spoiling the factorization.

It is worth mentioning at this point that in standard situations of 
soft gluon resummation not requiring the consideration of Coulomb 
singularities, the expansion in powers of $1-z$ and $\alpha_s$ can be 
considered separately. Thus the discussion of power-suppressed 
interactions is not necessary to any order in $\alpha_s$ as long 
as one drops subleading terms 
in $1-z$. In joint soft-gluon and Coulomb resummations 
the two expansions are linked by the counting $\alpha_s/\beta\sim 1$, 
so that $O(\beta)$ suppressed terms that would normally be referred 
to as power corrections now appear at the same NNLL order 
as $\alpha_s$ terms, see (\ref{eq:syst}). This is a complication 
characteristic of perturbative non-relativistic systems that 
have kinematic singularities which are stronger than logarithmic.

\vspace*{0.2cm}\noindent
{\em Hard effects.} Higher-order production operators 
such as the $P$-wave operator~\eqref{eq:pwave} can appear at ${\cal
  O}(\beta)$, which according to (\ref{eq:syst}) 
corresponds to a NNLL effect. These can include interactions with soft
gluons through a spatial covariant derivative that are not removed by
the decoupling transformation. However, since there is no interference between
$S$- and $P$-wave production for the total cross section, 
these corrections are at least ${\cal O}(\beta^2)$, beyond NNLL accuracy.  

\vspace*{0.2cm}\noindent
{\em Potential effects.} Beyond the leading-order 
Coulomb potential in (\ref{eq:NRQCD}), subleading effects 
lead to further potential interactions of the form 
\begin{equation}
\int d^3 \vec{r}\, 
\left[\psi^{\dagger}_a\psi_b\right]\!(\vec r\,)
\delta V_{abcd}(r)
\left [\psi_c^{\prime\,\dagger}\psi_d^\prime\right]\!(0)\,.
\end{equation}
The running of the strength of the Coulomb potential causes 
a NLL correction beginning with  $\alpha_s^2/\beta \times \log \beta$,
which is accounted for by evaluating the coupling in the 
leading-order Coulomb potential at a scale of order $m_{\rm
  red}\beta$, where $m_{\rm red}$ denotes the reduced mass 
of the heavy-particle system, or by an explicit logarithmic correction
to $\delta V_{abcd}(r)$. Genuine loop corrections to the 
colour Coulomb potential lead to the substitution 
\begin{equation} \label{eq:Coulomb_run_coup}
\frac{\alpha_s(1/r)}{r} \to 
\frac{\alpha_s(1/r)}{r} \left(1+\hat a_1 \frac{\alpha_s}{4\pi} + 
\ldots\right)
\end{equation}
in the Lagrangian (\ref{eq:NRQCD}).
Only the one-loop correction $\hat a_1$ is needed at NNLL, and 
contributes terms beginning with $\alpha_s^2/\beta$. In addition 
there exist the spin-dependent and independent non-Coulomb
potentials of form $\alpha_s^2/r^2$ and (summarily) $\alpha_s/r^3$,
see e.g.~\cite{Beneke:1999qg}, leading to NNLL terms beginning 
with $\alpha_s^2\log\beta$. Here the logarithm arises from 
non-relativistic factorization and is related to the fact that 
the short-distance functions $H_{ii'}^{S}(\mu)$ have potential 
infrared divergences in addition to the familiar soft and collinear 
divergences. These additional potential terms in the PNRQCD Lagrangian
do not involve soft gluon fields and the decoupling transformation can
be applied as for the leading potential. Therefore these terms can be 
included in the evaluation of the potential function $J$, as was done
in~\cite{Beneke:2009ye} to compute all $\log\beta $ and $1/\beta$
enhanced terms of the $t\bar t$-production cross section at
$\mathcal{O}(\alpha_s^4)$.

\vspace*{0.2cm}\noindent
{\em Subleading potential-soft interactions.}
We next discuss possible contributions to subleading terms
in~\eqref{eq:factform} arising from the higher-order couplings of soft gluons
to collinear and potential fields that are not decoupled by the field
redefinition. In a diagrammatic language, these arise from the
interference of subleading soft gluons coupling to the initial or 
final state with potential
gluon exchange, that could contribute NNLL terms beginning with
$\alpha_s/\beta\times\alpha_s \beta \log^{2,1}\beta$. As shown
in~\cite{Beneke:2009ye} using power-counting and rotational
invariance, the fixed-order NNLO corrections of this form vanish for
the total cross section.  We will now use the effective theory
language to show that the NNLL corrections arising from the
subleading SCET and PNRQCD Lagrangians vanish to all orders in the
strong coupling, so that the leading factorization
formula~\eqref{eq:fact-final} does not require modification 
at NNLL.

We begin with the ${\cal O}(\beta)$ suppressed interactions in 
the subleading PNRQCD 
Lagrangian~\cite{Voloshin:1978hc,Pineda:1997bj,Beneke:1999zr},
\begin{equation}
\label{eq:x-dot-e}
\begin{aligned}
\mathcal{L}_{\text{PNRQCD}}^{(1)}&=-g_s\left[
\psi^\dagger(x)\,\vec x\cdot \vec E_s(x^0,\vec 0)\psi(x)
+\psi^{\prime\dagger}(x)\,
\vec x\cdot \vec E_s(x^0,\vec 0)\psi^\prime(x)\right]\\
&\equiv -g_s\,\vec x\cdot \vec E^a_s(x^0,\vec 0)\mathcal{J}^a(x)
\end{aligned}
\end{equation}
with the bilinear product of potential fields
\begin{equation}
\label{eq:cal-j}
\mathcal{J}^a=\psi^{\dagger}{\bf T}^{(R)a}\psi+
     \psi^{\prime\dagger}{\bf T}^{(R')a}\psi^{\prime}\,.
\end{equation}
We treat the chromoelectric vertex perturbatively, i.e. we consider
contributions to the cross section where one of the matrix elements
contains an insertion of the vertex \mbox{$\vec x\cdot \vec E$} and evaluate
these matrix elements with the LO PNRQCD Lagrangian.  The first-order
correction to~\eqref{eq:matrix} from the
interaction~\eqref{eq:x-dot-e} is given by
\begin{equation}
\label{eq:matrix-chre}
\Delta\mathcal{A}^{(1)}(pp'\to HH'X) = 
\;C^{(0)}_{\{a;\alpha\}}(\mu)\;\int d^4 x\,
\langle H H'X |\mbox{T}[\mathcal{L}_{\text{PNRQCD}}^{(1)}(x)
  \mathcal{O}^{(0)}_{\{a;\alpha\}}(\mu)] |
p p'\rangle_{\rm EFT} \,.
\end{equation} 
The expectation value is evaluated with the leading-order
effective-theory Lagrangian, so the soft, collinear and potential fields
decouple after the field redefinition. 
Under this redefinition, the  chromoelectric interaction is 
transformed into
\begin{equation}
\psi^{\dagger}(x)
  \vec x\cdot \vec E_s(x_0,\vec 0)\psi(x)
=\psi^{\dagger(0)}(x)
  \vec x\cdot \vec {\mathcal E}_s(x_0,\vec 0)\psi^{(0)}(x)
\end{equation}
with $\mathcal{E}_s=S_w^\dagger E_s S_w$. Independent of the colour
representation of the heavy particles, due to the
identity~\eqref{eq:trafo-generator}, 
which also holds for the Wilson lines, 
the components of the redefined chromoelectric fields are given by
\begin{equation}
 \vec{\mathcal E}^a_{s}(x_0)=S_{w,ab}^{(8)} \vec E^b_{s}(x_0)\,.
\end{equation}

In analogy to~\eqref{eq:matrix_element}, the effective theory matrix
element factorizes and the correction to the scattering amplitude can
be written as
\begin{eqnarray}
\label{eq:matrix_element_chr}
&& \Delta\mathcal{A}^{(1)}(pp'\to HH'X) = \sum_i
\;C^{(0,i)}_{\{\alpha\}}(\mu)\;\braket{0|\phi^{(0)}_{c;j_1\alpha_1}|p}
\braket{0|\phi^{(0)}_{\bar c;j_2\alpha_2} |p'}
\nonumber \\
&& \hspace*{1cm} \times \int d^4x
\,\vec x
\braket{ X|\mbox{T}[
 \vec {\mathcal E}_s^a(x_0,\vec 0)\mathcal{S}^{(i)}_{\{j\}}(0)]| 0}
 \langle HH' |\mbox{T}[\mathcal{J}^{a(0)}(x)
 \psi^{(0)\dagger}_{j_3\alpha_3}\psi^{\prime (0)\,\dagger}_{j_4\alpha_4}]
|0\rangle \,.
\end{eqnarray}
Repeating the steps leading to the factorization
formula~\eqref{eq:factorization} and the subsequent discussion, we
obtain an analogous formula with the replacement
\begin{multline}
\label{eq:factorization-sub}
\int \frac{d q^0}{(2 \pi)}\;
\,J_{R_\alpha}(E-q^0)\,
 \int d z^0 e^{i q^0  z^0} \,
\hat W^{R_\alpha}_{ii'}(z^0,\mu)\\\Rightarrow
 \int \frac{d q^0}{(2 \pi)} \int \frac{d^4 k}{(2 \pi)^4} \,
J_{R_\alpha}^{a(1)}(E-q^0,k) \int d z^0 e^{i q^0  z^0} 
\int d^4 x \, e^{-i k\cdot x} \, \vec x\, \cdot
 \vec{\hat W}^{a,R_\alpha(1)}_{ii'}(z^0,x^0,\mu)
+\text{c.c}\,.
\end{multline}
Here we have defined the subleading soft function with an insertion of
the chromoelectric field according to
\begin{equation}
\vec{\hat W}^{a,R_\alpha (1)}_{ii'}(z^0,x^0,\mu)=
P^{R_\alpha}_{\{k\}}
 \braket{0|\overline{\mbox{T}}[\mathcal{S}^{(i')*}_{\{ijk_1k_2\}}(z^0)]
   \mbox{T}[\vec {\mathcal
     E}_s^a(x^0)\mathcal{S}^{(i)}_{\{ijk_3k_4\}}(0)]|0}
\end{equation}
as well as a subleading potential function with an insertion of the
bilinear~\eqref{eq:cal-j}:
\begin{equation}
J^{a (1)}_{\{\alpha\}R_\alpha}(q,k)
= P^{R_\alpha}_{\{a\}}\int d^4 z\, e^{iq\cdot z}  \int d^4 x\, e^{ik\cdot x}
\braket{0|[ \psi_{\alpha_1a_1}^{(0)}\psi_{\alpha_2a_2}^{'(0)} ](z) 
{\mbox T}[\mathcal{J}^{a(0)}(x)[
 \psi_{\alpha_3a_3}^{(0)\dagger} \psi_{\alpha_4a_4}^{'(0)\dagger}](0)]|0} \, .
\end{equation}
Since the matrix elements are evaluated with the spin-independent
leading order PNRQCD Lagrangian, the spin dependence simplifies in
analogy to~\eqref{eq:spin-trivial} as was used
in~\eqref{eq:factorization-sub}. In the terms denoted by ``c.c.''
in~\eqref{eq:factorization-sub} the operator $\mathcal{E}_s$ 
is inserted in the anti-time-ordered product in the subleading soft function and the potential function is defined with the insertion of $\mathcal{J}^{(0)}$ in an anti-time-ordered product with the annihilation fields.

Because the subleading soft function does not depend on $\vec x$, we can perform the integral over the spatial components of $x$ in~\eqref{eq:factorization-sub} and obtain the expression
\begin{equation}
 \int \frac{d^4 k}{(2 \pi)^4}\,\delta^{(3)}(\vec k)\,
\frac{\partial}{\partial\vec k}\,
J_{R_\alpha}^{a(1)}(E-q^0,k)=0\,,
\end{equation}
which vanishes since (in the partonic cms frame)
there is no three-momentum left that the integral over 
$ J^{(1)}$ can depend on. Hence we conclude that the
correction to the cross section due to a single insertion of the NLO
potential Lagrangian vanishes to all orders in the strong coupling
constant.  Similar to the diagrammatic argument
in~\cite{Beneke:2009ye}, an essential ingredient in the argument was
the multipole expansion of soft fields when multiplying potential
fields. Because of this, only $\vec E(x^0)$ appears in the subleading
PNRQCD interaction and, as a consequence, the subleading soft function
depends only on $x^0$ which allowed to perform the simplification in
the last step.  The second ingredient is rotational invariance in
combination with the independence of the potential function on
any external three-momentum.
 
\vspace*{0.2cm}\noindent
{\em Subleading collinear-soft interactions.}
The effects potentially relevant at NNLL arise from the
effective Lagrangians at the next-to-leading order in the kinematic
expansion. For the SCET Lagrangian of collinear quarks, the relevant
interactions are given by~\cite{Beneke:2002ph,Beneke:2002ni}
\begin{equation}
\mathcal{L}_{\xi}^{(1)}=\bar{\xi} \left( x_\perp^\mu n^\nu \, W_c 
\,gF_{\mu\nu}^{\rm 
s}W_c^\dagger \right) \frac{\not\!\bar n}{2} \, \xi\,,
\end{equation}
for $n$-collinear modes 
and similar terms involving transverse derivatives or factors of
$x_\perp$ for the couplings to collinear gluons, and an interaction 
\begin{equation}
\label{eq:softquarks} 
{\cal L}^{(1)}_{\xi q} =
    \bar{\xi} \,i\Slash{D}_{\perp c} W_{c}\, q_s + 
    \bar q_s \,W_{c}^\dagger i\Slash{D}_{\perp c} \,\xi
\end{equation}
involving soft quarks but only collinear gluons. As argued already 
in~\cite{Beneke:2009ye}, these vertices with transverse derivatives
or $x_\perp^\mu$ cannot contribute to the total cross section 
since the initial-state momenta can be chosen to have zero transverse 
momentum. Then loop integrals with transverse momentum factors vanish
by rotational invariance in the plane transverse to the beam axis. 
Eq.~(\ref{eq:softquarks}) contains a vertex that describes 
the collinear splitting of a quark into a gluon and a soft quark, 
which is also an ${\cal O}(\beta)$ term. However, a non-vanishing 
contribution to the squared amplitude requires two such vertices 
to contract the soft quark field, hence there is again no contribution
at NNLL. Note that insertions of power-suppressed interaction 
Lagrangians lead to collinear matrix elements that do not take 
the form of the standard parton densities. At this point it 
appears that one must distinguish the collinear expansion in 
powers of $k_\perp \sim M \sqrt{\lambda} \sim M\beta$ from the 
standard collinear expansion in $k_\perp\sim \Lambda_{\rm QCD}$. 
However, as mentioned above, this issue is not relevant for 
NLL and NNLL resummations.

\vspace*{0.2cm}
To summarize, the factorization formula~\eqref{eq:fact-final}
is valid at NNLL accuracy provided subleading Coulomb and non-Coulomb
potentials are included in the computation of the potential function
and the hard and soft functions are computed at NLO.

\subsubsection{Comparison to factorization at fixed invariant mass}
\label{sec:compare-s}

The result~\eqref{eq:fact-final} can be compared to the analogous
formula~\eqref{eq:factor-thresh} given in~\cite{Kidonakis:1997gm} for
the pair production of relativistic coloured particles with different
four-velocity vectors $w_1$,$w_2$, where potential degrees of freedom
do not appear.  In this case the potential function is absent in the
factorization formula and the soft function is given by (taking into
account our convention for the antiparticles):
\begin{equation} \label{eq:soft-simple}
 \hat S_{\{ab\}}(z)
=\langle 0|\overline{\mbox{T}}[S^\dagger_{n, ib_1} 
S^\dagger_{\bar{n},jb_2} S_{w_2,b_4l} S_{w_1,b_3k}](z)
\mbox{T}[ S_{\bar{n},a_2j} S_{n,a_1i}S^\dagger_{w_1,k a_3} 
S^\dagger_{w_2,l a_4}](0)|0\rangle \, .
\end{equation}
In our case the extraction of the potential function $J$ leads to a
more complicated colour structure of the soft function $W$, but to a
simpler kinematical structure due to the equal four-velocities of the
heavy particles.  Using the completeness relation of the projectors
and the definition of the components of the soft
function~\eqref{eq:project-soft-coulomb}, the limit
of~\eqref{eq:soft-simple} for the case of equal four velocities,
$w_1=w_2=w$, is formally related to the sum of the soft functions for the
different final-state representations:
\begin{equation}
\label{eq:soft-sum}
S_{ii'}|_{w_1=w_2}=\sum_{R_\alpha}  W^{R_\alpha}_{ii'} \, .
\end{equation}
The resummation coefficients used in the Mellin space approach to
threshold resummation have been calculated at the one-loop level for
various colour
representations~\cite{Kidonakis:1997gm,Kulesza:2008jb,Beenakker:2009ha}
from the threshold limit of the soft function $S$. This amounts to
taking the threshold limit after extracting the $1/\epsilon$ poles of
the eikonal diagrams contributing to $S$. Applying this approach at
the two-loop level~\cite{Ferroglia:2009ep} one encounters three-parton
colour correlations proportional to $\log\beta$ that arise from an
interference of soft and potential divergences, and lead to
off-diagonal contributions in the colour basis constructed
in~\cite{Beneke:2009rj} that diagonalizes the soft functions
$W^{R_\alpha}$.  In contrast, the effective theory approach
constructs an expansion in $\beta$ before taking the
$\epsilon\to 0$ limit, resulting in the tower
of higher dimensional hard, soft and potential
functions~\eqref{eq:factform}. In this approach, the $\log\beta$-dependent
divergences are attributed to higher-dimensional soft and potential
functions due to the subleading $\vec x\cdot \vec E$ vertex in the PNRQCD
Lagrangian and subleading potentials,  while the leading soft
function~\eqref{eq:project-soft-coulomb} is diagonal to all orders.
In order to resum all terms that are enhanced by powers of
$\log\beta$ and inverse powers of $\beta$ at a given accuracy one
should use the formula~\eqref{eq:factform},
while~\eqref{eq:factor-thresh} is appropriate for partonic thresholds
where the relative velocities of the final state particles are
relativistic, as considered e.g. in the recent resummation of the
invariant mass distribution of top-quark pairs
in~\cite{Ahrens:2010zv}.

\subsubsection{Soft-Coulomb factorization in Mellin moment space}
The enhancement due to single-Coulomb exchange has been included in
the Mellin moment space approach to resummation as part of the hard
function~\cite{Bonciani:1998vc}, which implicitly assumes the
factorization of soft and potential gluons. We now briefly discuss the
relation of our factorization formula~\eqref{eq:fact-final} to the
Mellin space formalism and show that the convolution of the soft and
potential function factorizes in Mellin space in the large $N$ limit,
which justifies the earlier treatment. We recall that the Mellin
moments of the cross section in the variable $z=4M^2/\hat s$ are
defined as:
\begin{equation}
  \hat \sigma(N,\mu)=\int_0^1dz\, z^{N-1}\hat\sigma(\hat s,\mu)\,.
\end{equation}
In order to Mellin-transform the factorized cross
section~\eqref{eq:fact-final}, we approximate $E=\sqrt{\hat s}-2
M\approx M(1-z)$ and make the transformation of variables
$\omega=2M(1-w)$. For stable particles and neglecting bound-state
contributions, the potential function is non-vanishing only for
$E-\omega/2\approx M(w-z)>0$ so we can write
\begin{equation}
\hat \sigma_{pp'}(\hat s,\mu) \approx 2M 
\sum_{S,i,i'}H_{ii'}^S
\;\int_z^1 \frac{d w}{w}\;
\sum_{R_\alpha}\,J_{R_\alpha}^S(M(1-\tfrac{z}{w}))\,
W^{R_\alpha}_{ii'}(2M(1-w),\mu)\,.
\end{equation}
Here we have used $z< w\lesssim 1$ so that $E-\omega/2\approx
M(w-z)\approx M(1-z/w)$ and introduced a factor $\frac{1}{w}\approx
1$.  Since a convolution of the form $\int \frac{dw}{w} f(w)g(z/w)$
factorizes under a Mellin transform, we conclude that, up to power
suppressed terms, our factorization formula implies multiplicative 
soft-Coulomb factorization of the cross section in Mellin space:
\begin{equation}
  \hat \sigma_{pp'}(N,\mu) \approx
\sum_{S,i,i'}H_{ii'}^S
\sum_{R_\alpha}\,J_{R_\alpha}^S(N)\,
S^{R_\alpha}_{ii'}(N,\mu)\,,
\end{equation}
where the soft function in Mellin space is given by 
\begin{equation}
S^{R_\alpha}_{ii'}(N,\mu)=2M\int_0^1dz\, z^{N-1} W^{R_\alpha}_{ii'}(2M(1-z),\mu)=
\int_{0}^{\infty} d \omega e^{-\frac{N}{2M} \omega}
\, W^{R_\alpha}_{ii'}(\omega,\mu)+\mathcal{O}(N^{-1})\,,
\end{equation}
and 
\begin{equation}
J_{R_\alpha}^S(N) = \int_0^1 dz\,z^{N-1}\,J_{R_\alpha}^S(M(1-z))\,.
\end{equation}
This shows that previous treatments that put the one-loop Coulomb
corrections into the hard
function~\cite{Bonciani:1998vc} can be extended to include 
the higher-order Coulomb corrections as well.

\subsubsection{Invariant mass distribution near threshold}

The combination of soft and Coulomb effects for the invariant mass
distribution of top-quark and gluino pairs near threshold has been
performed in several approximations {\em assuming} soft-Coulomb 
factorization~\cite{Hagiwara:2008df,Kiyo:2008bv,Hagiwara:2009hq}. 
We now show that a factorization formula
for the differential cross section $d\sigma/dM_{HH'}$ with
$M_{HH'}^2=(p_1+p_2)^2$, valid for $M_{HH'}$ near $2M$, can be derived
from our main result (\ref{eq:fact-final}).

We first introduce the parton luminosity 
\begin{equation}
\label{eq:lumi}
L_{p p^\prime}(\tau,\mu) = \int_0^1 dx_1
dx_2\,\delta(x_1 x_2 - \tau) \,f_{p/N_1}(x_1,\mu)f_{p^\prime/N_2}(x_2,\mu) 
\end{equation}
to express the hadron scattering cross section as 
\begin{equation}
\sigma_{N_1 N_2\to H H^\prime X} = \sum_{p,p^\prime=q,\bar q,g} 
\,\int_{4 M^2/s}^1 \!\!d\tau \,L_{p p^\prime}(\tau,\mu)\,
\hat\sigma_{p p^\prime}(\tau s,\mu) \, ,
\label{eq:sigmahad1}
\end{equation}
with $\hat\sigma_{p p^\prime}(\tau s,\mu)$ given by
(\ref{eq:fact-final}). Next we observe that the argument $E-\omega/2 =
\sqrt{\tau s}-2 M - \omega/2$ of the Coulomb function  
$J_{R_\alpha}^S$ in that equation corresponds to the non-relativistic 
energy of the $H H^\prime$ pair in the partonic cms frame.  In this 
frame the three momentum of the pair is soft and therefore makes a 
negligible contribution to the invariant mass, so the relation 
\begin{equation}
M_{H H^\prime} = 2 M+E-\frac{\omega}{2} +{\cal O}(M\beta^4) 
= \sqrt{\tau s}- \omega/2 +{\cal O}(M\beta^4) 
\end{equation} 
applies. We now change variables from $\omega$ to $M_{H H^\prime}$ 
in (\ref{eq:fact-final}) and interchange the $\tau$ and (implicit) $\omega$ 
integration in (\ref{eq:sigmahad1}) using
\begin{equation}
\int_{4 M^2/s}^1 \!\! d\tau\int_0^{2 E} \!\!d\omega 
 = 2 \int_{2 M}^{\sqrt{s}} dM_{H H^\prime} 
\int_{M_{HH^\prime}^2/s}^1 \!\!d\tau 
\end{equation}
and the fact that for stable particles, and neglecting bound-state
contributions, the Coulomb function has support only for positive
values of its argument, which fixes the upper limit $2 E$. This
results in
\begin{eqnarray}
\frac{d\sigma_{N_1 N_2\to H H^\prime X}}{dM_{HH^\prime}}
&\!=\!& \sum_{p,p^\prime=q,\bar q,g} 
\,\int_{M_{HH^\prime}^2/s}^1 \!\!d\tau \,L_{p p^\prime}(\tau,\mu)\,
\frac{d\hat\sigma_{p p^\prime}(\tau
  s,\mu)}{dM_{HH^\prime}} 
\nonumber\\
&& \hspace*{-2cm} 
=\,\sum_{p,p^\prime=q,\bar q,g} \sum_{S=|s-s'|}^{s+s'}
\sum_{i,i'}\,2H_{ii'}^S(\mu)
\sum_{R_\alpha}\,J_{R_\alpha}^S(M_{HH^\prime}-2 M)
\nonumber\\
&& \hspace*{-1.3cm} \times \,
\int_{M_{HH^\prime}^2/s}^1 \!\!d\tau 
\,L_{p p^\prime}(\tau,\mu)\,
W^{R_\alpha}_{ii'}(2(\sqrt{\tau s}-M_{HH^\prime}),\mu)
\label{eq:invdist}
\end{eqnarray}
with
\begin{equation}
\label{eq:invdistpartonic}
\frac{d\hat \sigma_{pp'}(\hat s,\mu)}{dM_{HH^\prime}} = 
\sum_{S=|s-s'|}^{s+s'}
\sum_{i,i'}\,2H_{ii'}^S(\mu)
\sum_{R_\alpha}\,J_{R_\alpha}^S(M_{HH^\prime}-2 M)\,
W^{R_\alpha}_{ii'}(2(\sqrt{\hat s}-M_{HH^\prime}),\mu)\,,
\end{equation}
which is the desired result. This shows that Coulomb and soft effects
factorize multiplicatively in the invariant mass distribution near
threshold, as assumed
in~\cite{Hagiwara:2008df,Kiyo:2008bv,Hagiwara:2009hq}.  Note that in
the hadronic cross section only the soft function is averaged with the
parton luminosity.  Eqs.~(\ref{eq:invdist}),
(\ref{eq:invdistpartonic}) apply unmodified to bound-state
contributions, in which case the invariant mass distribution extends
below the nominal threshold $2 M$, and unstable particles, where in
addition the Coulomb function is evaluated at complex argument
$M_{HH^\prime}-2 M+i (\Gamma_H+\Gamma_{H^\prime})/2$.

\section{Resummation of soft and Coulomb gluons}
\label{sec:resum}

The factorization formula~\eqref{eq:fact-final} provides the basis for
resummation of soft and Coulomb gluon effects in the soft function $W$
and the potential function $J$, respectively. The soft gluon
resummation is performed by solving evolution equations for the soft 
and hard
functions~\cite{Korchemsky:1992xv,Korchemsky:1993uz,Contopanagos:1996nh,Manohar:2003vb,Becher:2006nr},
while the resummation of potential effects can be performed using
results obtained for top-quark production in electron-positron
collisions~\cite{Fadin:1987wz,Hoang:2000yr,Beneke:1999qg}.  In this
section we provide the explicit results for the resummed soft and
potential functions.  In section~\ref{sec:colour-soft} we recall the colour
basis constructed in~\cite{Beneke:2009rj} that diagonalizes the soft
function to all orders in the strong coupling constant and quote the
result for the one-loop soft function. In section~\ref{sec:hard} we relate the
hard function to the colour- and spin decomposed Born cross section
and the colour decomposed one-loop amplitudes at threshold to
demonstrate that the
implementation of the resummation formula requires only a standard
calculation in fixed-order perturbation theory.  In section~\ref{sec:coulomb}
we provide the leading order potential function summing up multiple
Coulomb gluon exchange, as required for the NLL resummation performed
in section~\ref{sec:squarks}.  In section~\ref{sec:rges} we obtain evolution
equations for the short-distance coefficients $C$ and the soft
function $W$ that allow to resum soft-gluon effects. The anomalous
dimensions required up to NNLL have already been collected
in~\cite{Beneke:2009rj}.  The final result for the resummed cross
section using the momentum-space formalism
of~\cite{Becher:2006nr,Becher:2006mr,Becher:2007ty} is presented
in section~\ref{sec:sum-sigma} with the explicit expression of the resummation
exponent up to NLL relegated to appendix~\ref{app:evolution}.
Expansions of the resummed result to
order $\alpha_s$, as required for the matching to a fixed-order
calculation, are given in appendix~\ref{app:expand}.

\subsection{The  soft function }
\label{sec:colour-soft}

In the factorization formula~\eqref{eq:fact-final}, the hard- and soft
function are matrices in colour space in the basis of the tensors
$c^{(i)}$ introduced in~\eqref{eq:colour-wilson}. The soft function
can be diagonalized to all orders in the strong coupling for an arbitrary
heavy coloured particle production process using the
colour basis constructed in~\cite{Beneke:2009rj}. In
this basis, the components of the soft
function~\eqref{eq:project-soft-coulomb} can be expressed in terms of
the soft function for a single heavy final state particle in the
irreducible colour representation $R_\alpha$.  This gives a precise
meaning to the picture of soft gluon radiation resolving only the
total colour charge of a heavy-particle pair at
rest~\cite{Bonciani:1998vc}. We review this construction here and
quote the result of the one-loop soft function.

It is useful to  perform a
decomposition of the product of the representations of the initial
state and final state particles into irreducible representations:
\begin{equation}
\label{eq:irreducible}
  r\otimes r' =\sum_{\alpha} r_\alpha\;,\qquad
R\otimes R'=\sum_{R_\alpha} R_{\alpha}.
\end{equation}
It is intuitively clear that a final state pair in an irreducible
colour representation $R_\alpha$ can only be produced from an initial state
system in an equivalent representation. In order to implement this
picture formally, one forms pairs $P_i=(r_{\alpha}, R_{\beta})$ of
equivalent initial and final state representations $r_\alpha$ and
$R_\beta$ , treating multiple occurrences of equivalent
representations in the decomposition as distinct, e.g. in the case of
a symmetric or antisymmetric coupling of $8\otimes 8\to 8$.  It has
been shown in~\cite{Beneke:2009rj} that a basis respecting colour
conservation~\eqref{eq:gauge-basis} can always been chosen by forming
for every pair $P_i$ of equivalent representations the associated basis element
\begin{equation}
\label{eq:prod-basis}
  c_{\{a\}}^{(i)}=\frac{1}{\sqrt{\text{dim}(r_\alpha)}}\,
  C^{r_\alpha}_{\alpha a_1a_2} C^{R_{\beta}\ast}_{\alpha a_3a_4}, 
\end{equation}
where the $C^{R_\alpha}_{\alpha a_1a_2}$ are Clebsch-Gordan
coefficients implementing a unitary basis transformation from the
basis vectors of the tensor product space $R\otimes R'$ to basis
vectors of the irreducible representations $R_\alpha$, and analogously
for the initial state representations.  The Clebsch-Gordan
coefficients, basis elements and projectors for all squark and gluino
production processes at hadron colliders have been provided in
appendix~B of~\cite{Beneke:2009rj}.  For squark-antisquark production
from quark-antiquark annihilation, the allowed pairs of
representations are
\begin{equation}
  P_i\in\{(1,1),(8,8)\}
\end{equation}
and  the basis has been given already in~\eqref{eq:basis_33}. For the
gluon fusion channel the allowed pairs of representations are
\begin{equation}
\label{eq:gluon-fusion-reps}
  P_i\in\{(1,1),(8_S,8),(8_R,8)\}
\end{equation}
and the basis is given in~\eqref{eq:basis_83} below.  

Using properties of
Wilson lines and the Clebsch-Gordan coefficients it was shown
in~\cite{Beneke:2009rj} that the components of the soft
function~\eqref{eq:project-soft-coulomb} in the basis~\eqref{eq:prod-basis}
can be obtained from the soft function for  the production of a
\emph{single particle} in the representation $R_\alpha$
\begin{equation}
\label{eq:soft-R}
\hat W^{R_\alpha}_{\{a\alpha ,b\beta \}}(z,\mu)\equiv
\langle 0|\overline{\mbox{T}}[S^{R_\alpha}_{w,\beta \kappa}
S^\dagger_{\bar{n},jb_2} S^\dagger_{n, ib_1} ](z)
\mbox{T}[S_{n,a_1i}S_{\bar{n},a_2j}S^{R_\alpha\dagger}_{w,\kappa\alpha}](0)|
0\rangle 
\end{equation}
by contracting with appropriate Clebsch-Gordan coefficients
\begin{equation}
\label{eq:soft-coulomb-2}
 W^{R_\alpha}_{ii'}(\omega,\mu)=  
\frac{1}{\sqrt{\text{dim}(r_\alpha)\text{dim}(r_{\alpha'})}  }
\;\delta_{R_\alpha R_{\beta'}}\delta_{R_\alpha R_{\beta}}\;
  C^{r_\alpha}_{\{a\alpha\}}  W^{R_\alpha}_{\{a\alpha,b\beta\}}(\omega,\mu)
C^{r_\alpha'\ast }_{\{b\beta\}} \, .
\end{equation}
As indicated, the elements are non-vanishing only if the irreducible
representation $R_\alpha$ is identical to both final state
representations $R_\beta$ and $R_{\beta'}$ in the pairs
$P_i=(r_\alpha,R_{\beta})$ and
$P_{i'}=(r_{\alpha^\prime},R_{\beta^\prime})$ that define the tensors
$c^{(i)}$,~$c^{(i')}$.  This is intuitively plausible since we
project on a specific final state representation so only production
from initial states in an equivalent representation is possible.
This shows that the soft function is automatically block diagonal in
the basis~\eqref{eq:prod-basis} where off-diagonal elements are only
possible if several initial state representations
$r_\alpha$ are  equivalent.  For initial state quarks, antiquarks and gluons
this only happens for two gluons in the initial state that can be
combined into a symmetric and antisymmetric octet,
see~\eqref{eq:gluon-fusion-reps}.  Using Bose symmetry it is
furthermore possible to show that the symmetric and antisymmetric octet
production channels do not interfere so the soft function is diagonal,
i.e.
\begin{equation}
\label{eq:soft-diagonal}
 W^{R_\alpha}_{ii'}(\omega,\mu)=  W^{R_\alpha}_{i}(\omega,\mu)\;\delta_{ii'}\;
 \delta_{R_\alpha R_\beta}.
\end{equation}

The one-loop term in the loop expansion
of the soft function, 
\begin{equation}
\label{eq:soft-alpha}
   W_{i}^{R_\alpha}(\omega,\mu)=\sum_{n=0}^\infty
 \left(\frac{\alpha_s(\mu)}{4\pi}\right)^{n}  
W_{i}^{(n)R_\alpha}(\omega,\mu)\,,
\end{equation}
was calculated in~\cite{Beneke:2009rj} and depends only on the
quadratic Casimir operators $C_R$ of the representations of the initial
state particles and the final state pair, in agreement with
calculations for specific colour
representations~\cite{Kidonakis:1997gm,Kulesza:2008jb,Beenakker:2009ha}.
The result in position space can be written in terms of the variable
$L=2\ln\left(\frac{i z_0\mu e^{\gamma_E}}{2}\right)$ and is given by
\begin{equation}
 \hat W^{(1)R_\alpha}_{i}(L,\mu)=\left(
C_r+C_{r'}\right)\left(\frac{2}{\epsilon^2}+\frac{2}{\epsilon} L
  + L^2+\frac{\pi^2}{6}\right)
+2C_{R_\alpha}\left(\frac{1}{\epsilon}+ L+2\right)\, ,
\label{eq:soft-one}
\end{equation}
where the normalization is such that $ \hat W^{(0)R_\alpha}_{i}(L,\mu)=1$.
For a final state singlet and a quark-antiquark initial state this agrees with
the Drell-Yan Wilson line~\cite{Korchemsky:1993uz}, for a colour octet final state the result agrees with~\cite{Idilbi:2009cc}.  The Fourier transform of
this result enters the factorization formula~\eqref{eq:fact-final} and has
been computed in~\cite{Beneke:2009rj}.  Since in the momentum space
formalism the solution of the renormalization group
equation for the soft function is obtained using the Laplace transform of the
momentum space soft function that can be obtained from the position space
result by a simple replacement rule, the momentum space expression is not
explicitly needed in this paper.

\subsection{The hard function} 
\label{sec:hard}
In order to simplify the application of the factorization
formula~\eqref{eq:fact-final} we show how to bypass the matching
to the effective theory used in our derivation and express the hard
function~\eqref{eq:hard-colour} directly in terms of hard-scattering
amplitudes for the production of the heavy-particle pair.
The potential and soft functions are universal functions, depending
only on the colour quantum numbers. The only process dependent input for NLL
resummation are then the colour-separated leading-order production cross
sections for the process of interest.
For NNLL resummation, in addition the tree cross sections for
individual spin states of the heavy-particle pair and the colour 
separated NLO amplitudes at threshold are required.
We work in the colour basis that
diagonalizes the soft function and denote the
diagonal elements of the hard function by $H_i^S$.  

We recall that the $S$-wave 
matching coefficients $C^{(0,i)}_{\{\alpha\}}$ are simply
given by the scattering amplitudes at threshold, with polarization
vectors removed and projected on a specific colour channel 
according to 
${\cal A}_{\{\alpha\}}=\sum_i c^{(i)}_{\{\alpha\}} {\cal A}^{(i)}$.
Thus the scattering amplitude 
for a given colour configuration and 
fixed helicities and spins $\lambda_i$ can be written as
\begin{equation}
\label{eq:match-born}
  \mathcal{A}^{(i)}(p^{\lambda_1}p^{\prime \,\lambda_2}\to 
  H^{\lambda_3}H^{\prime \,\lambda_4})=
2\sqrt{m_{H}m_{H'}}\,
C_{\{\alpha\}}^{(0,i)}(\mu)\,\chi_{\alpha_1}^{\lambda_1}
\chi_{\alpha_2}^{\lambda_2}
\xi^{\lambda_3*}_{\alpha_3}\xi^{\lambda_3*}_{\alpha_4}\,,
\end{equation}
where  we  have accounted for the normalization factors $\sqrt{2
  m_H}$ in the definition of the nonrelativistic states.
Here the $\xi$ are the polarization spinors or vectors of the
non-relativistic particles satisfying the completeness relation
\begin{equation}
  \sum_{\lambda}\xi^{\lambda}_{\alpha}\xi^{\lambda\ast}_\beta
=\delta_{\alpha\beta} \, .
\end{equation}

Computing the spin averaged partonic tree cross section for the
production of a heavy-particle pair in a fixed colour state using the
expression of the amplitude in the effective theory~\eqref{eq:match-born} we
obtain
\begin{equation}
\label{eq:sigma-match}
\begin{aligned}
  \hat\sigma_{pp'}^{(0,i)}(\hat s)&= \frac{1}{2\hat s}
  \frac{\lambda^{1/2}(\hat s,m_H^2,m_{H'}^2)}{8\pi\hat s}
\frac{1}{4 D_r D_{r'}}
\sum_{\text{pol}}
|\mathcal{A}^{(i)}_{\text{Born}}(pp^\prime\to HH^\prime)|^2\\
&\hspace*{-0.3cm} \underset{\hat s\to 4M^2}{\approx}
 \frac{(m_Hm_{H'})^{3/2}}{M}\,\frac{\beta}{4\pi \hat s}
\frac{1}{4 D_r D_{r'}}
\sum_{\lambda_i}
(C_{\{\alpha\}}^{(0,i)}(\mu)\chi_{\alpha_1}^{\lambda_1}
\chi_{\alpha_2}^{\lambda_2})\delta_{\alpha_3\beta_3}\delta_{\alpha_4\beta_4}
(C_{\{\beta\}}^{(0,i)}(\mu)\chi^{\lambda_1}_{\beta_1}
\chi_{\beta_2}^{\lambda_2})^\ast\, .
\end{aligned} 
\end{equation}
In the second line we
have approximated the prefactor using $\lambda(\hat s,
m_H^2,m_{H'}^2)=(\hat s-4M^2)(\hat s-(m_H-m_{H'})^2) \approx (\hat
s-4M^2) 4m_Hm_{H'}$ and the $C_{\{\alpha\}}^{(0,i)}(\mu)$ should 
be computed in the tree approximation.
Comparing to the definition of the hard
function~\eqref{eq:hard-colour}, we obtain a simple relation between
the  leading order hard functions and
the spin averaged total cross section for a given colour channel:
\begin{equation}
\label{eq:sigma-hard}
  \hat\sigma_{pp'}^{(0,i)}(\hat s)
\underset{\hat s\to 4M^2}{\approx}
 \frac{(m_Hm_{H'})^{3/2}}{M}\,\frac{\beta}{2\pi} H^{(0)}_{ii} \, .
\end{equation}
The tree-level hard functions can therefore be obtained from the
threshold limit of the Born cross sections in a
specific colour channel.  With
the aim of extending the validity of the resummed expressions one
may consider defining an improved hard function by using 
full tree-level cross section instead of its threshold limit. In this way, 
some higher order terms in $\beta$ are included, though not 
systematically.

At NNLL spin-dependent hard functions are required in the 
tree approximation,
which can be obtained by a formula analogous to~\eqref{eq:sigma-hard}
from the cross section for a fixed colour and spin channel.  In the
framework of a standard computation of the leading order cross
section, the projection on the final-state spin can be performed
introducing scattering amplitudes for a fixed final state spin $S$ 
that can be obtained from the helicity amplitudes for
the production of the $HH'$ pair according to 
\begin{equation}
\label{eq:spin-amps}
  \mathcal{A}^{(i)}(p^{\lambda_1}p^{\prime\,\lambda_2}\to(HH')^{S,\lambda})=
\sum_{\lambda_3\lambda_4}N_S(\lambda|\lambda_3\lambda_4)
\mathcal{A}^{(i)}(p^{\lambda_1}p^{\prime\,\lambda_2}
\to H^{\lambda_3}{H'}^{\lambda_4}).
\end{equation}
The $N_S(\lambda|\lambda_3\lambda_4)$ are Clebsch-Gordan coefficients
that can be found for the case of spin one-half particles, 
e.g. in~\cite{Artoisenet:2007qm}.

Furthermore, at NNLL accuracy
also the one-loop hard function is required.  Since the spin-dependence of
the potential function is already an ${\cal O}(\alpha_s^2)$ effect, 
one may use the spin summed one-loop hard functions $
H_{i}^{(1)}$  at NNLL without formal loss of accuracy.
The one-loop hard function is
given by the interference of the Born and the one-loop amplitudes for
a given colour channel, evaluated directly at threshold:
\begin{eqnarray}
  H_i^{(1)}(\mu) &=& \frac{1}{2\hat s}
  \left[ N^{p}_{\alpha_1\beta_1} N^{p'}_{\alpha_2\beta_2}\,
    2\text{Re}\,(C_{\alpha_1\alpha_2\gamma\delta}^{(0,i) \rm 1-loop}(\mu)
    C_{\beta_1\beta_2\gamma\delta}^{(0,i) \rm tree\,*}(\mu))\right]
\nonumber \\
  && \hspace*{-1.5cm} = 
\frac{1}{8 m_Hm_{H'} \hat s}\frac{1}{4 D_r D_{r'}}
\sum_{\text{pol}}2\,
\text{Re}\left(\mathcal{A}^{(i)\ast}_{\text{Born}}(pp^\prime\to 
HH^\prime)\mathcal{A}^{(i)}_{\text{NLO}}(pp^\prime\to
HH^\prime)\right)
\, .
\label{eq:hard-nlo}
\end{eqnarray}
Here $\mathcal{A}_{\text{NLO}}$ is the UV-renormalized one-loop
amplitude evaluated directly at threshold with
IR singularities regularized dimensionally and subtracted in the
$\overline{\text{MS}}$-scheme. This is therefore the only process
independent input required for NNLL resummations and is far simpler to
compute than the full NLO cross section. Alternatively, 
the hard function can be extracted from the threshold 
expansion of the  NLO partonic cross sections in each 
colour channel~\cite{Beneke:2009ye}. These are available for 
$t\bar t$ production at hadron 
colliders~\cite{Czakon:2008cx} but not yet for
the production of squarks or gluinos.

\subsection{The potential function}
\label{sec:coulomb}

We now discuss the relation of the potential function $J$ to the
Coulomb Green function familiar from PNRQCD.  The momentum-space
potential function~\eqref{eq:fourier-potential} can be written in
terms of the correlation function
\begin{eqnarray}
J_{\{a\}}^{\{\alpha\}}(q)
&=&\int d^4z \,e^{iq\cdot z} \int d\Phi_1 d\Phi_{2}
\sum_{\mbox{\tiny pol,colour}}
\braket{0|[\psi_{a_2\alpha_2}^{\prime(0)} \psi_{a_1\alpha_1}^{(0)}](z) |HH'}
\braket{HH'|[\psi_{a_3\alpha_3}^{(0)\dagger}
\psi_{a_4\alpha_2}^{\prime(0)\dagger}](0)|0} 
\nonumber\\
&=&\int d^4z \,e^{iq\cdot z} 
\braket{0|[\psi_{a_2\alpha_2}^{\prime(0)} \psi_{a_1\alpha_1}^{(0)}](z) 
[\psi_{a_3\alpha_3}^{(0)\dagger}
\psi_{a_4\alpha_2}^{\prime(0)\dagger}](0)|0}\,,
\end{eqnarray}
where the matrix elements are
evaluated with the leading order \mbox{PNRQCD} Lagrangian.\footnote{As
discussed in section~\ref{sec:comments}, to reach
NNLL accuracy, the LO Lagrangian is to be supplemented by NLO Coulomb
and the leading non-Coulomb potentials.}  
In the second line we used that in
this approximation the soft gluon fields are decoupled from the fields
$\psi^{(0)}$ so we can replace the sum over the two-particle Hilbert space
by a sum over the full Hilbert space and use the completeness
relation.  The definition of $J$ given here is sufficient for $S$-wave
production of the heavy particles; for $P$-wave production also
expectation values of fields with derivatives have to be
considered. As explained in section~\ref{sec:subleading},
effects of the $\vec x\cdot \vec E$
vertex in the sub-leading PNRQCD Lagrangian are not included in the
calculation of the potential function $J$ at higher orders but lead to
additional contributions to the factorization formula with generalized
soft and potential functions, to be calculated with the leading
effective Lagrangians. As shown in section~\ref{sec:subleading} 
these terms are only relevant for the total cross section beyond the
NNLL order.

Defining a tensor product notation for the decoupled potential fields,
\begin{equation}
(\psi\otimes\psi')_{a_1a_2}(t,\vec r,\vec R)=\psi^{(0)}_{a_1}(t,\vec
  R+\tfrac{\vec r}{2}) \psi^{\prime (0)}_{a_2}(t,\vec R-\tfrac{\vec r}{2}) \, ,
\end{equation}
with the relative
coordinates $r$ and the cms coordinates $R$ and using the
projectors onto the irreducible representations $R_\alpha$ introduced 
in~\eqref{eq:potential_colour}, we can perform the colour decomposition
\begin{equation}
(\psi\otimes\psi')_{a_1a_2}=\sum_{R_{\alpha}}P^{R_\alpha}_{\{a\}}
\,(\psi\otimes\psi')_{a_3a_4}
\equiv\sum_{R_\alpha}(\psi\otimes\psi')^{R_\alpha}_{a_1a_2}\,.
\end{equation}
This is analogous to the singlet-octet decomposition discussed
in~\cite{Pineda:1997bj} (see also~\cite{Brambilla:2004jw}, in 
particular eq.~(48)). With this notation, the colour structure of the 
Coulomb potential simplifies to
\begin{equation}
\begin{aligned}
\left[\psi^{(0)\dagger}{\bf T}^{(R)a} \psi^{(0)}\right]
\left [\psi^{\prime(0)\dagger}{\bf T}^{(R')a}\psi^{\prime(0)}\right]
&=\sum_{R_\alpha, R_\beta}
(\psi^{\prime\,\dagger}\otimes\psi^{\dagger})_{b_4b_3}
P^{R_\beta}_{\{b\}} {\bf T}^{(R)b}_{b_1a_1}{\bf T}^{(R')b}_{b_2a_2}
P^{R_\beta}_{\{a\}}(\psi\otimes\psi^\prime)_{a_3a_4}\\
&=\sum_{R_\alpha}(\psi\otimes\psi^\prime)^{R_\alpha\dagger }D_{R_\alpha}
(\psi\otimes\psi^\prime)^{R_\alpha} \, .
\end{aligned}
\end{equation}
Here the coefficients of the Coulomb potential are defined by the relation
\begin{equation}
\label{eq:coulomb-coeff}
  {\bf T}^{(R)b}_{a_1c_1}{\bf T}^{(R')b}_{a_2c_2}P^{R_{\alpha}}_{c_1c_2a_3a_4}
  = D_{R_\alpha}P^{R_{\alpha}}_{\{a\}}
\end{equation}
and  we have used the projection property
\begin{equation}
\label{eq:project}
  P^{R_{\alpha}}_{a_1a_2b_1b_2}P^{R_{\beta}}_{b_1b_2c_1c_2}=
\delta_{R_{\alpha} R_{\beta}}
P^{R_\alpha}_{a_1a_2c_1c_2}\, .
\end{equation}

Using the simplification of the Coulomb potential, the leading order 
PNRQCD Lagrangian for the decoupled fields can be
written in the tensor-product notation as a sum over the irreducible
colour representations,
\begin{equation}
\label{eq:NRQCD-irrep}
\begin{aligned}
\mathcal{L}_{\mbox{\tiny PNRQCD}} =&
\,\sum_{R_\alpha}\,\Bigl\{
 (\psi\otimes {\psi'})^{R_\alpha\dagger}
 \left(i \partial_s^0+\frac{\vec{\partial}_r^2}{2 m_{red}^2}
 +\frac{\vec{\partial}_R^2}{2 (2M)^2}\right)
(\psi\otimes\psi')^{R_\alpha}\\
& +\int d^3 \vec{r}\, 
\left[(\psi\otimes \psi')^{R_\alpha\dagger}
  (t,-\vec r/2, 0)\right]
\left(\frac{\alpha_s D_{R_\alpha}}{r}\right)
\left [({\psi}\otimes \psi')^{R_\alpha}(t,\vec r/2, 0)\right]
\Bigr\}\,,
\end{aligned}
\end{equation}
 with the reduced mass $m_{\text{red}}=m_Hm_{H'}/(m_H+m_{H'})$.
The kinetic term can be written in terms of the tensor field as shown 
here after a projection on the two-particle sector of the 
theory~\cite{Pineda:1997bj,Brambilla:2004jw}.

For top-antitop and squark-antisquark production
the irreducible representations are given by $3\otimes\bar 3=1\oplus 8$
and the coefficients are familiar from quarkonium physics (recall 
our sign convention for the Coulomb potential):
\begin{eqnarray}
&& D_1=-C_F=-\frac{N_C^2-1}{2N_C}\quad\text{(attractive)},
\nonumber\\
&& D_8=-\left[C_F-\frac{C_A}{2}\right]=  \frac{1}{2N_C}
\quad\text{(repulsive)} \, .
\label{eq:coulomb-squark}
\end{eqnarray}
The explicit values of the coefficients for all remaining representations
relevant for the production of coloured SUSY particles are collected in
appendix~\ref{sec:susy-coulomb}.

Since the Lagrangian is now diagonal in colour and spin, 
the leading-order Coulomb Green function is of the form
\begin{equation}
J^{\{\alpha\}}_{\{a\}}(q)=\sum_{R_\alpha}P^{R_{\alpha}}_{\{a\}}\;
\delta_{\alpha_1\alpha_3}\delta_{\alpha_2\alpha_4}\;J^{R_{\alpha}}(q)
\end{equation}
with the correlation function
\begin{equation}
J^{R_{\alpha}}(q)=
\int d^4z \,e^{iq\cdot z} 
\braket{0|[\psi^{\prime(0)} \psi^{(0)}](z) 
[\psi^{(0)\dagger} \psi^{\prime (0)\dagger}](0)|0}\,.
\end{equation}
In this expression the fields carry no colour and spin indices 
any more and the correlation functions are to be calculated with the 
term in (\ref{eq:NRQCD-irrep}) corresponding to the representation 
$R_\alpha$ with Coulomb potential $\alpha_s D_{R_\alpha}/r$.  
Since the annihilation fields $\psi$
annihilate the vacuum, we can replace
$(\psi\psi')(z)(\psi^\dagger\psi^{'\dagger})(0)=[(\psi\psi')(z),
(\psi^\dagger\psi^{'\dagger})(0)]$
in the vacuum expectation value and express the correlation function in
terms of the imaginary part of a time ordered product
 \begin{equation}
\begin{aligned}
J^{R_{\alpha}}(q)
&= 
\int d^4z \,e^{iq\cdot z} \,2\,\text{Im}
\braket{0|\mathrm{T}[\psi^{'(0)} \psi^{(0)}](z) 
[\psi^{(0)\dagger} \psi^{'(0)\dagger}](0)|0}\\
&= 2\,\text{Im}\,G_{\rm C}^{R_\alpha (0)}(0,0;E) \, .
\end{aligned}
\label{eq:potential-green}
 \end{equation}
Here we introduced the zero-distance Coulomb Green
function of the Schr{\"o}dinger operator 
$-\vec{\nabla}^2/(2m_{\text{red}})-(-D_{R_\alpha})\alpha_s/r$, 
i.e. the Green function $G^{R_\alpha(0)}_{\rm C}(\vec{r}_1,\vec{r}_2;E)$
evaluated
at $\vec{r}_1=\vec{r}_2=0$.

Using the representation of the Green function 
given in~\cite{Wichmann:1961}, the
$\overline{\text{MS}}$-subtracted zero-distance Coulomb
Green function including all-order gluon 
exchange, reads as follows~\cite{Beneke:1999zr}:
\begin{eqnarray}
G_{\rm C}^{R_\alpha(0)}(0,0;E) &= & 
-\frac{(2m_{\text{red}})^2}{4\pi} \Bigg\{
    \sqrt{-\frac{E}{2m_{\text{red}}}}
    + (-D_{R_\alpha})\alpha_s \bigg[\,
      \frac{1}{2}\ln \bigg(\!
      -\!\frac{8\,m_{\text{red}}E}{\mu^2}\bigg)
\nonumber\\
&&    -\,\frac{1}{2}  +\gamma_E
    +\psi\bigg(1-\frac{(-D_{R_\alpha})\alpha_s}{
      2\sqrt{-E/ (2m_{\text{red}})}}\bigg)\bigg]\Bigg\}\,.
\label{coulombGF}
\end{eqnarray}
Here $\gamma_E$ is the Euler-Mascheroni constant and one should apply
the prescription $E\to E+i\delta$ for stable heavy particles and
 $E\to E+i(\Gamma_H+\Gamma_{H'})/2$ for unstable ones.

For positive values of $E$ the potential function evaluates to 
the Sommerfeld factor
\begin{equation}
\label{eq:sommerfeld}
J_{R_\alpha}(E)=
\frac{(2 m_{\text red})^2\pi D_{R_\alpha}\alpha_s }{2\pi}
\left(e^{\pi D_{R_\alpha}\alpha_s
\sqrt{\frac{2m_{\text{red}}}{E}}}-1\right)^{-1}
\,\qquad E>0.
\end{equation}
If the potential is attractive, $D_{R_\alpha}<0$, there is a 
sum of bound states below threshold given by 
\begin{equation}
J_{R_\alpha}^{\text{bound}}(E)=2
\sum_{n=1}^\infty \delta(E-E_n)
\left(\frac{2m_{\text{red}} (-D_{R_\alpha})\alpha_s}{2 n}\right)^3
\,\qquad E<0
\end{equation}
with bound-state energies
\begin{equation}
\label{eq:bound}
  E_n=-\frac{2m_{\text{red}}\alpha_s^2 D_{R_\alpha}^2}{4 n^2}\,.
\end{equation}
A series representation of the imaginary part of the Coulomb Green
function for finite widths and arbitrary energies can be found
in~\cite{Fadin:1990wx}.

The above results suffice for resummation with NLL accuracy as
performed in section~\ref{sec:squarks}. For NNLL accuracy in the
counting~\eqref{eq:syst} one has to resum in addition all
$\alpha_s\times (\alpha_s/\beta)^n$ corrections as well as the
non-relativistic logarithms of the form $\alpha_s^2\ln\beta$,
$\alpha_s^3\ln^2\beta,\dots$.  The fixed-order NNLO corrections of
this form have been obtained in~\cite{Beneke:2009ye}. An analytical
result resumming all $\alpha_s\times (\alpha_s/\beta)^n$ terms was
obtained in~\cite{Beneke:1999qg} and is quoted
e.g. in~\cite{Kiyo:2008bv}. Resummation of the non-relativistic
logarithms requires the generalization of results for top-quark pair
production obtained e.g. in~\cite{Hoang:2001mm,Pineda:2006ri} 
to arbitrary colour representations.

\subsection{Evolution equations of hard and soft functions}
\label{sec:rges}

In the momentum-space approach to threshold resummation one 
calculates the short-distance coefficients $C^{(i)}(\mu)$
at a hard scale $\mu_h\sim 2M$ and the soft function
$W(\omega,\mu)$ at a soft scale of the order of $\mu_s\sim M
\beta^2$.  Threshold logarithms $\log\beta$ are resummed by
using evolution equations to evolve both functions to an intermediate
factorization scale $\mu_f$.  In this subsection we will provide these
evolution equations in the colour basis~\eqref{eq:prod-basis} that
diagonalizes the soft function. In this case the structure of the
evolution equations is similar to those in the Drell-Yan 
process~\cite{Becher:2007ty,Ahrens:2008nc} and the
complications of matrix-valued anomalous dimensions depending on the
kinematics~\cite{Kidonakis:1997gm,Ahrens:2010zv} are avoided. The
resummed cross section obtained from solving the equations is given
in~\ref{sec:sum-sigma}.

The evolution equation of the hard coefficients has been obtained
in~\cite{Beneke:2009rj} from the results of~\cite{Becher:2009kw} for
the IR structure of general massive QCD amplitudes:
\begin{equation}
\label{eq:evolution-hard}
\frac{d}{d\ln\mu} C_{\{\alpha\}}^{(i)}(\mu) =\left(
\frac{1}{2}\,\gamma_{\text{cusp}}\left[
(C_r+C_{r'})\left(\ln\left(\frac{4M^2}{\mu^2}\right)-i \pi\right)
+i\pi C_{R_\alpha}\right]+\gamma^V_i \right) C_{\{\alpha\}}^{(i)}(\mu)  .
\end{equation}
Here  Casimir scaling was used 
to write the cusp anomalous dimension for a massless parton in the
representation $r$ in the form $\Gamma_{\text{cusp}}^r=C_r
\gamma_{\text{cusp}}$ which holds at least up to the three-loop order.
At least up to the two-loop level the anomalous dimension
$\gamma^V_i$ can be written in terms of single-particle anomalous dimensions:
\begin{equation}
\label{eq:gamma-v}
\gamma^V_i=\gamma^r+\gamma^{r'}+\gamma_{H,s}^{R_\alpha}.
\end{equation}
The anomalous dimension $\gamma_{H,s}^{R_\alpha}$ is related to a
massive particle in the final state representation $R_\alpha$ in the
pair $P_i=(r_\alpha',R_\alpha)$ defining the colour basis element
$c^{(i)}$ with index $i$.  While the one- and two-loop
anomalous-dimension coefficients $\gamma^r$ of massless quarks and
gluons have been available for some time, the two-loop results for the heavy
particle soft anomalous dimension $\gamma_{H,s}^{R_\alpha}$ have
become available only recently~\cite{Beneke:2009rj,Czakon:2009zw}.
The one-loop coefficients of the cusp and soft anomalous dimensions
are simply $\gamma_{\text{cusp}}^{(0)}=4$ and
$\gamma_{H,s}^{R_\alpha(0)}=-2C_{R_\alpha}$.  The anomalous dimensions
$\gamma^r$ related to the light partons do not obey Casimir scaling
already at one loop.  The explicit one- and two-loop results for all
remaining anomalous dimensions appearing in this section are available
in the literature and have been collected in~\cite{Beneke:2009rj}.  We
observe that, as noted for the production of a colour
octet final state particle in~\cite{Idilbi:2009cc}, the imaginary part
in the evolution equation~\eqref{eq:evolution-hard} can not simply
absorbed in the argument of the logarithm by the continuation $M^2\to
-(M^2+i0)$ which complicates the resummation of ``$\pi^2$-enhanced''
terms compared to colour singlet final states as in
Higgs production~\cite{Ahrens:2008qu,Ahrens:2008nc}. We will not
consider such a resummation of constant terms here.  The evolution
equation of the  hard-functions is obtained
from~\eqref{eq:evolution-hard} as
\begin{equation}
\label{eq:evolution-hard-2}
\frac{d}{d\ln\mu} H_{i}^{S}(\mu) =\left(
\gamma_{\text{cusp}}
(C_r+C_{r'})\ln\left(\frac{4M^2}{\mu^2}\right)
+2\gamma^V_i  \right) H_{i}^{S}(\mu)  .
\end{equation}

As discussed in~\cite{Beneke:2009rj} and section~\ref{sec:subleading},
starting from the two-loop order there are IR divergent contributions to the
short-distance coefficients that are related to UV divergences of
insertions of the non-Coulomb PNRQCD potentials in the extended
factorization formula~\eqref{eq:factform}. These divergences 
cause additional scale dependence, not included in 
(\ref{eq:evolution-hard-2}), which is cancelled by a non-trivial 
scale-dependence of the potential function including the 
non-Coulomb potential insertions. The factorization scale 
dependence in the separation of $H$ and $J$ is 
related to additional non-relativistic $\log\beta$ terms 
of the NNLL order. Since in this paper we do not
consider the resummation of these non-relativistic logarithms, in
order to obtain the evolution equation of the soft function, these
contributions to the scale dependence of the hard function have to be dropped 
as done in (\ref{eq:evolution-hard-2}). Beyond NNLL further 
complications can arise from the structure of  the extended
factorization formula~\eqref{eq:factform} including terms 
with higher-dimensional soft functions, but the discussion 
of resummation beyond NNLL is beyond the scope of this paper. 

The evolution equation of the soft function can be obtained from that
of the hard function using the factorization scale independence of the
hadronic cross section.  Consistent with our discussion above, we
consider the potential function to be scale independent which is
appropriate for the NLO potential function quoted
in~\cite{Kiyo:2008bv}. Scale invariance of the total cross section and
the known factorization scale dependence of the PDFs implies the
evolution equation of the partonic cross section
\begin{equation}
\label{eq:dmu-sigma}
  \frac{d}{d\ln\mu}\hat \sigma_{pp'}(z,\mu)
  =-\sum_{\tilde p,\tilde p'}\int_z^1\frac{dx}{x}\left(P_{p/\tilde
      p}(x)+ P_{p'/\tilde p'}(x)\right)
  \hat \sigma_{\tilde p \tilde p'}(z/x,\mu),
\end{equation}
where $P_{p/\tilde p\,}(x)$ are the Altarelli-Parisi 
splitting functions for the
splitting of a parton $p$ into a parton $\tilde p$.  
Consistent with the derivation
of the factorization formula in section~\ref{sec:factorize} from the on-shell
scattering process $pp'\to HH'X$ at production threshold, we use the $x\to 1$
limit of the splitting functions for a parton $p$ in the colour representation
$r$,
\begin{equation}
\label{eq:splitting}
P_{p/\tilde p\,}(x)
=\left(2 \Gamma^r_{\text{\text{cusp}}}(\alpha_s)
\frac{1}{[1-x]_+}+2 \gamma^{\phi,r}(\alpha_s) \delta(1-x)\right)
\delta_{p\tilde p}+\mathcal{O}(1-x) \, .
\end{equation}
As discussed in section~\ref{sec:subleading}, subleading collinear
terms could potentially be enhanced by the Coulomb singularity, so
care in dropping higher order corrections in equations like this is
required. In the present case, however, the corrections
to~\eqref{eq:splitting} are of the order $1-z=\beta^2$ 
so they they are not relevant at NNLL accuracy.  We recall the
property of the plus distribution,
\begin{equation}
  \int_z^1 dx\,f(x)\,\left[\frac{1}{(1-x)}\right]_+= 
  \int_z^1dx\,\frac{f(x)-f(1)}{(1-x)}-\int_0^z dx\,\frac{f(1)}{(1-x)}.
\end{equation}

Inserting the factorized form of the partonic cross section 
into~\eqref{eq:dmu-sigma} and making use of the evolution equation of the
short-distance coefficients~\eqref{eq:evolution-hard} leads to an evolution
equation of the soft function.  Using the relations $\hat s\approx 4M^2$,
$(1-z/x)\approx (1-z)-(1-x)$ valid in the $x\to 1$, $z\to 1$ limits and
introducing the variable $\omega=4M(1-x)$, the resulting equation can be
written in the same form as the equation found for the Drell-Yan
process~\cite{Becher:2007ty}:
\begin{align}
 \frac{d}{d\ln\mu}
 W_{i}^{R_\alpha}(\omega,\mu)
  =&-2\left[\left(\Gamma^r_{\text{\text{cusp}}}
+\Gamma^{r'}_{\text{\text{cusp}}}\right)\ln\left(\frac{\omega}{\mu}\right)
     +2\gamma^{R_\alpha}_{W,i}\right]
 W_{i}^{R_\alpha}(\omega,\mu)
\nonumber\\
&-2\left(\Gamma^r_{\text{\text{cusp}}}
+\Gamma^{r'}_{\text{\text{cusp}}}\right)
\int_0^\omega d\omega'
\frac{ W_{i}^{R_\alpha}(\omega',\mu)- 
W_{i}^{R_\alpha}(\omega,\mu)}{\omega-\omega'},
\label{eq:evolution-soft}
\end{align}
with the  anomalous-dimension coefficient
\begin{equation}
\label{eq:relation-soft-gamma}
\gamma_{W,i}^{R_\alpha}=\gamma^V_i+\gamma^{\phi,r}+\gamma^{\phi,r'}.
\end{equation}
These results hold for the colour basis (\ref{eq:prod-basis}) 
that diagonalizes the soft
function to all orders~\cite{Beneke:2009rj}, where only the diagonal
elements $W_i=W_{ii}$ of the soft function have to be considered.
Analogously to the anomalous dimension $\gamma^V_i$ of the hard
function~\eqref{eq:gamma-v}, at least up to the two-loop level the
anomalous-dimension coefficient of the soft
function~\eqref{eq:relation-soft-gamma} can be written in terms of
separate single-particle contributions
\begin{equation}
\label{eq:gamma-W}
\gamma_{W,i}^{R_\alpha}=\gamma_{H,s}^{R_\alpha}+\gamma^r_s+\gamma_s^{r'}
\end{equation}
with 
\begin{equation}
\label{eq:gamma-s-massless}
 \gamma^r_s=\gamma^r+\gamma^{\phi,r}.
\end{equation}
The anomalous-dimension coefficients $\gamma^r_s$ vanish at one-loop
level so that
$\gamma_{W,i}^{R_\alpha(0)}=\gamma_{H,s}^{R_\alpha(0)}=-2C_{R_\alpha}$
.  Taking the Fourier transform of the evolution equation of the
position space soft function given in~(3.30)
of~\cite{Beneke:2009rj} one obtains the momentum-space evolution
equation quoted in~\eqref{eq:evolution-soft}.  In order to obtain this
result, the logarithmic term multiplied by $\Gamma_{\text{cusp}}$ has
to be treated with care, for instance by writing $\ln
x=\lim_{\epsilon\to 0}\frac{1}{\epsilon}(1-x^{-\epsilon})$ (see
also~\cite{Korchemsky:1992xv}).

The terms proportional to the cusp anomalous dimensions in the
evolution equations are related to the resummation of double
logarithms $\ln^2\beta$, while the coefficients $\gamma^V_i$ and
$\gamma^{R_\alpha}_{W,i}$ are related to single logarithms. For 
LL summation one needs the one-loop cusp anomalous dimension while all
other quantities enter at LO. For an NLL resummation the required
ingredients are the two-loop cusp anomalous dimensions and the
one-loop anomalous dimensions $\gamma^V_i$ and
$\gamma^{R_\alpha}_{W,i}$. For a NNLL resummation one needs in
addition to the three-loop cusp anomalous dimensions and the two-loop
anomalous dimensions, the one-loop soft function~\eqref{eq:soft-one} and the
one-loop hard function~(\ref{eq:hard-nlo}).

\subsection{Resummed cross section}
\label{sec:sum-sigma}

\begin{figure}[t]
  \begin{center}
    \includegraphics[width=0.55\textwidth]{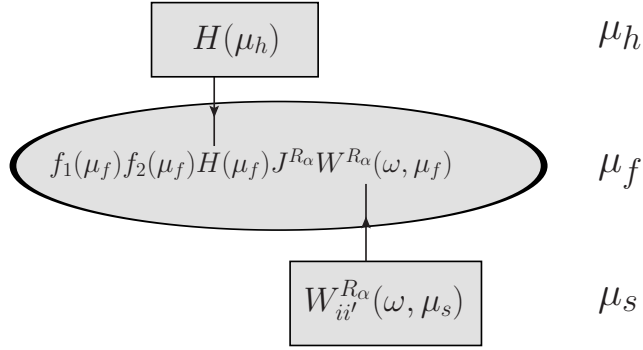}
  \caption{Sketch of the resummation of soft gluon corrections using RGEs. }
  \label{fig:running}
  \end{center}
\end{figure}

As mentioned in the beginning of section~\ref{sec:rges}, the
resummation of soft-gluon corrections in the approach 
of~\cite{Becher:2006nr,Becher:2006mr,Becher:2007ty} is
performed by calculating the hard and soft functions at scales $\mu_h$
and $\mu_s$ that minimize the radiative corrections to these
quantities and subsequently using the renormalization group 
equations~\eqref{eq:evolution-hard-2}
and~\eqref{eq:evolution-soft} to evolve to a common scale $\mu_f$
to compute the partonic cross section~\eqref{eq:fact-final} and
perform the convolution with the PDFs evaluated at the same scale (see
figure~\ref{fig:running}). 

Since we have determined a colour basis that
diagonalizes the soft function to all orders in perturbation theory for
all hadron collider processes of interest, the evolution equations
have the same form as those for the Drell-Yan 
process~\cite{Becher:2007ty}. 
The resummed hard function solving the evolution
equation~\eqref{eq:evolution-hard-2} is given by
\begin{equation}\label{eq:coeff_resummed}
H_i^S(\mu)=\exp[4 S(\mu_h,\mu)-2a_i^{V}(\mu_h,\mu)]
\left(\frac{4 M^2}{\mu_h^2}\right)^{-2a_\Gamma(\mu_h,\mu)} 
H_i^S(\mu_h)\, ,
\end{equation}
with the functions $S$, $a_i^{V}$ and $a_\Gamma$ defined
as \cite{Becher:2007ty}
\begin{eqnarray} 
\label{eq:res_funct_def}
S(\mu_h,\mu) &=& -\int_{\alpha_s(\mu_h)}^{\alpha_s(\mu)}d \alpha_s 
\frac{\Gamma^r_{\text{\text{cusp}}}(\alpha_s)
  +\Gamma^{r'}_{\text{\text{cusp}}}(\alpha_s)}{2\beta(\alpha_s)}
\int_{\alpha_s(\mu_h)}^{\alpha_s}\frac{d \alpha_s^{\prime}}
{\beta(\alpha_s^{\prime})} 
\, ,\nonumber\\
a_{\Gamma}(\mu_h,\mu) &=& -\int_{\alpha_s(\mu_h)}^{\alpha_s(\mu)} d \alpha_s 
\frac{\Gamma^{r}_{\text{\text{cusp}}}(\alpha_s)
+\Gamma^{r'}_{\text{\text{cusp}}}(\alpha_s)}{2\beta(\alpha_s)}
\, , \nonumber\\
a^{V}_i (\mu_h,\mu) &=& -\int_{\alpha_s(\mu_h)}^{\alpha_s(\mu)} d \alpha_s 
\frac{\gamma_i^{V}(\alpha_s)}{\beta(\alpha_s)}.
\end{eqnarray}
Here $\alpha_s(\mu)$ represents the QCD coupling in the 
$\overline{\rm MS}$ scheme and
$\beta(\alpha_s)$ the corresponding $\beta$-function.
Explicit results for the functions $\beta$, $a_\Gamma$ and the Sudakov
exponent $S$ up to the NLL order as needed for
section~\ref{sec:squarks}  are collected in
appendix~\ref{app:evolution}; expressions up to the N$^3$LL order
can be found in~\cite{Becher:2007ty}. 

The evolution equation of the soft function~\eqref{eq:evolution-soft}
can be solved in momentum space~\cite{Becher:2006nr,Becher:2006mr} by
introducing the Laplace-transform with respect to the variable $s =
1/(e^{\gamma_E} \mu e^{\rho/2})$,
\begin{equation}
\tilde{s}^{R_\alpha}_{i}(\rho,\mu) =\int_{0_-}^{\infty} d \omega e^{-s \omega}
\, \overline W_{i}^{R_\alpha}(\omega,\mu) \,,
\end{equation}
where we have defined the $\overline{\text{MS}}$-renormalized soft
function $\overline W^{R_\alpha}_i$. In practice, it is not necessary
to perform the Laplace transform explicitly, since the soft function in
position space $\hat W^{R_\alpha}_i(z_0,\mu)$ depends on the arguments
solely through the
combination~\cite{Korchemsky:1992xv,Korchemsky:1993uz}
\begin{equation}
\label{eq:def-L}
\frac{iz_0\mu e^{\gamma_E}}{2}\equiv e^{L/2}\,.
\end{equation}
It is then easy to see that the function $\tilde s_i^{R_\alpha}(\rho)$
is obtained by
simply replacing $L\to -\rho$ in the $\overline{\text{MS}}$-renormalized
result for the soft function in position space. From the results of the
tree-level and one-loop soft function in the diagonal
basis~\eqref{eq:prod-basis} quoted in~\eqref{eq:soft-one} we obtain,
discarding the $1/\epsilon$ poles due to the $\overline{\rm MS}$ subtraction,
\begin{align}
 \tilde  s^{(0)R_\alpha}_{i}(\rho,\mu)&=1, \\
 \tilde s^{(1)R_\alpha}_{i}(\rho,\mu)&=\left(
C_r+C_{r'}\right)\left(
   \rho^2+\frac{\pi^2}{6}\right)
- 2C_{R_\alpha}\left( \rho-2\right) .
\label{eq:soft-laplace}
\end{align}

The resummed soft function can be written
in terms of the Laplace transform with the argument $\rho$ replaced by
the derivative with respect to an auxiliary variable $\eta$ that is set
to $\eta = 2 a_{\Gamma}(\mu_s,\mu)$ after performing the
derivatives:
\begin{equation}
\label{eq:w-resummed}
W^{R_\alpha,\text{res}}_{i}(\omega,\mu)=
\exp[-4 S(\mu_s,\mu)+2 a^{R_\alpha}_{W,i}(\mu_s,\mu)]
\tilde{s}_{i}^{R_\alpha}(\partial_\eta,\mu_s) 
\frac{1}{\omega} \left(\frac{\omega}{\mu_s}\right)^{2 \eta} \theta(\omega)
\frac{e^{-2 \gamma_E \eta}}{\Gamma(2 \eta)}\, .
\end{equation}
Here the functions $a^{R_\alpha}_{W,i}$ are defined in analogy to the
functions~\eqref{eq:res_funct_def} with the obvious replacement of the
anomalous dimensions $\gamma_i^{V}\to\gamma_{W,i}^{R_\alpha}$.  For
$\eta <0$ the solution~\eqref{eq:w-resummed} should be understood in
the distributional sense~\cite{Becher:2006mr} as will be discussed
below.
The evolution equation of the soft function~\eqref{eq:evolution-soft} can
also be solved in Mellin moment space~\cite{Korchemsky:1993uz} and 
the standard resummation formula in Mellin space is 
reproduced.  The precise relation between
the quantities in the momentum space approach to resummation and the
Mellin moment space formalism is discussed
in~\cite{Becher:2006mr,Becher:2007ty,Ahrens:2008nc,Beneke:2009rj}.

Using the solutions of the RGEs to evolve the hard and soft function in the
factorized expression for the partonic cross
section~\eqref{eq:fact-final} to the common factorization scale
$\mu_f$ we obtain the result for the resummed cross section
\begin{align}
\hat\sigma^{\text{res}}_{pp'}(\hat s,\mu_f)=&
\sum_{S=|s-s'|}^{s+s'}\sum_{i}
H^S_{i}(\mu_h)  \;
 U_i(M,\mu_h,\mu_s,\mu_f)
 \left(\frac{2M}{\mu_s}\right)^{-2\eta}\nonumber\\
&\times
\tilde{s}_{i}^{R_\alpha}(\partial_\eta,\mu_s)
\frac{e^{-2 \gamma_E \eta}}{\Gamma(2 \eta)}\,
\int_0^\infty d \omega \;
\frac{ J^S_{R_{\alpha}}(E-\tfrac{\omega}{2})}{\omega} 
\left(\frac{\omega}{\mu_s}\right)^{2 \eta}\nonumber\\
 =&\sum_{S=|s-s'|}^{s+s'}\sum_{i}
H^S_{i}(\mu_h)  \;
 U_i(M,\mu_h,\mu_s,\mu_f)\nonumber\\
&\times
\int_0^\infty d \omega \;
\frac{ J^S_{R_{\alpha}}(E-\tfrac{\omega}{2})}{\omega} 
\left(\frac{\omega}{2M}\right)^{2 \eta}
\tilde{s}_{i}^{R_\alpha}
\left(2\ln\left(\frac{\omega}{\mu_s}\right)+\partial_\eta,\mu_s\right)
\frac{e^{-2 \gamma_E \eta}}{\Gamma(2 \eta)}\,,
\label{eq:fact-resum}
\end{align}
where have defined the evolution function
\begin{eqnarray}
U_i(M,\mu_h,\mu_f,\mu_s )&=&
\left(\frac{4M^2}{\mu_h^2}\right)^{-2a_\Gamma(\mu_h,\mu_s)}\,
\left(\frac{\mu_h^2}{\mu_s^2}\right)^{\eta}
\times \,\exp\Big[4  (S(\mu_h,\mu_f)-S(\mu_s,\mu_f))
\nonumber\\[0.1cm]
&&
 -\,2a_i^{V}(\mu_h,\mu_s) +2 a^{\phi,r}(\mu_s,\mu_f)+
2 a^{\phi,r'}(\mu_s,\mu_f)\Big]
\label{eq:def-u}
\end{eqnarray}
and now $\eta = 2 a_{\Gamma}(\mu_s,\mu_f)$. The sum over the final state
representations $R_\alpha$ in the factorization
formula~\eqref{eq:fact-final} has disappeared in the colour
basis~\eqref{eq:prod-basis} since there is a unique final state representation
for each term in the sum over $i$ (see the 
expression~\eqref{eq:soft-diagonal} for the diagonal soft function).
The evolution function $U_i$ has been simplified using  the 
identity
\begin{equation}
  a_\Gamma(\mu_h,\mu_f)+a_\Gamma(\mu_f,\mu_s)=a_\Gamma(\mu_h,\mu_s) \, .
\end{equation}
It could be further simplified using the 
identity~\cite{Becher:2006mr} 
\begin{equation}
\label{eq:s-id}
  S(\mu_h,\mu_f)- S(\mu_s,\mu_f)= 
  S(\mu_h,\mu_s)-a_{\Gamma}(\mu_s,\mu_f)\ln\frac{\mu_h}{\mu_s}
\end{equation}
to obtain
\begin{eqnarray}
U_i(M,\mu_h,\mu_f,\mu_s) &=&
\left(\frac{4M^2}{\mu_h^2}\right)^{-2a_\Gamma(\mu_h,\mu_s)}
\times 
\exp\Big[4  S(\mu_h,\mu_s)
\nonumber\\[0.1cm]
&& -\,2a_i^{V}(\mu_h,\mu_s) +2 a^{\phi,r}(\mu_s,\mu_f)
+2 a^{\phi,r'}(\mu_s,\mu_f)\Big],
\end{eqnarray}
which makes it clear that the $\mu_f$ dependence is related to 
the parton density functions. Eq.~(\ref{eq:s-id}) is 
formally satisfied at whatever N$^k$LL accuracy employed, but not
strictly valid in an expansion in the strong coupling since $S$ and
$a_\Gamma$ are evaluated at different orders. 
In the numerical evaluation we do not use this simplification 
but use (\ref{eq:def-u}).

A similar convolution of the resummed Coulomb Green function as
in~\eqref{eq:fact-resum} was encountered in~\cite{Actis:2008rb} where
the mixed Coulomb and one-loop soft corrections to $W$-pair production
were considered. The formula~\eqref{eq:fact-resum} generalizes the
result in~\cite{Actis:2008rb} to QCD and by resumming soft-gluons to
all orders. An analogous formula was already used
in~\cite{Falgari:2009zz} to resum QED corrections of the form
$\ln(\Gamma_W/M_W)$ for $W$-pair production in electron-positron
collisions.

At the NLL order only the trivial, LO ${\cal O}(\alpha_s^0)$ soft function, the
spin-independent LO potential function and the leading-order hard functions
are required so that the resummed cross section is given by
\begin{align}
\hat\sigma^{\text{NLL}}_{pp'}(\hat s,\mu_f)
 =&\sum_{i}H^{(0)}_{i}(\mu_h)\,
 U_i(M,\mu_h,\mu_s,\mu_f)
\frac{e^{-2 \gamma_E \eta}}{\Gamma(2 \eta)}\,
\int_0^\infty \!d \omega \,
\frac{ J_{R_{\alpha}}(E-\tfrac{\omega}{2})}{\omega} 
\left(\frac{\omega}{2M}\right)^{2 \eta}\,.
\label{eq:resum-NLL}
\end{align}

As mentioned before, for $\eta<0$ the kernel
$\frac{1}{\omega}\left(\frac{\omega}{2M}\right)^{2\eta}$ should be
understood in the distribution
sense~\cite{Becher:2006mr}. Neglecting heavy particle decay and
bound-state effects, the potential function $J_{R_\alpha}$ vanishes for
$E<0$ and the $\omega$ integral is defined as the
star-distribution~\cite{Becher:2006mr}
\begin{eqnarray}
 \int_0^{2E} d\omega\,f(\omega)
 \left[\frac{1}{\omega}\left(\frac{\omega}{2M}\right)^{2\eta}\right]_*
&=& \int_0^{2E}d\omega\,
\frac{f(\omega)-f(0)-\omega f'(0)}{\omega}
\left(\frac{\omega}{2M}\right)^{2\eta}\nonumber\\
&&+\,\left[\frac{f(0)}{2\eta}
+\frac{2E}{2\eta+1}f'(0)\right]\left(\frac{E}{M}\right)^{2\eta}\,,
\label{eq:star-dist}
\end{eqnarray}
where the double subtraction is sufficient to render the
integral convergent for $\eta >-1$. The poles of the original integrand at
$\eta=0$ and $\eta=-\frac{1}{2}$ show up explicitly in the second line
but are cancelled by the overall prefactor $1/\Gamma(2\eta)$ in 
(\ref{eq:fact-resum}) and (\ref{eq:resum-NLL}).

Note that the all-order solution~\eqref{eq:fact-resum} to the RGEs
does not depend on the hard and soft scales $\mu_h$ and $\mu_s$, but
truncating the perturbative expansion at a finite order in $\alpha_s$
introduces a residual dependence on these scales, which is of
higher order in $\alpha_s$.   As
pointed out in~\cite{Becher:2006mr,Becher:2007ty} the standard
resummation formula in Mellin space
following~\cite{Sterman:1986aj,Catani:1989ne} corresponds to the
implicit scale choices $\mu_h=\mu_f$ and $\mu_s=M/N$ for the $N$-th
Mellin moment.  The explicit appearance of these scales in the
momentum-space formalism allows to use the residual scale dependence
of the resummed cross section to estimate the remaining uncertainties
from uncalculated higher order corrections. Our choices for these
scales and the estimates for the remaining scale uncertainties are
discussed in section~\ref{sec:squark-soft}.

\subsection{Matching to the fixed-order NLO calculation}
\label{sec:match}

For the implementation of the NLL resummed cross section, we
match to the
fixed-order NLO calculation valid outside the threshold region. The
matched result for the partonic cross section is then given by
\begin{equation}
  \begin{aligned}
  \hat\sigma^{\text{matched}}_{pp'}(\hat s)
  &=\left[\hat\sigma^{\text{NLL}}_{pp'}(\hat s)-
    \hat\sigma^{\text{NLL}(1)}_{pp'}(\hat s)\right]
  \theta(\Lambda-[\sqrt{\hat s}-2 M])+ \hat\sigma^{\text{NLO}}_{pp'}(\hat s)
\\[0.1cm]
&\equiv  \Delta \hat\sigma^{\text{NLL}}_{pp'}(\hat s,\Lambda)+
 \hat\sigma^{\text{NLO}}_{pp'}(\hat s)ß \,,
  \end{aligned}
\label{eq:cross_matched}
\end{equation}
where $\hat\sigma^{\text{NLO}}_{pp'}(\hat s)$ is the fixed-order NLO
cross section obtained in standard perturbation theory and
$\hat\sigma^{\text{NLL}(1)}_{pp'}$ is the resummed cross section
expanded to NLO. The result is given explicitly in~\eqref{eq:NLLexp}
in appendix~\ref{app:expand}.  Since the resummed NLL cross section is
expected to be a good approximation to the total cross section only
near the threshold, we allowed for a cutoff $E=\sqrt{\hat
  s}-2M<\Lambda$ to switch off the resummation outside the threshold
region, as was done in~\cite{Moch:2008qy}. In the calculation of 
squark-antisquark production we choose, however, not to introduce this cut-off 
as discussed in section~\ref{sec:squark-soft}.

The  total hadronic cross section at NLL is then obtained by
convoluting~\eqref{eq:cross_matched} with the parton luminosity, 
\begin{equation}
\sigma^{\text{matched}}_{N_1 N_2\to HH^\prime X}(s)=
\sum_{p,p'=q,\bar q,g}\,\int_{4 M^2/s}^1 \!d\tau\,L_{pp'}(\tau,\mu_f)
\,\hat\sigma^{\text{matched}}_{pp'} (s \tau,\mu_f)\,,
\label{eq:match-had}
\end{equation}
where the parton luminosity is defined in terms of the PDFs 
in (\ref{eq:lumi}).


\section{NLL resummation for squark-antisquark production}
\label{sec:squarks}
In this section we use the results of section~\ref{sec:resum} to
perform the NLL resummation for squark-antisquark production 
in proton-proton and proton-antiproton collisions. 
According to the systematics introduced in~\eqref{eq:syst}, NLL
accuracy requires the combined resummation of Coulomb gluons and soft
gluons.  Therefore our results extend previous ones on 
squark-antisquark production at higher
order~\cite{Kulesza:2008jb,Kulesza:2009kq,Langenfeld:2009eg,Beenakker:2009ha},
that treated soft and Coulomb resummations separately or used a 
fixed-order expansion.  In section~\ref{sec:squark-eft} we perform the
matching to the effective theory, and compare the threshold
approximation of the LO and NLO cross sections to the full results. 
In~\ref{sec:squark-soft} we discuss our choices for the hard, soft 
and Coulomb scales appearing in the resummed cross section and 
present predictions for squark-antisquark production at the Tevatron 
and the LHC at cms energies of $7, 10$ and $14$~TeV. We 
also compare to results from soft gluon resummation in the 
Mellin-moment approach~\cite{Beenakker:2009ha}.

\subsection{Threshold expansion of the squark-antisquark cross section}
\label{sec:squark-eft}

In the following we perform the matching from the MSSM to an
effective theory containing only (anti-)collinear partons and
non-relativistic squarks and antisquarks, and derive the hard functions
for the partonic subprocesses.  As discussed in section~\ref{sec:hard}, the
hard functions required for NLL resummation can be inferred from the
leading-order cross sections for the colour-singlet and octet
production channels~\cite{Langenfeld:2009eg,Kulesza:2009kq}. Therefore
the explicit matching to the EFT is not strictly necessary, 
but is included here in some detail in
order to provide an illustration for the somewhat abstract discussion
in section~\ref{sec:factorize}. We also use the general result for
the threshold expansion at NLO~\eqref{eq:NLOexp} to reproduce the 
known threshold behaviour of the NLO cross section~\cite{Beenakker:1996ch}.

\begin{figure}[t]
  \begin{center}
     \includegraphics[width=0.46\textwidth]{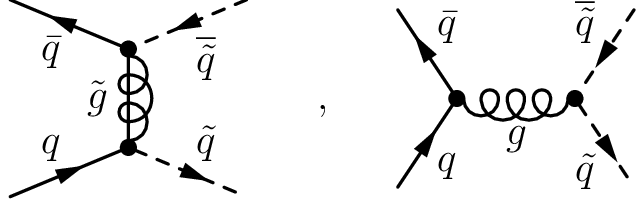} \\[0.5cm]
     \includegraphics[width=\textwidth]{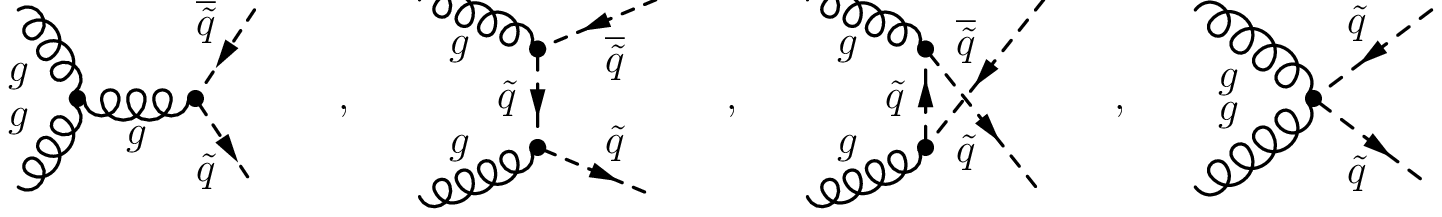}
\caption{Leading-order Feynman diagrams for the quark-antiquark (top) and
      gluon-fusion (bottom) induced production of squark-antisquark
      pairs.}
    \label{fig:squark-diags}
  \end{center}
\end{figure}

We consider squark-antisquark production at hadron colliders, which at
leading order proceeds through quark-antiquark and gluon-induced
subprocesses:
\begin{equation}
\label{eq:squark-antisquark}
\begin{aligned}
q_i(k_1)\bar q_j(k_2)&\to \tilde
q_{\sigma_1k}(p_1)\overline{\tilde q}_{\sigma_2 l}(p_2)\\
g(k_1)g(k_2)&\to \tilde
q_{\sigma_1i}(p_1)\overline{\tilde q}_{\sigma_2 j}(p_2) \,  ,
\end{aligned}
\end{equation}
where $i, j, k, l$ denote the quark/squark flavours and
$\sigma_{1,2}=L/R$ label the scalar partners of the left/right-handed
quarks.  The relevant Feynman diagrams are shown in
figure~\ref{fig:squark-diags}.  We follow the setup
of~\cite{Beenakker:1996ch} and take all squarks
to be mass degenerate and do not consider top squarks.  The
first restriction is done for simplicity (and was also used in other
recent works on higher-order soft-gluon
effects~\cite{Kulesza:2008jb,Kulesza:2009kq,Langenfeld:2009eg,Beenakker:2009ha})
and is not essential in our formalism, since the results in
section~\ref{sec:resum} include the case of different final state
masses. The application to top squarks would require an extension of
our framework, since in the quark-antiquark production channel they
are produced in a $P$-wave~\cite{Bigi:1991mi}. Resummation for top
squarks was performed recently in~\cite{Beenakker:2010nq} assuming
that the resummation formalism for $S$-wave production can be applied
to $P$-wave dominated processes.
Since, in the present work, we are concerned with the total cross
section, there are no sizable corrections from the finite squark width
$\Gamma$ as long as $\Gamma\ll M$ unlike the case of the invariant
mass distribution near threshold. Contributions to the cross section
from below the nominal production threshold will be included through
the bound-state contribution to the Coulomb Green function.  From a
recent study of threshold effects in gluino
production~\cite{Hagiwara:2009hq} we expect additional finite-width
effects at most of the order of these bound state corrections.

To obtain the leading-order
production operators we evaluate the tree-level scattering amplitudes
for the processes~\eqref{eq:squark-antisquark} at the production
threshold, $(k_1+k_2)^2=4m_{\tilde q}^2$, and determine the
short-distance coefficients from the matching
condition~\eqref{eq:match} in order to compute the hard
functions~\eqref{eq:hard-colour}.
 
\subsubsection{Production from quark-antiquark fusion}

The tree-level scattering
amplitude for quark-antiquark induced squark-antisquark production at
threshold is dominated by $t$-channel gluino exchange, if quark
and squark (and antiquark and antisquark) flavours are
identical.  Near threshold the full-theory tree-level scattering 
amplitude for the quark-antiquark channel is
\begin{equation}
\label{eq:qq-sq-amp}
 i \mathcal{A}^{(0)}_{\{a\}}
(q_i \bar q_j\to\tilde q_{kL} \overline{\tilde q}_{lR})
 =-\frac{2ig_s^2m_{\tilde g}}{{m_{\tilde q}^2+m_{\tilde g}^2}}
 \delta_{ik}\delta_{jl}
 T^b_{a_2 a_4}T^b_{a_3 a_1}\bar v(m_{\tilde q}\bar n)
 P_{L}u(m_{\tilde q}n)\,.
\end{equation}
Here we use the usual projectors $P_{R/L}=(1\pm\gamma^5)/2$.
There are analogous expressions with left and right labels exchanged.

\begin{figure}[t]
  \begin{center}
    \includegraphics[width=0.49\textwidth]{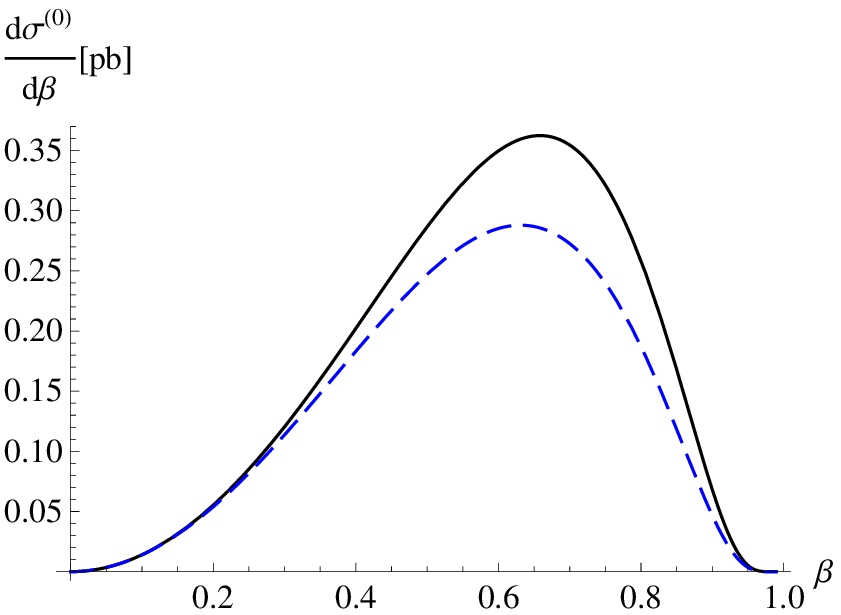} 
    \includegraphics[width=0.49\textwidth]{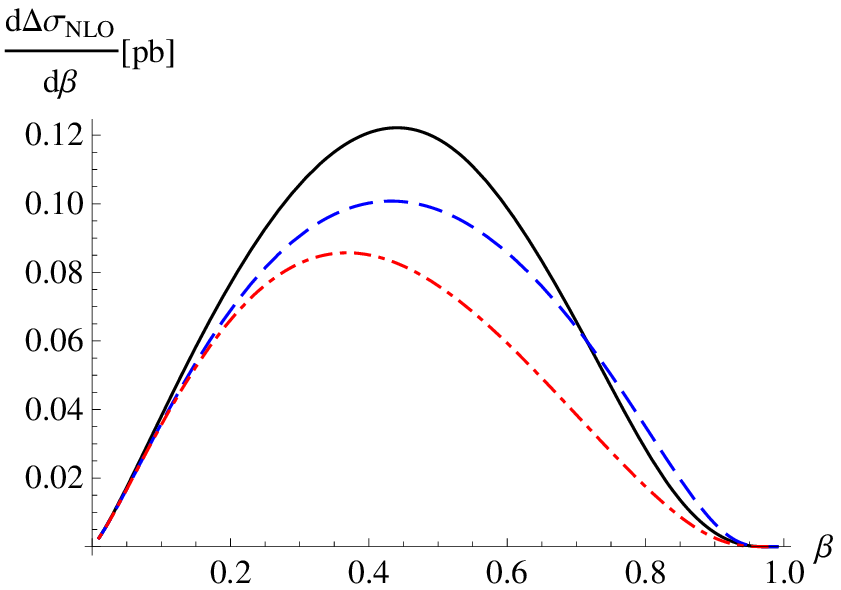}
    \caption{Partonic cross sections for $q_i\bar q_j\to\tilde
      q_k\bar{\tilde q_l}$, multiplied with the quark-antiquark luminosity
      and summed over quark and squark flavours, for $m_{\tilde q}=1$
      TeV and $m_{\tilde g}=1.25$~TeV at the $14$ TeV LHC.  Left
      panel: exact leading order result (black/solid) and threshold
      approximation (blue/dashed). Right panel: exact NLO result
      (black/solid) and the approximation based on~\eqref{eq:NLOqq}
      with $h^{(1)}_{q\bar q} (\mu)=0$, using the threshold
      approximation of the tree (red/dot-dashed) and the full tree
      (blue/dashed).}
    \label{fig:partonqq}
  \end{center}
\end{figure}

In the effective theory we introduce fields $\psi_{i}$
($\psi'_{i}$) that annihilate a left squark (right antisquark) of
flavour $i$. The quark fields are given by the SCET
fields~\eqref{eq:quark-scet}.  We can then match the amplitude onto
the effective theory according to the condition~\eqref{eq:match}, with
the production operator given by (no sum over flavours is implied)
\begin{equation}
\label{eq:squark-prod-q}
\mathcal{O}^{(0)}_{\{a;\alpha\}}= 
\left[(\bar{\xi}_{j,\overline{c};\alpha_2}W_{\overline{c}})_{a_2}
  (W^\dagger_{c}  \xi_{i,c;\alpha_1})_{a_1}\right]
\left(\psi^{\dagger}_{i;a_3}\psi^{'\dagger}_{j;a_4}\right)\,.
\end{equation}
Here $ \xi_{i,c}$ and $\xi_{i,\bar c}$ are the SCET fields describing 
collinear and anticollinear  quarks with flavour $i$. 
From the matching condition~\eqref{eq:match} we can read off the
matching coefficient, taking the non-relativistic normalization factor
$2m_{\tilde q}$ into account:
\begin{equation}
   C_{q\bar q, \{a;\alpha\}}=-\frac{g_s^2m_{\tilde g}}{
m_{\tilde q}(m_{\tilde q}^2+m_{\tilde g}^2)}
 T^b_{a_2 a_4}T^b_{a_3 a_1} (P_{L})_{\alpha_2 \alpha_1}\,.
\end{equation}
We now introduce the decomposition of the coefficients into a colour
basis~\eqref{eq:colour-wilson}, but leave the spin indices open. For the
case at hand, the colour basis has been given already
in~\eqref{eq:basis_33}. We find for the matching coefficients
for the two colour channels: 
\begin{eqnarray}
  C_{q\bar q, \{\alpha\}}^{(1)} &=& 
(-C_F)\,\frac{4\pi\alpha_s m_{\tilde g}}{m_{\tilde q}(
m_{\tilde q}^2+m_{\tilde g}^2)} (P_L)_{\alpha_2 \alpha_1}\,,
\nonumber\\
  C_{q\bar q, \{\alpha\}}^{(2)}&=&\sqrt{\frac{C_F}{2N_C}}\,
  \frac{4\pi\alpha_s m_{\tilde g}}{m_{\tilde q}(m_{\tilde q}^2+m_{\tilde g}^2)}
  (P_L)_{\alpha_2 \alpha_1}\,.
\label{eq:squark-match-q}
\end{eqnarray}
From the definition~\eqref{eq:hard-colour} and the polarization
sum~\eqref{eq:nq} we obtain the diagonal elements of the hard
function\footnote{For simplicity, in \eqref{eq:H_qq} and
  \eqref{eq:H_gg} we set $1/(2 \hat{s}) \sim 1/(8
  m_{\tilde{q}}^2)$. However, in our numerical implementation of the
  resummed cross section \eqref{eq:resum-NLL} this prefactor is kept
  unexpanded.},
\begin{equation} \label{eq:H_qq}
  H_{ii}^{(0)}=\frac{1}{8 m_{\tilde q}^2}\, h_i^{(0)}
\left(\frac{4\pi\alpha_s m_{\tilde g} }{
    m_{\tilde q}(m_{\tilde q}^2+m_{\tilde g}^2)}\right)^2 
\frac{m_{\tilde q}^2}{N_c^2}
\tr\left[\frac{\slash n}{2}\frac{\slash{\bar n}}{2}P_L \right] 
=2 h_i^{(0)} \left(\frac{\pi\alpha_s m_{\tilde g} }{N_c
    m_{\tilde q}(m_{\tilde q}^2+m_{\tilde g}^2)}\right)^2 \, ,
\end{equation}
with colour factors
\begin{equation}
   h_1^{(0)}=C_F^2 , \qquad \quad h_2^{(0)}=\frac{C_F}{2N_c}
\end{equation}
for the two colour channels. 
From~\eqref{eq:sigma-hard} we obtain the total partonic Born cross
section at threshold
\begin{equation}
\hat\sigma_{q_i\bar q_j\to \tilde q_{iL}
\overline{\tilde q}_{jR}}^{(0)}|_{\hat s=4m_{\tilde q}}
     =\frac{C_F \beta}{2\pi N_c}
 \left(\frac{\pi\alpha_s m_{\tilde g} }{
   (m_{\tilde q}^2+m_{\tilde g}^2)}\right)^2 \, .
\end{equation}
Adding the same expression with left and right squark labels exchanged
reproduces the result given e.g. in~\cite{Beenakker:1996ch}.  In
figure~\ref{fig:partonqq} we compare the partonic cross section at
threshold to the full partonic LO cross
section~\cite{Beenakker:1996ch}, for a squark mass of $1$~TeV at the
LHC with $14$~TeV cms energy. We plot the integrand of the convolution
in the formula for the hadronic cross section~\eqref{eq:sigmahad1},
i.e. the product of the partonic cross section and the parton
luminosity, as a function of $\beta$,
taking the Jacobian $\frac{\partial \tau}{\partial\beta}$
into account:
\begin{equation}
\label{eq:sigma-beta}
\frac{d \sigma_{q\bar q\to \tilde q_L
\overline{\tilde q}_R}^{(0)}}{d\beta}=
\frac{8\beta m_{\tilde q}^2}{s(1-\beta^2)^2}
L_{q\bar q}(\beta,\mu_f)\hat\sigma_{q\bar q\to \tilde q_L
\overline{\tilde q}_R}^{(0)}.
\end{equation}
In the plot we use the MSTW2008LO PDFs~\cite{Martin:2009iq}
with $\mu_f=m_{\tilde{q}}$.
It can be seen that the
threshold approximation is applicable for $\beta\lesssim 0.3$, while
the main contribution to the cross section comes from the region
$\beta\approx 0.6$. The results shown here are for a gluino-squark
mass ratio of $1.25$. For other gluino masses the discrepancies in
the region $\beta >0.3$ can be somewhat bigger.

An approximation to the NLO cross section can be obtained by
expanding the resummed cross section~\eqref{eq:fact-resum} to order
$\alpha_s^3$. The resulting expression can be written in the form
$\Delta\sigma^{(1)}_{pp'}=\sigma^{(0)}_{pp'}\frac{\alpha_s}{4\pi}
f^{(1)}_{pp'}+\mathcal{O}(\beta)$, where the scaling functions
$f^{(1)}_{pp'}$ for an arbitrary colour channel are given explicitly
in~\eqref{eq:NLOexp}.  Summing up the two colour channels we obtain the
threshold expansion of the scaling function for the quark-antiquark channel
\begin{eqnarray}
f^{(1)}_{q\bar q} &\!\!=\!\!& 
\frac{\pi^2 (N_c^2-2)}{N_c}\frac{1}{\beta} +
8 C_F\bigg[\ln^2\left(\frac{8 m_{\tilde q}\beta^2}{\mu}\right)
+8 -\frac{11 \pi^2}{24}\bigg]
- \,4 \,\frac{4N_c^2-3}{N_c}\,
\ln\left(\frac{8 m_{\tilde q}\beta^2}{\mu}\right)\nonumber\\
&&   
+ \frac{12}{N_c} +h^{(1)}_{q\bar q} (\mu)+ 
{\cal O}(\beta)\,.
\label{eq:NLOqq}
\end{eqnarray}
The logarithmically enhanced terms and the Coulomb correction agree
with the results of~\cite{Beenakker:1996ch}.  The one-loop hard
coefficient $h^{(1)}$ is currently unknown in analytical form. In the
right-hand plot in figure~\ref{fig:partonqq} we plot the full NLO
corrections to the partonic cross section obtained from the
parameterization given in~\cite{Langenfeld:2009eg} (solid/black) to
the threshold approximation~\eqref{eq:NLOqq} with the constant
$h^{(1)}_{q\bar q}$ set to zero. In the red/dot-dashed curve we
multiply the correction~\eqref{eq:NLOqq} with the threshold
approximation of the tree, in the blue/dashed curve we use the full
tree cross section. It can be seen that the NLO corrections are peaked
closer to threshold than the tree cross section, as might be expected.
Multiplying the scaling function by the full tree leads to an
improved agreement with the full NLO result. Overall the approximation
of the full NLO result by the threshold expansion is 
quite good, in the sense that integrating the approximated partonic 
cross section captures the bulk part of the exact result.

\subsubsection{Production from gluon fusion}

For the gluon induced process the $s$-channel diagram is again
$P$-wave suppressed, so the dominant contribution comes from the $t$-
and $u$-channel diagrams and the quartic vertex. These diagrams give
\begin{equation}
  i \mathcal{A}^{(0)}_{\{a\}}(gg\to\tilde q_{iR} \,\overline{\tilde q}_{jR})
= i g_s^2 \delta_{ij}\{T^{a_1},T^{a_2}\}_{a_3a_4}(\epsilon_1^\mu 
g^\perp_{\mu\nu}\epsilon^\nu_2)\, ,
\end{equation}
with the transverse metric
$g^\perp_{\mu\nu}=g_{\mu\nu}-\frac{1}{2}(n_\mu\bar n_\nu+\bar n_\mu
n_\nu)$.  This form of the scattering amplitude at threshold is
reproduced by the matrix element of the effective theory operator (no
sum over flavours is implied)
\begin{equation}
\label{eq:squark-prod-g}
O^{(0)}_{\{\mu;a\}}=
\mathcal{A}^\perp_{c;a_1\mu_1} \mathcal{A}^\perp_{\bar c;a_2\mu_2}
\psi^{\dagger}_{i;a_3}\psi^{'\dagger}_{i;a_4}\,.
\end{equation}
Here it was used that the polarization vector corresponding to a free SCET
gluon fields $\mathcal{A}^\perp$ is given by $\epsilon^\mu_\perp
=g^{\perp\, \mu\nu}\epsilon_\nu$. This can be seen from~\eqref{eq:gluon-scet}
using a Fourier transformation after setting the strong coupling to zero, so
that $W_c=1$ and the non-abelian terms in the field strength vanish.

\begin{figure}[t]
  \begin{center}
    \includegraphics[width=0.49\textwidth]{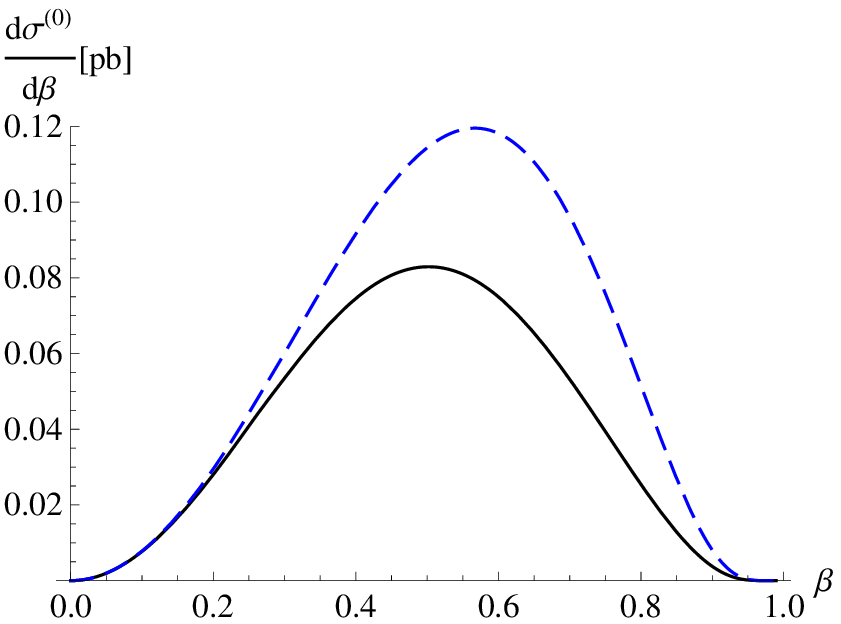} 
    \includegraphics[width=0.49\textwidth]{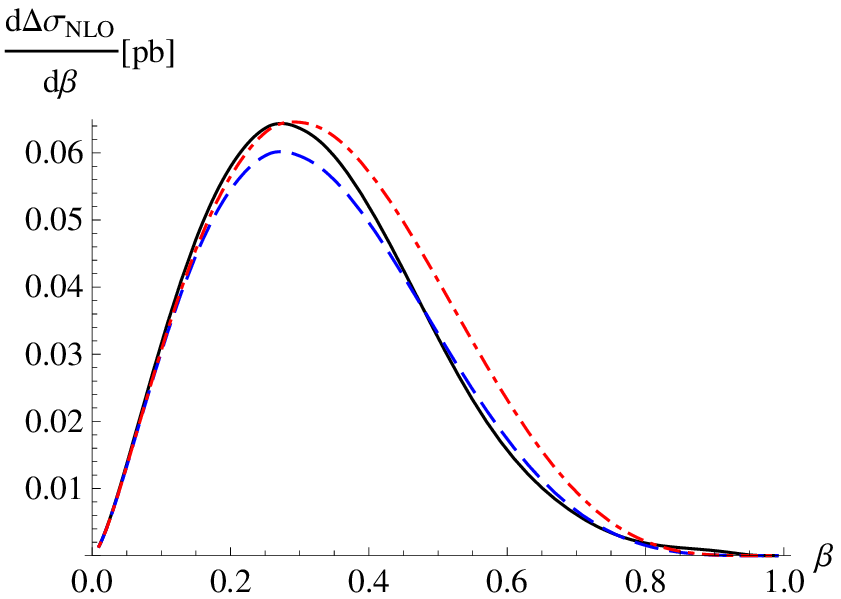}
    \caption{Partonic cross sections for $gg\to\tilde q\bar{\tilde q}$
      multiplied with the gluon-gluon luminosity for $m_{\tilde q}=1$
      TeV at the $14$ TeV LHC.  Left panel: exact leading-order result
      (black/solid) and threshold approximation (blue/dashed). Right
      panel: exact NLO result (black/solid) and the
      approximation~\eqref{eq:NLOgg} with $h^{(1)}_{q\bar q} (\mu)=0$,
      using the threshold approximation of the tree
      (red/dot-dashed) and the full tree (blue/dashed).}
    \label{fig:partongg}
  \end{center}
\end{figure}

To introduce the colour basis, note that the gluon-gluon system admits
the decomposition $8 \otimes 8=1 \oplus 8_s \oplus 8_a \oplus 10
\oplus \overline{10} \oplus 27$. The squark-antisquark pair is again either
in a singlet or octet state, but this time the latter can be produced
by two different initial colour-octet states. The colour basis
obtained for this case from the prescription~\eqref{eq:prod-basis}
agrees with the one given in~\cite{Kidonakis:1997gm}, 
\begin{eqnarray} \label{eq:basis_83}
c^{(1)}_{\{a\}} &=& \frac{1}{\sqrt{N_c D_A}}  \delta_{a_1a_2}\delta_{a_3a_4}
\nonumber\\
c^{(2)}_{\{a\}} &=& \frac{1}{\sqrt{2 D_A B_F}}  d^{ba_2a_1} T^b_{a_3a_4}
\nonumber\\
c^{(3)}_{\{a\}} &=& 
\sqrt{\frac{2}{N_cD_A}} F^b_{a_2a_1} T^b_{a_3a_4}  \, ,
\end{eqnarray}
where $d^{abc}$ are the usual symmetric invariant tensors of $SU(3)$
and $F^b_{a_1a_2}=if^{a_1b a_2}$. We also defined the
coefficients $D_A=N_c^2-1$ and $B_F=\frac{N_C^2-4}{4 N_C}=\frac{5}{12}$. 
Thus, the matching coefficients for the three colour channels are 
(including the non-relativistic normalization $2m_{\tilde q}$)
\begin{eqnarray}
 C_{gg,\{\mu\}}^{(0,1)}&=&\sqrt {2 C_F}\, \frac{2\pi\alpha_s}{m_{\tilde q}}\,
 \; g^\perp_{\mu_1\mu_2}\,,
\nonumber\\
 C_{gg,\{\mu\}}^{(0,2)}&=&\sqrt{2 D_A B_F}\,\frac{2\pi\alpha_s}{m_{\tilde q}}\,
 g^\perp_{\mu_1\mu_2}\,,
\nonumber\\
 C_{gg,\{\mu\}}^{(0,3)}&=&0\,.
\label{eq:squark-match-g}
\end{eqnarray}
The diagonal elements of the hard function are given by
\begin{equation}\label{eq:H_gg}
  H_{ii}^{(0)}=\frac{1}{8 m_{\tilde q}^2}\, h_i^{(0)}\,
  \left(\frac{2\pi\alpha_s}{m_{\tilde q}}\right)^2
  \frac{1}{4D_A^2} 
  g^{\perp\,\mu}_\mu=\left(\frac{\pi\alpha_s}{2D_Am_{\tilde q}^2}\right)^2
  \, h_i^{(0)}\,
\end{equation}
with the colour factors
\begin{align}
     h_1^{(0)}&=2 C_F ,&  h_2^{(0)}&=2 D_A B_F ,&  
    h_3^{(0)}&=0\,.
\end{align}
Using $\sum_ih_i^{(0)}=C_F(N_c^2-2)$ gives for the total partonic cross
section at threshold
\begin{equation}
 \hat\sigma_{g\bar g\to \tilde q_{iL}\overline{\tilde q}_{iR}}^{(0)}|_{\hat 
s=4m_{\tilde q}}
 =\frac{\pi\alpha_s^2 \beta}{16 m_{\tilde q}^2}\frac{N_c^2-2}{N_c(N_c^2-1)}\,. 
\end{equation}
Adding the same expression for left and right squark labels exchanged
again reproduces the result of~\cite{Beenakker:1996ch}.  In
figure~\ref{fig:partongg} we compare the partonic cross section at
threshold to the full partonic LO cross
section~\cite{Beenakker:1996ch}, for a squark mass of $1$~TeV at the
LHC with $14$~TeV cms energy. As for the quark-antiquark induced
subprocess we plot the product of the partonic cross section with
the parton luminosity and the Jacobian $\frac{\partial
  \tau}{\partial\beta}$. Again the threshold approximation is adequate
up to $\beta\sim
0.3$ while the maximum of the integrand occurs around $\beta\approx 0.5$.

For the threshold expansion of the NLO cross section we obtain
\begin{eqnarray}
f^{(1)}_{gg} &\!\!=\!\!& 
 \frac{\pi^2(N_c^2+2) }{N_c(N_c^2-2)}\frac{1}{\beta} +
8N_c \bigg[\ln^2\left(\frac{8m_{\tilde q}\beta^2}{\mu}\right)
+8 -\frac{11 \pi^2}{24}\bigg]
\nonumber\\
&&  
- \,4 N_c\,\frac{9N_c^2-20}{N_c^2-2} \,
\ln\left(\frac{8 m_{\tilde q}\beta^2}{\mu}\right) 
+ 12 N_c\,\frac{N_c^2-4}{N_c^2-2} +h^{(1)}_{gg} (\mu)+ 
{\cal O}(\beta)\,.
\label{eq:NLOgg}
\end{eqnarray}
The $\beta$-dependent terms agree with~\cite{Beenakker:1996ch}. Again,
the hard corrections $h_{gg}^{(1)}$ are currently not available analytically.  In the
right plot of figure~\ref{fig:partongg} we compare the full NLO correction
in the parameterization of~\cite{Langenfeld:2009eg} (black/solid) to
the threshold approximation of the NLO correction~\eqref{eq:NLOgg} with
$h_{gg}^{(1)}=0$, using the threshold approximation of the tree
(red/dot-dashed) or the full tree cross section (blue/dashed) as
pre-factor. In this case, the threshold approximation gives a 
better agreement with the full NLO result than for the quark-antiquark
initial state. Using the full tree
further improves the agreement with the full NLO correction in the
full $\beta$-range. Once again the integrated threshold approximation 
provides a very good approximation to the exact NLO correction, 
especially when multiplying (\ref{eq:NLOgg}) by the full Born 
cross section. This can be compared to the Born cross section itself, 
for which the leading term in the $\beta$-expansion is typically a
less reliable approximation.

 \subsection{Results for squark-antisquark production at the LHC and Tevatron}
\label{sec:squark-soft}

We are now ready to perform the combined soft and Coulomb NLL
resummation for squark-antisquark production. We first discuss our
choice for the soft, hard and Coulomb scales entering the resummed
cross section, and then present numerical results for the LHC and the
Tevatron. We also compare to the results of a soft gluon resummation
in Mellin space presented in~\cite{Beenakker:2009ha}.

\subsubsection{Scale choices}
Evaluating the NLL resummed cross section~\eqref{eq:resum-NLL}
requires a choice for the soft, hard and Coulomb scales.  As can be
seen from the Sudakov logarithm in the evolution equation of the hard
function~\eqref{eq:evolution-hard-2}, the natural scale of the hard
corrections is of the order $\mu_h\sim 2 M$. Therefore we will choose
the default value of the hard scale to be $\tilde \mu_h=2 m_{\tilde
  q}$.  Analogously, the form of the approximate NLO
corrections~\eqref{eq:NLOqq} and~\eqref{eq:NLOgg} implies that the
scale eliminating large logarithms in the soft corrections to the
\emph{partonic} cross section is given by $\mu_s\sim 8E\approx
8M\beta^2$. However, it was argued
in~\cite{Becher:2006mr,Becher:2007ty} that this choice leads to
similar problems as the inversion of Mellin-space resummation formulae
for the partonic cross section~\cite{Catani:1996dj}, i.e. an
ill-defined convolution with the parton luminosity. Here we
follow~\cite{Becher:2007ty} and choose a soft scale such that it
minimizes the relative fixed-order one-loop soft correction to the
\emph{hadronic} cross section.  More precisely, we vary the scale in
the PDFs and the soft correction and determine the value $\tilde
\mu_s$ that minimizes the relative soft corrections:
\begin{equation}
\label{eq:def-mus}
0
=\frac{d}{d\tilde\mu_s}
\sum_{p,p'}\int_{4m_{\tilde q}^2/s}^1 d\tau \,L_{pp'}(\tau,\tilde\mu_s)
\frac{\hat\sigma^{(1)}_{pp',\text{soft}} (\tau s,\tilde\mu_s)}{
\sigma^{(0)}_{N_1N_2}(s,\tilde \mu_s)}\,.
\end{equation}
Here the fixed-order NLO soft correction 
$\hat\sigma^{(1)}_{pp'\text{soft}}$ is obtained
from~\eqref{eq:NLOqq} and~\eqref{eq:NLOgg}, 
setting the Coulomb correction and the hard corrections $h_i$  to zero,
and we divide by the leading-order hadronic cross section.
This procedure results  in values 
\begin{equation}
\tilde{\mu}_s
= 123-455 \,\text{GeV} \quad \text{for} \qquad m_{\tilde{q}}=200-2000
\,\text{GeV}
\end{equation}
for the LHC with $\sqrt s=14\,\text{TeV}$.  The fact that the soft
scale is over half of the mass for light squarks, but considerably
smaller for larger masses, indicates that the threshold resummation is
more important for heavier squarks, as expected.  We can approximately
fit the numerical results for the soft scale by a function of the form
proposed in~\cite{Becher:2007ty}, 
\begin{equation}
 \tilde\mu_s^{\rm LHC}\approx\frac{m_{\tilde q}(1-\rho)}{\sqrt{3.8+149\rho}} \, ,
\end{equation}
with $\rho=4 m_{\tilde q}^2/s$, which is accurate to better than $5\%$
for $m_{\tilde q}>450$ GeV. We find that for different 
cms energies $\sqrt{s}$ the ratio  $\tilde\mu_s^{\rm LHC}/m_{\tilde
  q}$ is to a good approximation a function of $\rho$ only, that is,
for a squark of mass 500 GeV at 7~TeV cms energy 
the ratio is approximately the same as
for a 1~TeV squark at 14~TeV. In our numerical NLL results, 
however, we use the numerical values for the
soft scale obtained directly by the condition~\eqref{eq:def-mus}. 
At the Tevatron the values for the soft scale can be fitted by the function
\begin{equation}
 \tilde\mu_s^{\rm TeV}\approx\frac{m_{\tilde q}(1-\rho)}
{\sqrt{1.7+81.3\rho}} \, ,
\end{equation}
and representative values for the default soft 
scale are
\begin{equation}
\tilde\mu_s^{\rm TeV}(200\text{GeV})=84\text{GeV}\;,\quad
\tilde\mu_s^{\rm TeV}(400\text{GeV})=86\text{GeV}\;,\quad
\tilde\mu_s^{\rm TeV}(600\text{GeV})=65\text{GeV}.
\end{equation}
Note that the soft scale begins to decrease above a certain value of 
$\rho$.

\begin{figure}[t]
 \begin{center}
   \includegraphics[width=0.49\textwidth]{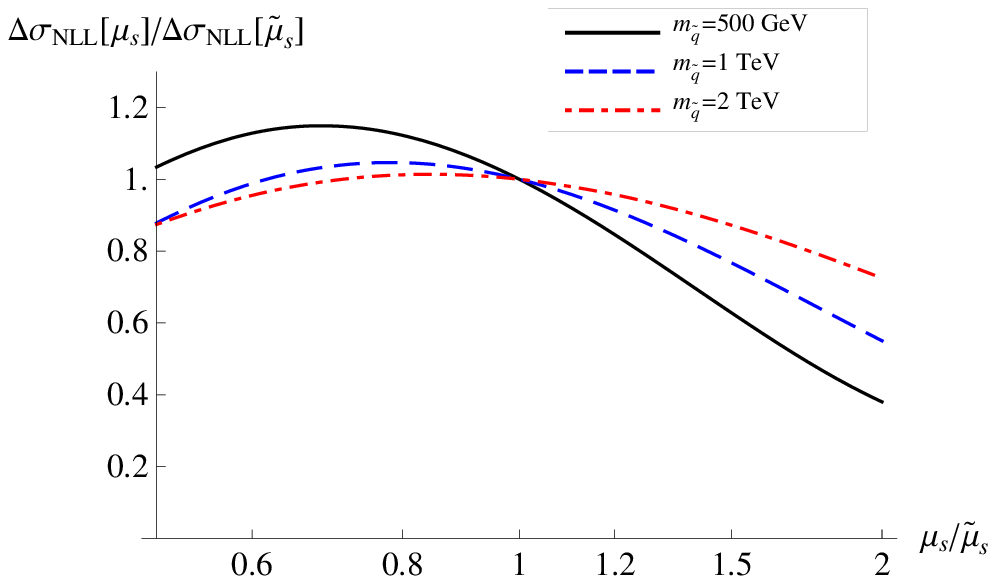} 
   \includegraphics[width=0.49\textwidth]{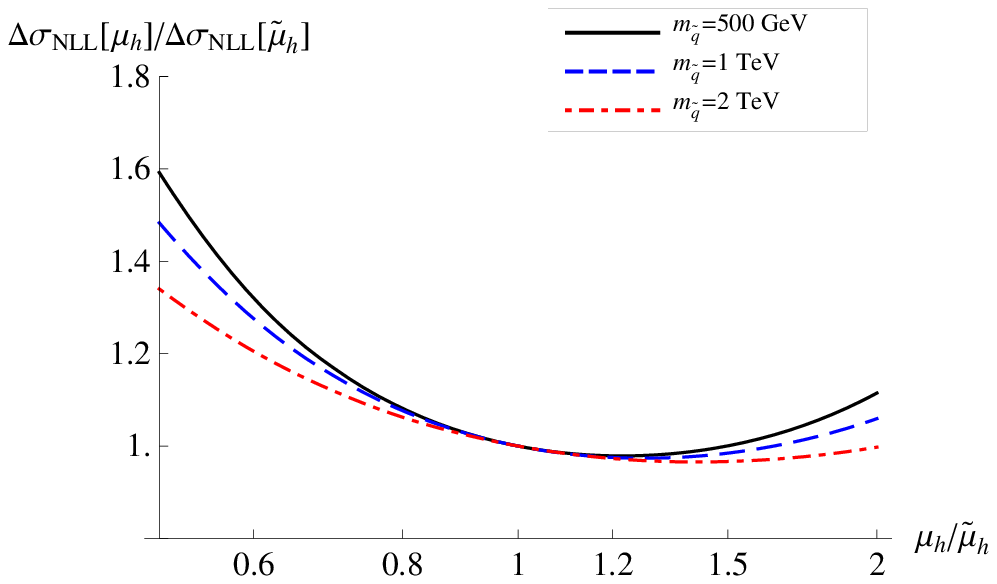}
   \caption{Dependence of the NLL corrections at the $14$ TeV LHC on
     the soft scale (left panel) and the hard scale (right panel),
     normalized to the corrections at the default scales $\tilde
     \mu_s$ determined from~\eqref{eq:def-mus} and $\tilde \mu_h=2
     m_{\tilde q}$. Black (solid): $m_{\tilde q}=500$ GeV, blue
     (dashed): $m_{\tilde q}=1$ TeV, red (dot-dashed): $m_{\tilde
       q}=2$ TeV }
   \label{fig:soft_scale}
 \end{center}
\end{figure}

Figure~\ref{fig:soft_scale} shows the dependence of the NLL
corrections to the hadronic cross section at the LHC with $14$ TeV on
the soft and hard scales, varied around the default values
$\tilde\mu_s$ and $\tilde \mu_h=2m_{\tilde q}$. Here by ``NLL
corrections'' we mean the convolution of the expression $\Delta
\hat\sigma^{\text{NLL}}_{pp'}(\hat s,\Lambda)$ in
(\ref{eq:cross_matched}) with the parton luminosity, i.e.~the
corrections on top of the exact NLO cross section due to the
higher-order soft-gluon and Coulomb terms (bound-state corrections are
not included in these curves, and we set the cutoff $\Lambda$ to the
maximum possible value). In the left panel, the hard scale is fixed to
the default value $\mu_h=\tilde\mu_h$ and we vary the soft scale in
the interval $0.5\tilde \mu_s \dots 2 \tilde \mu_s$. Analogously, in
the right panel the soft scale is fixed to the default value and we
vary the hard scale in the interval $m_{\tilde q}\dots 4 m_{\tilde
  q}$. The ambiguity due to the choice of the soft scale becomes
smaller for larger masses, in agreement with the expectation that
soft-gluon resummation is better justified for larger masses, where
the contribution from the threshold region to the total cross section
is more important.

\begin{figure}[t]
 \begin{center}
   \includegraphics[width=0.49\textwidth]{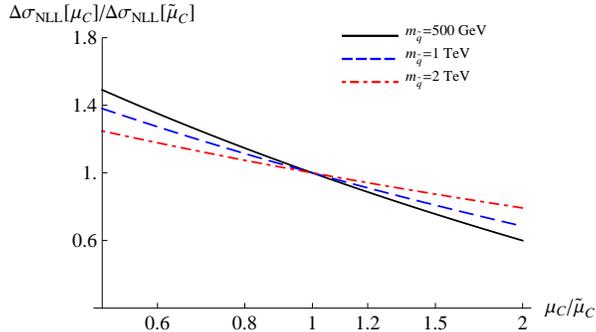}
   \caption{Dependence of the NLL corrections at the $14$ TeV LHC on
     the Coulomb scale, normalized to the corrections at the default
     scale $\tilde \mu_C= \text{max}\{2\tilde m_q \beta, \mu_B\}$
     . Black (solid): $m_{\tilde q}=500$ GeV, blue (dashed):
     $m_{\tilde q}=1$ TeV, red (dot-dashed): $m_{\tilde q}=2$ TeV }
   \label{fig:Coulomb_scale}
 \end{center}
\end{figure}

We now address the choice of the scale used for the strong coupling
constant in the Coulomb Green function. The only scale dependence of
the imaginary part of the leading-order Coulomb Green
function~\eqref{coulombGF} enters through the coupling constant of the 
leading-order Coulomb potential, so the potential function is separately
renormalization-group invariant at NLL, and the coupling constant can be evaluated at a scale 
$\mu_C$ different from the factorization scale.
Since the necessary truncation of the perturbative series for the running coupling constant 
introduces a residual higher-order scale dependence, the scale $\mu_C$ should be chosen such that higher-order corrections are minimized.
Since the
Coulomb corrections are related to the exchange of potential gluons
with momentum transfer $|\vec k|\sim M\sqrt{\lambda}$, a natural scale
choice $\mu_C$ is expected to be of the order
\begin{equation}
 \mu_C\sim M\sqrt\lambda =M\beta\sim M\alpha_s\, ,
\end{equation}
where we have used the NRQCD counting $\beta \sim \alpha_s$.  Indeed, as
mentioned in section~\ref{sec:subleading}, the effect of the
scale-dependent 
strength of the Coulomb potential can be incorporated using the choice
$\mu_C= 2 m_{\tilde q} \beta$. The choice $\mu_C\sim 
{\cal O}(M\beta)$ is in fact required to sum correctly all NLL terms, since 
$\mu_C\sim M$ would miss terms such as 
$\alpha_s^2/\beta\times \ln\beta$, which arise in part from the 
small virtuality of Coulomb gluons. However, very small values of 
$\beta$ are integrated over in the convolution of the partonic
cross section with the PDFs, and with the choice $\mu_C\propto \beta$ 
the strong coupling $\alpha_s(\mu_C)$ would hit
the Landau pole.  On the other hand, the
relevant scale for bound-state effects is set by the inverse Bohr
radius of the first $HH'$ bound state, $1/r_B=C_F m_{\tilde{q}}
\alpha_s/2\equiv \mu_B/2$.   Therefore, as our
default scale choice for the Coulomb Green function we use
\begin{equation}
\label{eq:mu-c} 
\tilde \mu_C= \text{max}\{2 m_{\tilde{q}} \beta, \mu_B\} \, ,
\end{equation}
where we solve the equation $\mu_B= C_F m_{\tilde{q}}\alpha_s(\mu_B)$ 
iteratively. The Bohr scale was used in recent studies
of threshold effects in top-pair or gluino
production~\cite{Kiyo:2008bv,Hagiwara:2008df,Hagiwara:2009hq}, while
in the recent calculation~\cite{Kulesza:2009kq} of squark-antisquark and
gluino production the Coulomb scale was set
equal to the factorization scale. The dependence of the NLL
corrections at the LHC with $14$ TeV cms energy on the Coulomb scale (with all 
other scales fixed to their default values) is shown 
in Figure~\ref{fig:Coulomb_scale}, where we vary the Coulomb scale in
the interval $0.5\tilde \mu_C \dots 2 \tilde \mu_C$.
The scale dependence is of a similar magnitude as for the soft and hard scales, and again improves for larger masses.
  In the results given below, we estimate the
uncertainty due to the scale choices by varying all scales $\mu_f$,
$\mu_s$, $\mu_h$ and $\mu_C$ from one-half to twice their default
value and add the uncertainties in quadrature.

\subsubsection{Results}

Having fixed the default scale values, we present numerical
results\footnote{There are some differences to preliminary results
  presented in~\cite{Beneke:2009nr,Beneke:2010gm} due to a different
  choice of hard scale ($\mu_h=m_{\tilde q})$ in the NLL results, and
  because the $qg$ initiated subprocess was not included in the NLO
  cross section.}  for the combined NLL resummation of soft and
Coulomb effects obtained by inserting the potential
function~\eqref{eq:sommerfeld} into the cross
section~\eqref{eq:resum-NLL}.  We use the MSTW2008 
PDF set~\cite{Martin:2009iq} and the associated value of the strong
coupling. The results for the fixed-order NLO cross section are
obtained using the program {\sc Prospino}, which is based on the calculation
of~\cite{Beenakker:1996ch}. In contrast to the default setting of {\sc
  Prospino}, we include contributions from initial-state bottom quarks
using $b$-quark PDFs.  Unless stated otherwise, we use the squark-gluino
mass ratio of $m_{\tilde g}=1.25 m_{\tilde q}$.  As discussed in
section~\ref{sec:squark-eft}, we consider the squarks to be
mass-degenerate and do not include top squark final states.  
Our default value for the factorization scale is $\tilde \mu_f=m_{\tilde q}$.

In table~\ref{tab:pdfs} we compare the {\sc Prospino} NLO predictions
obtained using the MSTW08 and the CTEQ6.6M~\cite{Nadolsky:2008zw} 
PDF set  for the Tevatron and the LHC at $14$~TeV. While we find good
agreement for small squark masses, the differences between the two
sets grow to over 10\% for large masses.  Since we expect very
similar effects for the results including NLL resummation, we present
our predictions using the MSTW PDFs below, but the uncertainties due to
differences between current PDF sets at large $x$ should be taken into
account in interpreting our results.
\begin{table}[p]
\begin{center}
\begin{tabular}{|c|l|l|}
\hline&\multicolumn{2}{c|}{
$\sigma(p\bar p\to \tilde q \bar{\tilde q})$(pb),
$\sqrt{s}=1.96$ TeV}
\\\hline
$m_{\tilde q}[\mbox{GeV}]$ &  NLO$_{\text{MSTW}}$   & NLO$_{\text{CTEQ}}$
\\[0.1cm]\hline
$200$&$ 11.6^{+1.7}_{-1.8}$&
$11.6^{+1.2}_{-1.7}$  \\[0.1cm]\hline
$300$&$6.68^{+1.19}_{-1.16}\times 10^{-1}$&
$6.93^{+1.19}_{-1.18}\times 10^{-2}$ \\[0.1cm]\hline
$400$&$4.31^{+0.94}_{-0.85}\times 10^{-2}$&
$4.64^{+0.98}_{-0.90}\times 10^{-2}$ \\[0.1cm]\hline
$500$&$2.39^{+0.63}_{-0.53}\times 10^{-3}$&
$2.73^{+0.69}_{-0.59}\times 10^{-3}$ \\[0.1cm]\hline
\end{tabular}
\\[0.5cm]
\begin{tabular}{|c|l|l|}
\hline&\multicolumn{2}{c|}{
$\sigma(p p\to \tilde q \bar{\tilde q})$(pb),
$\sqrt{s}=14$ TeV}
\\\hline
$m_{\tilde q}[\mbox{GeV}]$ &  NLO$_{\text{MSTW}}$   & NLO$_{\text{CTEQ}}$
\\[0.1cm]\hline
$500$&
$14.4^{+1.5}_{-1.6}$ &  $ 13.7^{+1.4}_{-1.5}$ 
 \\[0.1cm]\hline
$1000$&
$2.58^{+0.29}_{-0.31} \times 10^{-1}$ &  $2.49^{+0.27}_{-0.29} \times 10^{-1}$ 
 \\[0.1cm]\hline
$1500$& 
$1.35^{+0.17}_{-0.18} \times 10^{-2}$&$1.38^{+0.16}_{-0.20} \times 10^{-2}$
 \\[0.1cm]\hline
$2000$&
$9.97^{+1.43}_{-1.44}\times 10^{-4}$&$1.13^{+0.15}_{-0.16}\times 10^{-3}$
\\[0.1cm]\hline
\end{tabular}
\caption{\label{tab:pdfs}
NLO results for the squark-antisquark production cross section at
the Tevatron and the LHC at $\sqrt{s}= 14$~TeV
for $m_{\tilde g}=1.25  m_{\tilde q}$
obtained with {\sc Prospino}, using MSTW08NLO and CTEQ6.6M PDFs.
 The error estimate is
 obtained by varying the factorization scale in the interval
 $m_{\tilde q}/2\leq\mu_f\leq 2 m_{\tilde q}$}
\end{center}
\end{table}

\begin{figure}[p]
  \begin{center}
    \includegraphics[width=0.49\textwidth]{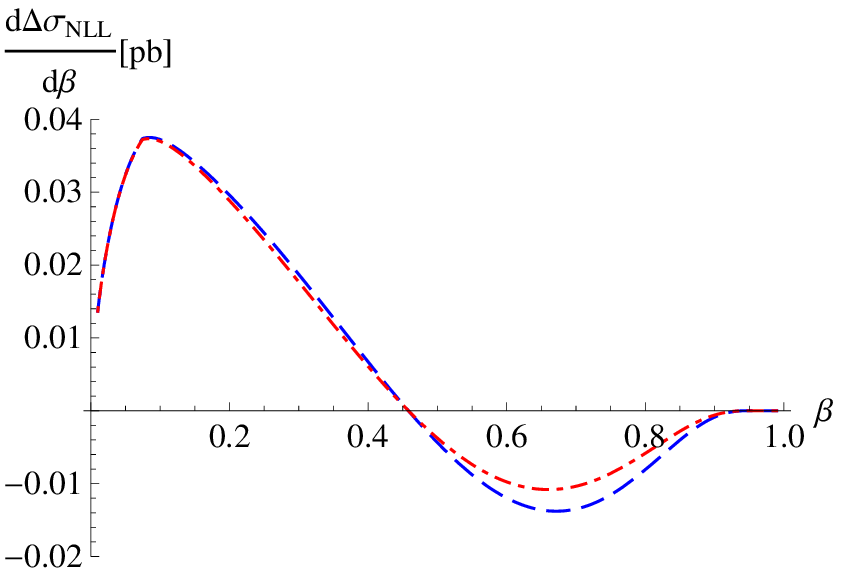}
    \includegraphics[width=0.49\textwidth]{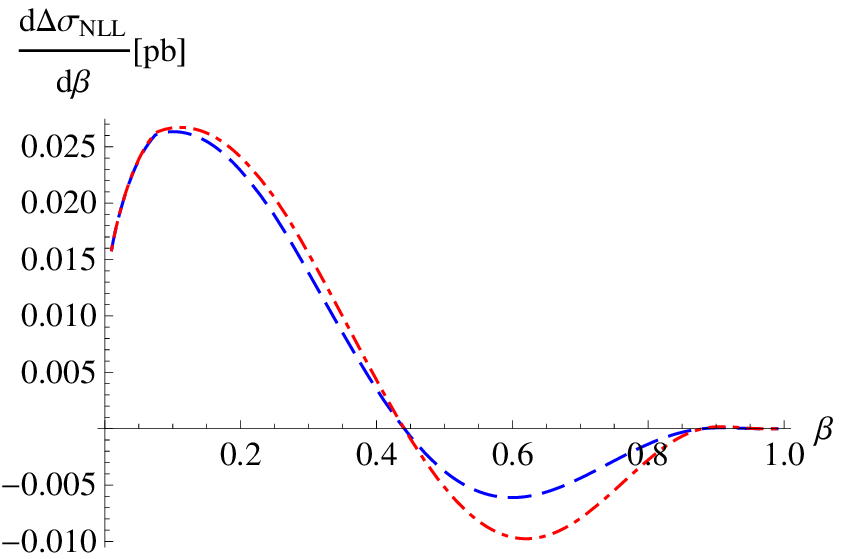} 
\end{center}
\caption{NLL corrections to the partonic cross sections, 
multiplied with the  parton luminosity, for $q_i\bar
  q_j\to\tilde q_k\bar{\tilde q}_l$ (left) and $gg\to \tilde q_k
  \bar{\tilde q}_l$ at the $14$ TeV LHC 
  for $m_{\tilde q}=1$ TeV and $m_{\tilde
    g}=1.25$~TeV,  using the threshold approximation of the tree
  (red, dot-dashed) and the full tree (blue/dashed).}
    \label{fig:partonNLL}
\end{figure}

In figure~\ref{fig:partonNLL} we show the NLL corrections to the
partonic cross sections multiplied by the parton luminosities and the
Jacobian $\frac{\partial \tau}{\partial\beta}$. It can be seen that the
dominant contributions arise from the threshold region
$\beta\leq 0.2$, and that there is only a small difference between
using the hard function obtained from the tree at threshold, as
determined in section~\ref{sec:squark-eft}, or from the full tree, as
mentioned in section~\ref{sec:hard}.  Thus, in the results presented below
we use the hard functions determined from the tree at threshold,
and set the cutoff $\Lambda$ in the matching formula~\eqref{eq:match}
to the maximal value, so that the resummed correction is applied in the
full phase space.  Another plausible choice is to switch off the NLL
corrections at the point where they turn
negative~\cite{Moch:2008qy}. In the example shown in
figure~\ref{fig:partonNLL} this occurs at $\beta\approx 0.4$, where the
validity of the threshold approximation is not immediately clear.
Since the integrated contributions from the region $\beta> 0.4$ are
moderate and negative, our procedure gives a conservative estimate of
the NLL corrections.  We stress that this choice depends on a study of
the behaviour of the resummed corrections at larger values of $\beta$
and that other choices might be required for other processes.

\begin{figure}[t]
 \begin{center}
   \includegraphics[width=0.55\textwidth]{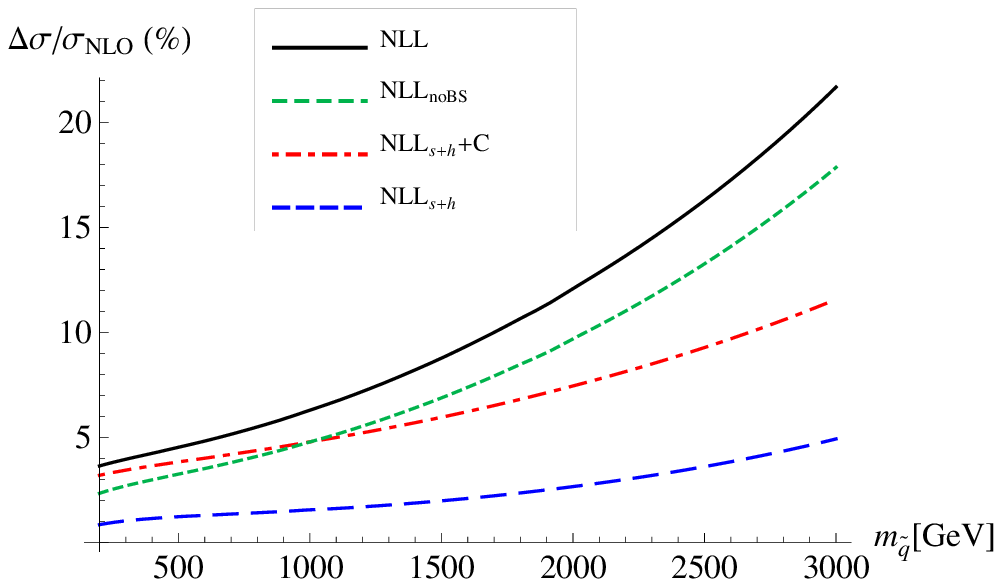}\\[0.5cm]
   \includegraphics[width=0.55\textwidth]{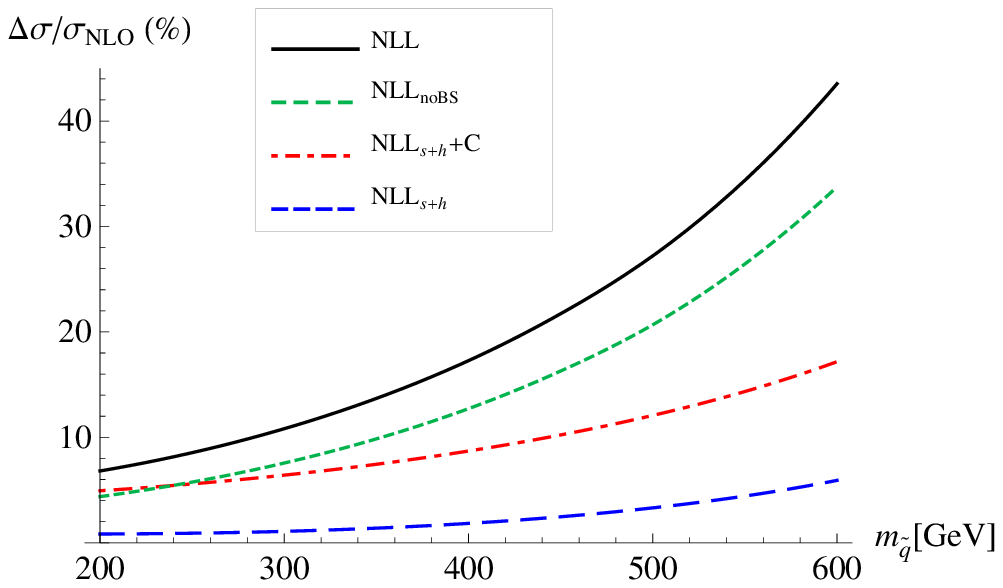}
 \caption{NLL corrections to the squark-antisquark cross section at
   the LHC with $\sqrt{s}=14$~TeV (above) and the Tevatron (below)
   relative to the NLO cross sections in various approximations. For
   explanations see text.}
 \label{fig:sigma-msq}
 \end{center}
\end{figure}

In figure~\ref{fig:sigma-msq} we plot 
the $K$-factor relative to the NLO corrections,
\begin{equation}
\Delta K_{\text{NLL}}=
\frac{\sigma_{\text{NLL}}-\sigma_{\text{NLO}}}{\sigma_{\text{NLO}}} \, ,
\end{equation}
 for the $\sqrt{s}=14$ TeV LHC and for the Tevatron, as a function of
 the squark mass, and consider various approximations to our NLL result
 to study the effect of the Coulomb and soft corrections
 separately. The different curves in the figure are defined as
 follows:
 \begin{description}
\item[NLL] (black, solid): Results for the full NLL soft and Coulomb resummation
   using the resummed cross section~\eqref{eq:resum-NLL} in
  the matching formula~\eqref{eq:cross_matched}, including bound-state
  effects obtained from the potential function below
  threshold~\eqref{eq:bound}.
\item[NLL$_{\rm noBS}$] (green, dashed): Bound-state effects are omitted from the full NLL result. 
\item [NLL$_{\text{s+h}}$] (blue, dashed): NLL resummation
 without Coulomb corrections, i.e. the cross
 section~\eqref{eq:resum-NLL} with the trivial potential function
 $J^{(0)}(E)=\frac{m_{\tilde q}^2}{2\pi}\sqrt{\frac{E}{m_{\tilde q}}}$.
\item [NLL$_{\text{s+h}}$+C] (red, dot-dashed): The pure Coulomb
  corrections, obtained by using the trivial soft function
  $W^{(0)}=\delta(\omega)$ and the resummed potential
  function~\eqref{eq:sommerfeld} without bound state effects, are
  added to the NLL$_{s+h}$ approximation.
 \end{description}

It can be seen that the soft-Coulomb interference effects, given by
the difference of the NLL$_{\rm noBS}$ and NLL$_{\text{s+h}}$+C
curves, are sizable, in particular for large squark masses. Most of
this effect is due to the interference of the first Coulomb correction
with the resummed soft corrections; for instance at the $14$ TeV LHC
the relative correction at $m_{\tilde q}=2\,$TeV in this approximation
is $\Delta K_{\text{NLL}}=9.0\%$ as opposed to $\Delta
K_{\text{NLL}}=9.7\%$ in the full NLL soft plus Coulomb
resummation. The NLL$_{\text{s+h}}$ corrections are of a similar
magnitude as the soft NLL corrections obtained
in~\cite{Kulesza:2008jb,Kulesza:2009kq,Beenakker:2009ha} from
resummation in Mellin space. A quantitative comparison to these
previous results will be performed below.  The Coulomb corrections are
larger than those observed in~\cite{Kulesza:2009kq}. This can be
traced to the choice of the Coulomb scale, which was taken to be
$\mu_C=\mu_f$ in that reference, while we use the smaller
scale~\eqref{eq:mu-c} required to sum all NLL terms, as discussed
above.

\begin{figure}[t]
\begin{center}
\includegraphics[width=0.49\textwidth]{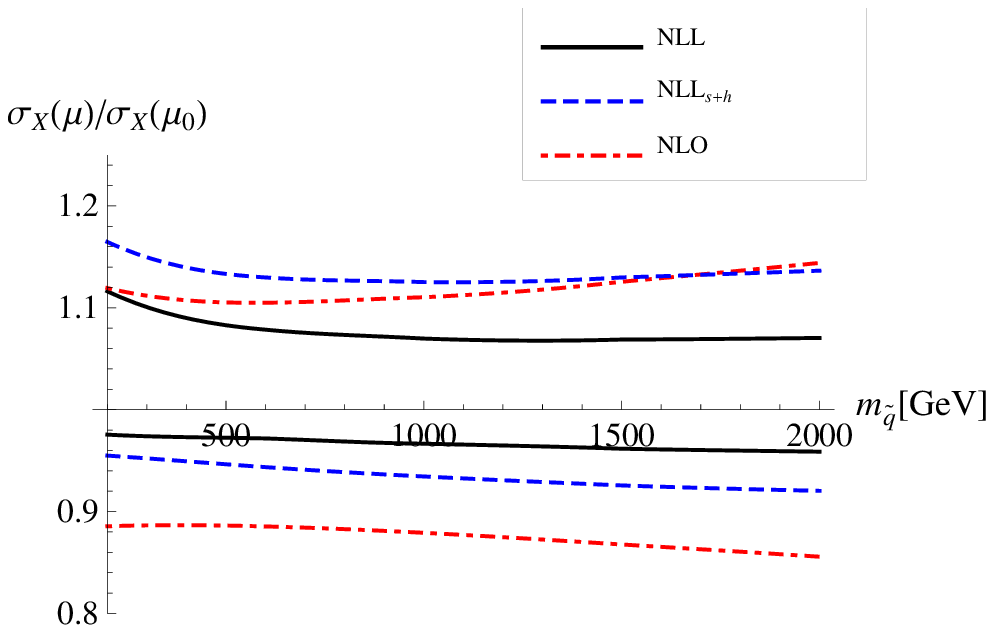}
\includegraphics[width=0.49\textwidth]{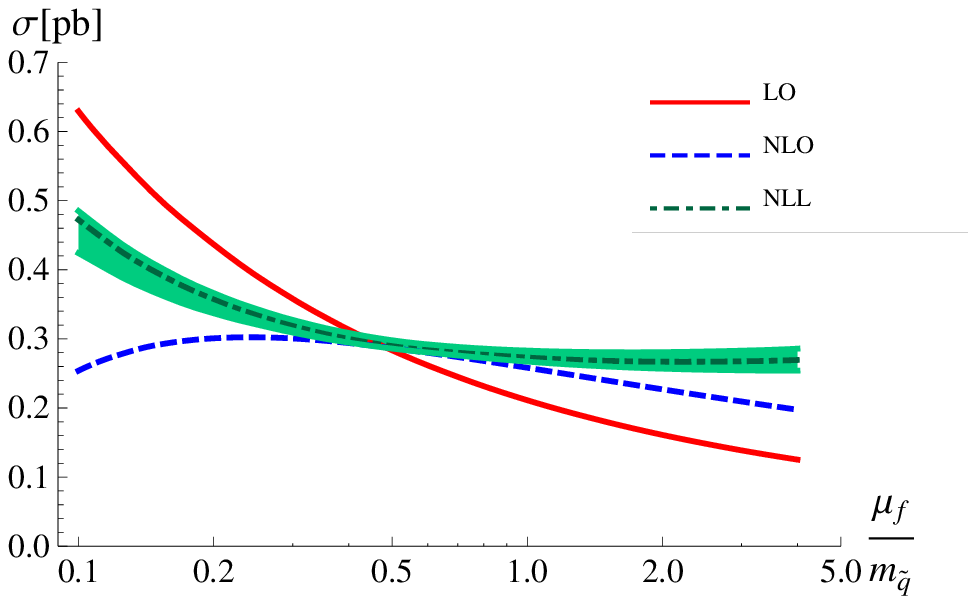}
\vspace*{0.5cm}
\includegraphics[width=0.49\textwidth]{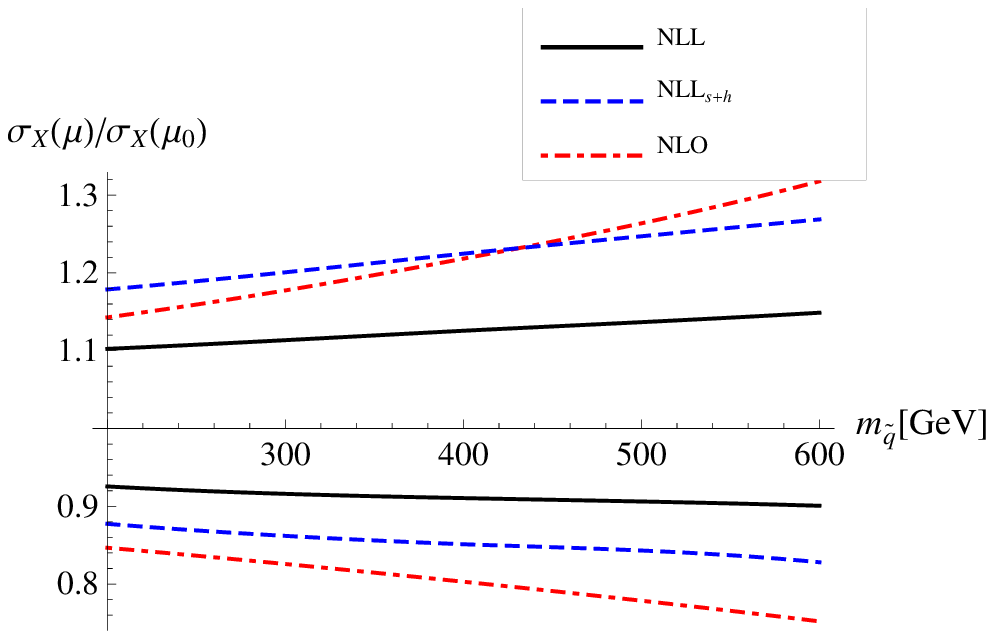}
\includegraphics[width=0.49\textwidth]{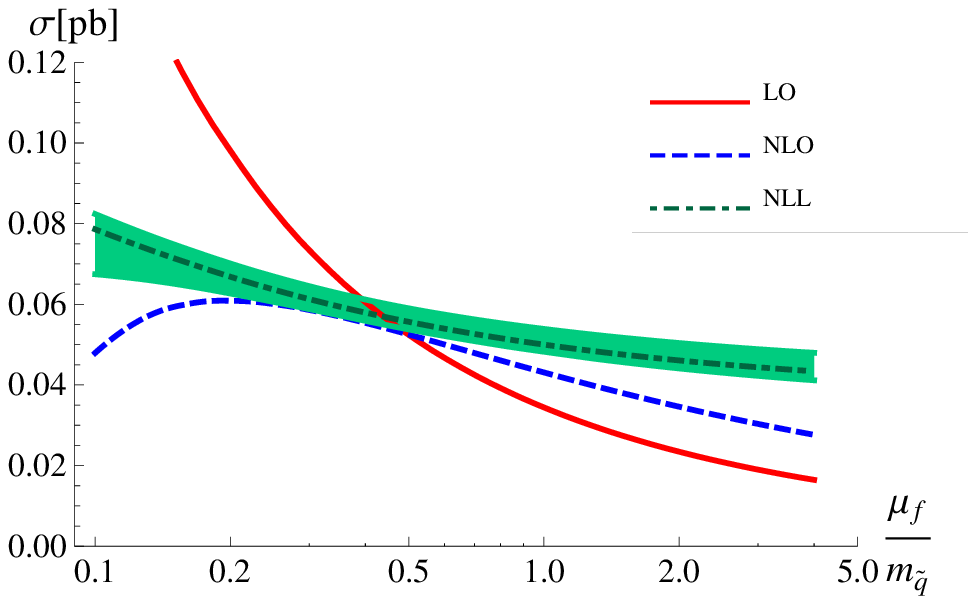}
\caption{Left: Scale dependence of the NLO, NLL and NLL$_{\text{s+h}}$
  approximations at the LHC with $\sqrt{s}=14$ TeV (top) and the
  Tevatron with $\sqrt{s}=1.96$ TeV (bottom).  The corrections are
  normalized to the value at the default scales.  Right: Dependence of
  the LO, NLO and NLL result on the factorization scale, for $m_{\tilde q}=1$ TeV at the $14$ TeV
  LHC (top) and for $m_{\tilde q}=400$ GeV at the Tevatron (bottom). 
  The green band indicates the uncertainty
  on the NLL result from the choice of the soft, hard and Coulomb scales. }
 \label{fig:scales}
 \end{center}
\end{figure}
\begin{table}[p]
\begin{center}
\begin{tabular}{|c|l|l|l|c|}
\hline&
\multicolumn{2}{c}{ $\sigma(p\bar p\to \tilde q \bar{\tilde q})$(pb),
$\sqrt{s}=1.96$ TeV}&&
\\\hline
$m_{\tilde q}[\mbox{GeV}]$ &LO &  NLO   &NLL & $\Delta K_{\text{NLL}}$
\\[0.1cm]\hline
$200$&$9.59^{+4.36}_{-2.76}$&$ 11.6^{+1.7}_{-1.8}$&
$12.4^{+1.3}_{-0.9}$&$6.9\%$  \\[0.1cm]\hline
$300$&$5.44_{-1.65}^{+2.64}\times 10^{-1}$&$6.68^{+1.19}_{-1.16}\times 10^{-1}$&
$7.41^{+0.84}_{-0.62}\times 10^{-1}$&$10.9\%$
  \\[0.1cm]\hline
$400$& $3.44^{+1.80}_{-1.10}\times 10^{-2}$&$4.31^{+0.94}_{-0.85}\times 10^{-2}$&
$5.06^{+0.64}_{-0.45}\times 10^{-2}$&
$17.3\%$ \\[0.1cm]\hline
$500$&$1.87^{+1.06}_{-0.63}\times 10^{-3}$&$2.39^{+0.63}_{-0.53}\times 10^{-3}$&
$3.04^{+0.41}_{-0.28}\times 10^{-3}$&$27.3\%$ \\[0.1cm]\hline
$600$& $6.93^{+4.27}_{-2.47}\times 10^{-5}$&$9.09^{+2.89}_{-2.25}\times 10^{-5}$&
$1.30^{+0.19}_{-0.13}\times 10^{-4}$&
$43.4\%$
 \\[0.1cm]\hline
\end{tabular}

\vskip0.5cm
\begin{tabular}{|c|l|l|l|c|}
\hline&
\multicolumn{2}{c}{ $\sigma(pp\to \tilde q \bar{\tilde q})$(pb),
$\sqrt{s}=7$ TeV}&&
\\\hline
$m_{\tilde q}[\mbox{GeV}]$ &LO &  NLO   &NLL & $\Delta K_{\text{NLL}}$
\\[0.1cm]\hline
$200$&$1.83^{+0.65}_{-0.44}  \times 10^{2}$ & $2.40^{+0.30}_{-0.31}\times 10^{2}$
&$2.52^{+0.27}_{-0.08}\times 10^{2}$ & $5.0\%$ \\[0.1cm]\hline
$400$&$4.47^{+1.66}_{-1.12} $ 
&$5.58^{+0.69}_{-0.73}  $  &$5.95^{+0.51}_{-0.21} $
& $6.5\%$
\\[0.1cm]\hline
$600$&$3.33^{+1.29}_{-0.86}\times 10^{-1}$  &$4.01^{+0.53}_{-0.56}\times 10^{-1}$
  &$4.35^{+0.34}_{-0.17}\times 10^{-1}$  
&8.5\%  \\[0.1cm]\hline
$800$&$3.69^{+1.49}_{-0.99} \times 10^{-2}$
 &  $4.32^{+0.63}_{-0.64}\times 10^{-2}$ &  $4.79^{+0.37}_{-0.20}\times 10^{-2}$
&$11.0\%$ \\[0.1cm]\hline
$1000$& $4.92^{+2.05}_{-1.35} \times 10^{-3}$& 
$5.52^{+0.90}_{-0.88} \times 10^{-3}$ &  $6.30^{+0.50}_{-0.29} \times 10^{-3}$ &
$14.1\%$  \\[0.1cm]\hline
$1200$& $7.08^{+3.07}_{-2.0} \times 10^{-4}$& 
$7.53^{+1.35}_{-1.28} \times 10^{-4}$&$8.89^{+0.71}_{-0.43} \times 10^{-4}$
& $17.9\%$ \\[0.1cm]\hline
\end{tabular}

\vskip0.5cm
\begin{tabular}{|c|l|l|l|c|}
\hline&
\multicolumn{2}{c}{ $\sigma(pp\to \tilde q \bar{\tilde q})$(pb),
$\sqrt{s}=10$ TeV}&&
\\\hline
$m_{\tilde q}[\mbox{GeV}]$ &LO &  NLO   &NLL & $\Delta K_{\text{NLL}}$
\\[0.1cm]\hline
$200$&$4.32^{+1.4}_{-1.0}\times 10^2$ &$5.73^{+0.70}_{-0.69}\times 10^2$ &
$5.97^{+0.66}_{-0.16}\times 10^2$ &$4.2\%$ \\[0.1cm]\hline
$400$&$1.39^{+0.47}_{-0.33}\times 10^1$ &$1.77^{+0.20}_{-0.21}\times 10^1$ &
$1.86^{+0.16}_{-0.06}\times 10^1$ &$5.1\%$ \\[0.1cm]\hline
$600$&$1.41^{+0.50}_{-0.34}$  & $1.75^{+0.20}_{-0.22}$  & $1.86^{+0.14}_{-0.06}$
 & $6.1\%$ 
\\[0.1cm]\hline
$800$&$2.25^{+0.82}_{-0.56}\times 10^{-1}$&$2.72^{+0.33}_{-0.35}\times 10^{-1}$&
$2.92^{+0.21}_{-0.11}\times 10^{-1}$  &  $7.4\%$ 
\\[0.1cm]\hline
$1000$&$4.48^{+1.68}_{-1.14}\times 10^{-2}$&$5.32^{+0.69}_{-0.72}\times 10^{-2}$&
$5.79^{+0.42}_{-0.22}\times 10^{-2}$ & $8.9\%$ 
 \\[0.1cm]\hline
$1200$& $1.02^{+0.39}_{-0.26}\times 10^{-2}$&$1.18^{+0.17}_{-0.17}\times 10^{-2}$&
$1.30^{+0.10}_{-0.05}\times 10^{-2}$&$10.6\%$ 
\\[0.1cm]\hline
$1400$&$ 2.48^{+0.98}_{-0.66}\times 10^{-3}$&$2.79^{+0.42}_{-0.42}\times 10^{-3}$&
$3.15^{+0.23}_{-0.13}\times 10^{-3}$& $12.7\%$ 
 \\[0.1cm]\hline
\end{tabular}
\vskip0.3cm

\caption{\label{tab:sigmatev} 
Results for the squark-antisquark production cross section at
the Tevatron and the LHC at $\sqrt{s}= 7$ TeV and 10 TeV 
for $m_{\tilde g}=1.25  m_{\tilde q}$ at leading order, 
next-to-leading order (NLO) and NLO plus
 resummed soft and Coulomb corrections (NLL). 
 The error estimates of the LO and NLO results are
 obtained by varying the factorization scale in the interval
 $m_{\tilde q}/2\leq\mu_f\leq 2 m_{\tilde q}$, the error estimate of
 the NLL result is obtained by varying
 $\mu_i\in\{\mu_f,\mu_h,\mu_s,\mu_C\}$ in the interval $\tilde
 \mu_i/2\leq\mu_i\leq \tilde 2\mu_i$ around the default values and
 adding the variations of the cross section in quadrature.} 
\end{center}
\end{table}

\begin{table}[p]
\begin{center}
\begin{tabular}{|c|l|l|l|c|}
\hline&
\multicolumn{2}{c}{ $\sigma(pp\to \tilde q \bar{\tilde q})$(pb),
$\sqrt{s}=14$ TeV}&&
\\\hline
$m_{\tilde q}[\mbox{GeV}]$ &LO &  NLO   &NLL & $\Delta K_{\text{NLL}}$
\\[0.1cm]\hline
$200$&$0.91^{+0.27}_{-0.2}  \times 10^{3}$ & $1.21^{+0.14}_{-0.14}\times 10^{3}$
&$1.25^{+0.15}_{-0.03}\times 10^{3}$ & $3.6\%$ \\[0.1cm]\hline
$400$&$3.48^{+1.1}_{-0.78}  \times 10^{1}$ 
&$4.53^{+0.49}_{-0.51}  \times 10^{1}$  &$4.73^{+0.43}_{-0.13}  \times 10^{1}$
& $4.2\%$
\\[0.1cm]\hline
$600$&$4.27^{+1.39}_{-0.98}$  &$5.43^{+0.57}_{-0.62}$   &$5.69^{+0.45}_{-0.16}$  
&4.8\%  \\[0.1cm]\hline
$800$&$0.84^{+0.28}_{-0.2} $  &  $1.04^{+0.11}_{-0.12}$ &  $1.10^{+0.08}_{-0.03}$
&$5.5\%$ \\[0.1cm]\hline
$1000$& $2.11^{+0.72}_{-0.5} \times 10^{-1}$& 
$2.58^{+0.29}_{-0.31} \times 10^{-1}$ &  $2.74^{+0.19}_{-0.09} \times 10^{-1}$ &
$6.3\%$  \\[0.1cm]\hline
$1500$& $1.15^{+0.41}_{-0.28} \times 10^{-2}$& 
$1.35^{+0.17}_{-0.18} \times 10^{-2}$&$1.47^{+0.10}_{-0.06} \times 10^{-2}$
& $8.8\%$ \\[0.1cm]\hline
$2000$& $8.92^{+3.38}_{-2.3}\times 10^{-4}$  &
$9.97^{+1.43}_{-1.44}\times 10^{-4}$&$1.12^{+0.08}_{-0.05}\times 10^{-3}$&
$12.1\%$\\[0.1cm]\hline
$2500$& $7.84^{+3.16}_{-2.09}\times 10^{-5}$ &
$8.16^{+1.33}_{-1.29}\times 10^{-5}$&$9.49^{+0.69}_{-0.42}\times 10^{-5}$&
$16.3\%$\\[0.1cm]\hline
$3000$&  $7.01^{+2.92}_{-1.94}\times 10^{-6}$  &
$6.54^{+1.21}_{-1.12}\times 10^{-6}$ & $7.96^{+0.60}_{-0.37}\times 10^{-6}$&
$21.7\%$\\\hline
\end{tabular}
\end{center}
\caption{\label{tab:sigma14}
Results for the leading order, fixed-order NLO and resummed
 soft + Coulomb corrections (NLL) to squark-antisquark production at
 the LHC at $\sqrt{s}= 14$ TeV, setup as in
 table~\ref{tab:sigmatev}.}
\end{table}
\begin{table}[p]
\begin{center}
\begin{tabular}{|c|c|c@{}c|c@{}c|c@{}c|}
\hline&
\multicolumn{6}{c}{ $\sigma(pp\to \tilde q \bar{\tilde q})$(pb),
$\sqrt{s}=14$ TeV}&
\\\hline
$m_{\tilde q}$[GeV]&  NLO
& NLL$_{\text{Mellin}}$& (ref.~\cite{Beenakker:2009ha})& NLL$_s$&&
     NLL &
\\\hline 
$200$ &$1.3\times 10^3$&$1.31\times 10^3$ &$(1\%)$ &$1.31\times
10^3$&$(1\%)$&$1.34\times 10^3$ &$(3.4\%)$\\[0.1cm]\hline 
$500$&$1.6\times 10^1$&$1.61\times 10^1$ &$(1.2\%)$&
$1.62\times 10^1$&$(1.3\%)$&
                $1.67\times 10^1$&$(4.2\%)$\\[0.1cm]\hline 
$1000$&$2.89\times 10^{-1}$&$2.93\times 10^{-1}$&$ (1.7\%)$&
$2.94\times 10^{-1}$&$ \,(1.7\%)$&$3.06\times 10^{-1}$&$ (5.8\%)$\\[0.1cm]\hline 
$2000$&$1.11\times 10^{-3}$&$1.14\times 10^{-3}$& $(3.4\%)$&
                 $1.14\times 10^{-3}$ &$\,(3.1\%)$ &$1.24\times 10^{-3}$ &$(11\%)$\\[0.1cm]\hline 
$3000$&$7.13\times 10^{-6}$&$7.59\times 10^{-6}$&$(6.4)\%$
                  &$7.54\times 10^{-6}$&$\,(5.8\%)$&$8.61\times 10^{-6}$&
$(21\%)$\\\hline 
\end{tabular}
\\[0.5cm]
\begin{tabular}{|c|c|c@{}c|c@{}c|c@{}c|}
\hline&
\multicolumn{6}{c}{ $\sigma(p\bar p\to \tilde q \bar{\tilde q})$(pb),
$\sqrt{s}=1.96$ TeV}&
\\\hline
$m_{\tilde q}$[GeV]&  NLO
& NLL$_{\text{Mellin}}$& (ref.~\cite{Beenakker:2009ha})& NLL$_s$&&
     NLL &
\\\hline 
$200$ &$1.29\times 10^1$&$1.30\times 10^1$ &$(1.6\%)$ &$1.30\times
10^1$&$(1\%)$&$1.37\times 10^1$ &$(6.5\%)$\\[0.1cm]\hline 
$300$&$7.35\times 10^{-1}$&$7.55\times 10^{-1}$ &$(2.6\%)$&
$7.47\times 10^{-1}$&$\,(1.5\%)$&
                $8.11\times 10^{-1}$&$(10\%)$\\[0.1cm]\hline 
$400$&$4.70\times 10^{-2}$&$4.91\times 10^{-2}$&$ (4.5\%)$&
$4.83\times 10^{-2}$&$\,(2.7\%)$&$5.48\times 10^{-2}$&$ (17\%)$\\[0.1cm]\hline 
$500$&$2.58\times 10^{-3}$&$2.77\times 10^{-3}$& $(7.1\%)$&
                 $2.70\times 10^{-3}$ &$\,(4.6\%)$ &$3.26\times 10^{-3}$ &
$(27\%)$\\[0.1cm]\hline 
$600$&$9.79\times 10^{-5}$&$10.9\times 10^{-5}$&$(11)\%$
                  &$10.6\times 10^{-5}$&$\,(7.8\%)$&$13.9\times 10^{-5}$&
$(42\%)$\\\hline 
\end{tabular}
\vskip0.4cm
\caption{Comparison to the results of~\cite{Beenakker:2009ha} for the
 LHC at $14$ TeV (above) and the Tevatron (below) for $m_{\tilde g}=m_{\tilde q}$. In brackets we quote
  the corrections $\Delta K_{\text{NLL}}$ relative to
 the NLO result.}
\label{tab:compare}
\end{center}
\end{table}

In tables~\ref{tab:sigmatev} and~\ref{tab:sigma14} we compare the NLL
resummed results pertaining to $p\bar p$ collisions at the Tevatron
cms energy $\sqrt{s}=1.96\,$TeV and $pp$ collisions at LHC with
$\sqrt{s}=7,10$ and $14$~TeV to the LO and NLO results obtained from
{\sc Prospino}.  For the LO predictions we have used the MSTW2008LO
set of PDFs, for the NLO and resummed predictions the MSTW2008NLO set.
One observes a sizable reduction in the scale dependence for larger
squark masses. This improvement can also be seen in
figure~\ref{fig:scales}.  In the left-hand plots the scale dependence
of the NLO cross section, the full NLL and the NLL$_{\text{s+h}}$
correction are plotted as a function of the squark mass for the
Tevatron and the $14$~TeV LHC. For the NLL results, the soft, hard,
Coulomb scales and the factorization scale are varied and the
uncertainties added in quadrature. Clearly, a significant reduction of
the scale dependence requires the inclusion of soft-Coulomb
interference.  In the right-hand plots in figure~\ref{fig:scales} we plot
the LO, NLO and NLL cross section as a function of
the factorization scale for $m_{\tilde q}=1$ TeV at the $14$~TeV LHC
and for $m_{\tilde q}=400$ GeV at the Tevatron.  One observes a
stabilization of the result with respect to variations of the
factorization scale in going from the NLO to the NLL approximation if
the factorization scale is varied in the usual interval $0.5 m_{\tilde
  q}\leq\mu_f\leq 2 m_{\tilde q}$, even after taking the variation of
the soft, hard and Coulomb scales (shown as the green band) into account.

In table~\ref{tab:compare} we compare to the NLL result
of~\cite{Beenakker:2009ha} obtained from threshold resummation in
Mellin space. In that reference no Coulomb resummation was
considered. In order to compare to their result we use the
approximation NLL$_{\text{s+h}}$ defined above, but in addition set
the hard scale to $\mu_h=\mu_f$. As can be seen from the comparison,
this approximation, denoted by NLL$_{\text{s}}$, is in good agreement
with the result from~\cite{Beenakker:2009ha} at the $14$~TeV LHC,
while the full NLL corrections including Coulomb effects are
considerably larger at higher masses.  At the Tevatron the
NLL$_{\text{s}}$ approximation results in smaller corrections than the
ones obtained in the Mellin space approach, while again the Coulomb
corrections and mixed soft/gluon corrections are sizable (note that,
due to statistical fluctuations, some of the NLO results from {\sc
  Prospino} differ from the numbers quoted in~\cite{Beenakker:2009ha}
in the last digit).

\section{Conclusion}

The production of pairs of massive coloured particles is an integral 
part of hadron collider physics programmes, since such particles, 
the top quark for certain, may be produced in large numbers. Precise 
calculations of the production cross sections are hampered by 
potentially large quantum corrections originating from the suppression 
of soft gluon emission and the exchange of Coulomb gluons near the 
partonic production threshold, especially for large masses of the 
produced particles. The techniques for summing 
soft gluon effects to all orders in perturbation theory have been 
available for some time, and have been used extensively to improve 
Drell-Yan production and related processes. Likewise, the summation of 
Coulomb effects is well established in the context of calculations 
of top quark pair production near threshold in $e^+ e^-$ collisions. 
But the interplay of both effects as relevant to hadronic pair production 
had not been studied to all orders up to now. 

Joint soft-gluon and Coulomb resummation is a non-trivial issue, since 
the energy of soft gluons is of the same order as the kinetic energy 
of the heavy particles produced. Soft-gluon lines may therefore connect 
to the heavy-particle propagators in between the Coulomb ladders, as well 
as to the Coulomb gluons itself, impeding the standard factorization 
arguments that assume either no couplings to the final state, or 
to energetic particles. In this paper we employed a combination of
soft-collinear and non-relativistic effective field theory techniques 
to show that, despite these apparent complications, soft gluons can 
be decoupled from the non-relativistic final state. Their effect is 
then contained in a more complicated soft function including two 
time-like Wilson lines for the final state particles, and in a 
convolution in energy of the soft function with the final state Coulomb 
function, since the latter depends on the energy but not the momentum 
taken away by the radiated gluons. Our main result is therefore 
a new factorization formula for the partonic cross section 
of heavy-particle pair production near threshold in hadron collisions 
given by~(\ref{eq:fact-final}), and the corresponding 
equation~(\ref{eq:invdistpartonic}) for the invariant mass
distribution, together with the resummation
formula~(\ref{eq:fact-resum}) for the partonic cross section. 

The result is naturally formulated in momentum space. When expressed 
in Mellin moment space, the factorization is multiplicative in 
each separate colour channel, and given by the product of a hard, 
soft and Coulomb (or potential) function. This justifies previous 
treatments of Coulomb effects at the one-loop order and extends them 
to all orders. The resummation of Coulomb effects to all orders is 
in fact formally required even at the leading-logarithmic level, though 
numerically the two- and higher-loop Coulomb correction is rather 
small. Starting from the NNLL level the combined soft-gluon and 
Coulomb resummation might have involved subleading soft gluon couplings 
from the effective Lagrangians (beyond the standard eikonal approximation 
in the collinear sector), making the resummation much more difficult. 
We applied symmetry considerations to show that these couplings 
do not contribute at NNLL. A new effect at this order is, however, 
that the potential function acquires factorization-scale dependence 
that cancels with the hard function, resulting in additional 
threshold logarithms on top of the ones from soft gluon emission. 
Eq.~(\ref{eq:fact-final}) is thus valid at NNLL, but beyond that 
order further, more complicated soft functions are most likely required. 

Factorization and resummation are largely model-independent. The details 
of the hard production mechanism enter only in the momentum-independent 
short-distance coefficients. We exemplified our theoretical approach 
by considering squark-antiquark production in $pp$ and $p\bar p$ 
collisions at LHC and Tevatron energies with NLL accuracy. We find 
that the effect of resummation on top of the fixed-order NLO cross 
section is around (12--27)\% for squark masses that give cross sections 
at the femtobarn level, and increases with mass as expected. The scale 
uncertainty is considerably reduced after resummation 
suggesting a reduction of 
the remaining theoretical error from which experimental limits or 
determinations of squark masses would benefit. We find resummation 
effects larger than previous estimates in the literature, since 
we normalize the Coulomb function at the scale $M\beta$ to 
sum all NLL terms, and include the interference terms between soft 
gluon and Coulomb exchange, as well as  squark-antisquark 
bound-state production in the total cross section. A more extensive 
phenomenological analysis of sparticle 
production covering all final states at the NNLL level is 
therefore of interest.

\subsubsection*{Acknowledgements}
We thank U.~Langenfeld and S.~Moch for providing interpolating
functions of squark production cross sections for performing 
numerical checks, G.~Watt for discussions on the MSTW PDFs, 
J.R.~Andersen for computing advice, and S.~Klein for helpful discussions. 
M.B. thanks the CERN theory
group for its hospitality, while part of this work was done. The work
of M.B. is supported in part by the DFG
Sonder\-for\-schungs\-bereich/Trans\-regio~9 ``Computergest\"utzte
Theoreti\-sche Teilchenphysik'';  the work of P.F. 
by the grant ``Premio Morelli-Rotary
2009" of the Rotary Club Bergamo.

\appendix

\section{Relation of collinear matrix elements to PDFs}
\label{app:pdfs}
In this appendix we show that collinear matrix elements of the form
$\braket{p|\phi_c(x)\phi_c(0)|p}$ can be expressed in terms of the
parton distribution functions (PDFs) and a polarization sum of the 
external wavefunctions, as stated
in~\eqref{eq:pdf-general}.  We recall the operator definitions of the
PDFs of quarks
and gluons in full QCD~\cite{Collins:1981uw}:
\begin{align}
  f_{q/N}(x_1,\mu) &=
  \frac{1}{2 \pi} \int dt \,e^{-i x_1 t \bar{n} \cdot P_1} 
\langle N|[\bar{\psi}(t \bar{n})
 \frac{\slash{\bar{n}}}{2}W_{\bar n}(t\bar n,0) \psi(0)|N \rangle\,,
\label{eq:q-pdf-qcd} \\
 f_{g/N}(x_1,\mu)& =
  \frac{1}{2 \pi x_1 (\bar n\cdot P_1)}
  \int dt \,e^{-i x_1 t \bar{n} \cdot P_1} \langle N|
  (\bar n_\mu F^{\mu\nu}(t \bar{n})W_{\bar n}(\bar n t,0)
  (\bar n^\rho F_{\nu\rho}(0))|N\rangle \,,
\label{eq:g-pdf-qcd}
\end{align}
where $\psi$ is the quark field in QCD, $F_{\mu\nu}$ the full gluon
field strength and an average over the physical polarization states 
is understood. $W_{\bar n}(x_+,0)$ denotes a Wilson line with the full
gluon field $\bar n\cdot A$ extending from $0$ to $x_+$,
\begin{equation}
W_{\bar n}(x_+,0) =
\mbox{P} \exp \left[i g_s \int_{0}^{x_+}d t 
\, \bar n \cdot A(\bar n t) \right]\,.
\end{equation}
 The PDFs in SCET have been introduced
in~\cite{Bauer:2002nz} and are collected in~\cite{Becher:2009th} in the SCET
conventions used here. The purpose of the present discussion is 
to fix our conventions and identify all the numerical
prefactors. 

In order to relate the collinear matrix elements to the PDFs, we
decompose the collinear and anticollinear matrix elements into a basis
of spin structures $\Gamma^i$, so that (suppressing colour indices)
\begin{equation}
\braket{p|\phi_{c;\beta}^\dagger\phi_{c;\alpha}|p}
=\sum_i\bar\Gamma^i_{\alpha\beta}\braket{p|\phi_c^\dagger\Gamma^i\phi_c|p}
\end{equation}
with $\text{tr}[\bar\Gamma^i\Gamma^i]=1$.
We now discuss the cases of quarks and gluons separately.  For
$n$-collinear quarks the fields are given
by $\phi_c=W_c^\dagger\xi_c$ . The identities $\slash n\xi_c=0$ and
$(\slash n\slash{\bar n}/4)\xi_c=\xi_c$ satisfied by the collinear
spinors imply~\cite{Bauer:2002nz} that the independent spin
structures for the decomposition of a matrix element
$\braket{p|\bar\xi_{c}\Gamma\xi_{c}|p}$ can be taken as
$\Gamma^i\in\{\slash{\bar n},\slash{\bar n}\gamma^5,\slash{\bar
  n}\gamma^\mu_\perp \}$, up to normalization.  Since we sum over the
initial state parton spin, only the $\slash{\bar n}$ term contributes.  In
colour space the product of $\xi^\dagger$ and $\xi$ decomposes into a
singlet and an octet component but the sum over the initial state
parton colours projects on the colour singlet. We therefore obtain for
the collinear quark matrix element
\begin{align}
  \langle p(P_1)|\phi^{(0)\dagger}_{c;k\beta}(z)
&    \phi^{(0)}_{c;i\alpha}(0)|p(P_1)\rangle |_{\text{avg.}} \nonumber\\
&=
\frac{1}{2N_c}\delta_{ki} 
\left(\frac{\slash n}{2}\gamma^0\right)_{\alpha\beta} \!
\langle p|[\bar{\xi}^{(0)}_{c} W^{(0)}_{c}](z)
 \frac{\slash{\bar{n}}}{2}[W^{(0)\dagger}_{c} \xi^{(0)}_{c}](0)|p \rangle 
|_{\text{avg.}}
\nonumber\\
&=  N^{q}_{\alpha\beta}(P_1)\,\delta_{kj} \int_0^1 dx_1 \,
  e^{i x_1  (z\cdot P_1)}\,f_{q/p}(x_1,\mu) \,.
\end{align}
In the last step we have identified the quark PDF defined in terms
of the SCET fields
\begin{equation}
  \label{eq:q-pdf}
  f_{q/p}(x_1,\mu) =
  \frac{1}{2 \pi} \int dt \,e^{-i x_1 t \bar{n} \cdot P_1} 
\,\langle p|[\bar{\xi}^{(0)}_{c} W^{(0)}_{c}](t \bar{n})
 \frac{\slash{\bar{n}}}{2}[W^{(0)\dagger}_{c} \xi^{(0)}_{c}](0)|p \rangle
 |_{\text{avg.}}\,,
\end{equation}
which is consistent with the definition (\ref{eq:pdf-general}) in the 
main text, since for quarks the spin-dependent normalization factor, 
identified as a polarization sum of collinear quark spinors with 
momentum $P_1^\mu=(\bar n\cdot P_1)/2 n^\mu$,
\begin{equation}
   N^q_{\alpha\beta}(P_1)
= \frac{\bar n\cdot P_1}{2N_c } 
  \left(\frac{\slash n}{2}\gamma^0\right)_{\alpha\beta} 
  = \frac{1}{2N_c} \sum_\lambda u_\alpha^\lambda(P_1)u_\beta^{\lambda*}(P_1)\,,
\label{eq:nq}
\end{equation}
satisfies $N^q(x_1 P_1) = x_1 N^q(P_1)$. 
(The factor $\gamma^0$ arises because of the definition of the
conjugate quark and antiquark fields as discussed
below~\eqref{eq:quark-scet}.)  Eq.~\eqref{eq:q-pdf} is of the
same form as the PDF in full QCD~\eqref{eq:q-pdf-qcd} for external
quark states, apart from the fact that the collinear quarks and Wilson
lines appear instead of the full QCD fields.  Since we are considering
external massless on-shell partons with vanishing transverse momentum,
the collinear region is the only one contributing to the PDFs.  As the
SCET Lagrangian for a single collinear direction  is equivalent to
full QCD~\cite{Beneke:2002ph}, 
one can replace the collinear fields in~\eqref{eq:q-pdf} by
the full QCD fields and identify the PDF with~\eqref{eq:q-pdf-qcd}.
For a matrix element with collinear antiquarks we obtain an analogous 
expression with the antiquark  PDF
\begin{equation}
  f_{\bar q/p}(x_1,\mu) =
  \frac{1}{2 \pi} \int dt \,e^{-i x_1 t \bar{n} \cdot P_1}\,
 \langle p|\tr\left\{\frac{\slash{\bar{n}}}{2}
[W^{(0)\dagger}_{c}\xi^{(0)}_{c}](t \bar{n})
[ \bar{\xi}^{(0)}_{c}W^{(0)}_{c}](0)\right\}|p \rangle|_{\text{ spin avg.}}
\end{equation}
and the normalization factor 
\begin{equation}
N^{\bar q}_{\alpha\beta}(P_1)=
 \frac{\bar n\cdot P_1}{2N_c} 
  \left(\gamma^0\frac{\slash n}{2}\right)_{\beta\alpha} =
\frac{1}{2N_c}
\sum_\lambda \bar u_\beta^{\lambda *}(P_1)\bar u_\alpha^{\lambda}(P_1)\,.
\end{equation}

For the matrix element of $n$-collinear gluons, the only available 
transverse symmetric 
tensor is $g_{\mu\nu}^\perp$ so that~\cite{Bauer:2002nz,Becher:2009th}
\begin{align}
\langle p|\mathcal{A}_{c;k \mu}^{\perp}(z_+)
  \mathcal{A}^{\perp}_{c;i \nu}(0)|p\rangle|_{\text{avg.}}& =
\frac{1}{2 (N_c^2-1)}\, g^\perp_{\mu\nu}\,\delta_{ki}
 \langle p|\mathcal{A}_{c,j}^{\perp\mu}(z_+)
  \mathcal{A}^{\perp}_{c;j\mu}(0)|p\rangle|_{\text{avg.}} 
\nonumber\\
&= N^g_{\mu\nu}(P_1) \,\delta_{ki}\int_{0}^1 \frac{dx_1}{x_1} f_{g/p}(x_1,\mu) 
  e^{i x_1 ( z\cdot P_1)}
\end{align}
with the gluon PDF in SCET
\begin{align}
  f_{g/p}(x_1,\mu)& =-
  \frac{x_1 (\bar n\cdot P_1)}{2 \pi}
  \int dt \,e^{-i x_1 t \bar{n} \cdot P_1} \,\langle p|
  \mathcal{A}_c^{\perp\mu}(t \bar{n})
  \mathcal{A}^{\perp}_{c,\mu}(0)|p\rangle |_{\text{avg.}}\,,
\end{align}
which is consistent with the definition (\ref{eq:pdf-general}) in the 
main text. The normalization factor is 
\begin{equation}
 N^g_{\mu\nu}(P_1)
= \frac{-g_{\mu\nu}^\perp}{2(N_c^2-1)}
=\frac{1}{2(N_c^2-1)} 
\sum_\lambda\epsilon_\mu^\lambda(P_1)\epsilon_\nu^{\lambda\ast}(P_1)\,.
  \end{equation}
For the last identity we use the polarization sum of the gluon
polarization vectors
  \begin{equation}
\sum_\lambda\epsilon_\mu^\lambda(P_1)\epsilon_\nu^{\lambda\ast}(P_1)=
-g_{\mu\nu}+\frac{q^\mu P_1^\nu+P_1^\mu q^\nu }{P_1\cdot q}=
-g_{\mu\nu}+\frac{q^\mu n^\nu+n^\mu q^\nu }{n\cdot q}
      \end{equation}
and choose the arbitrary light-like vector $q$ proportional to $\bar n$.

The equivalence of the SCET and QCD definitions of the gluon PDF 
is established most easily in light-cone gauge $\bar n\cdot A=0$.
In this gauge the SCET gluon operators~\eqref{eq:gluon-scet} reduce 
to the collinear gluon fields: $\mathcal{A}_c^\perp=A_c^\perp$, while the
QCD definition~\eqref{eq:g-pdf-qcd} becomes:
\begin{equation}
\begin{aligned}
  f_{g/p}(x_1,\mu)& =-
 \frac{1}{2 \pi x_1 (\bar n\cdot P_1)}
  \int dt \,e^{-i x_1 t \bar{n} \cdot P_1} \,\langle p|
  (\bar n\cdot \partial A^\nu(t \bar{n})
  (\bar n\cdot\partial A_{\nu}(0))|p \rangle |_{\text{avg.}}\,.
\end{aligned}
\end{equation}
Upon integration by parts and Fourier transformation one sees 
that this agrees with the SCET
definition.

\section{Coulomb potential for gluinos and squarks}
\label{sec:susy-coulomb}
Here we collect the coefficients $D_{R_\alpha}$ of the Coulomb
potential relevant for the production of coloured SUSY particles
defined by the relation~\eqref{eq:coulomb-coeff}. This relation can be
solved for arbitrary representations in terms of the quadratic Casimir
operators of the representations of the two heavy particles and the
final state system 
\begin{equation}
\label{eq:dralpha}
  D_{R_\alpha}=\frac{1}{2}(C_{R_\alpha}-C_R-C_{R'})\,,
\end{equation}
as can be deduced from the generator of the product representation ${\bf
  T}^{(R\otimes R')}=\mathbf{T}^{(R)}\otimes
\mathbf{1}^{(R')}+\mathbf{1}^{(R)}\otimes \mathbf{T}^{(R')}$ and by
projecting the identity
\begin{equation}
  {\bf T}^{(R)b}\otimes {\bf T}^{(R')b}=\frac{1}{2}\left[
    {\bf T}^{(R\otimes R') b}{\bf T}^{(R\otimes R') b}-(C_R+C_{R'})
\mathbf{1}^{(R)}\otimes \mathbf{1}^{(R')}\right]
\end{equation}
on the irreducible representation. The explicit results for gluino pair production are long known~\cite{Goldman:1984mj} while all the remaining
 cases have recently been collected in~\cite{Kats:2009bv}. We
provide these results here for completeness. Note that our different sign
convention implies that negative values of $D_{R_\alpha}$ correspond to an 
attractive Coulomb force, positive values to a repulsive one.

\vspace*{0.3cm}
\noindent
{\em Squark-antisquark production.}
The coefficients for squark-antisquark production have been quoted
already in~\eqref{eq:coulomb-squark}:
\begin{align}
  D_1&=-C_F=-\frac{N_C^2-1}{2N_C}=-\frac{4}{3}\,,
& D_8&=\frac{C_A}{2}-C_F=  \frac{1}{2N_C}=\frac{1}{6}\, .
\end{align}

\vspace*{0.2cm}
\noindent
{\em Squark-squark production.}
In squark-squark production $q q \rightarrow \tilde q \tilde q$
the final state system is
either in the $\bar 3$ or $6$ representation and
the coefficients of the Coulomb potential are
\begin{align}
  D_{\bar 3}&=-\frac{1}{2}\left(1+\frac{1}{N_C}\right)=-\frac{2}{3}\,,
& D_6&=
\frac{1}{2}\left(1-\frac{1}{N_C}\right)=\frac{1}{3}\, .
\end{align}

\vspace*{0.2cm}
\noindent
{\em Gluino-squark production.}
For gluino-squark production $qg\to \tilde q\tilde g$
the final-state representations appear in
the decomposition $3\otimes 8=3+\bar 6 +15$.
The coefficients read
\begin{align}
   D_{3}&=-\frac{N_c}{2}=-\frac{3}{2} \,,
& D_{\bar 6}&=
-\frac{1}{2}\,,
& D_{15}&=
+\frac{1}{2}\, .
\end{align}

\vspace*{0.2cm}
\noindent
{\em Gluino-pair production.}
For gluino pair production the final state representations appear in
the decomposition  $8
\otimes 8=1 \oplus 8_s \oplus 8_a \oplus 10 \oplus \overline{10} \oplus 27$
with the coefficients of the Coulomb potential:
\begin{align}
&D_{1}=-3\,,
&D_{8_S}=D_{8_A}=-\frac{3}{2}\, ,\\
&D_{10}=0 \, ,
&D_{27}=1\, .
\end{align}
Note that only the singlet and octet states can be produced from a
quark-antiquark initial state while all the states appear in
gluon fusion.

\section{Evolution functions}
\label{app:evolution}

For convenience we quote here the explicit expressions of the
resummation functions $S$, $a_i^{V}$ and $a_\Gamma$ given
in~\eqref{eq:res_funct_def} up to NLL.  The expressions for the
corresponding functions for the Drell-Yan process up to N$^3$LL order
have been given in ~\cite{Becher:2007ty}.  Up to NLL the relevant
expressions read
\begin{eqnarray} 
S(\mu_h,\mu_s) &=&\frac{\Gamma^{(0),r}_{\text{\text{cusp}}}
+\Gamma^{(0)r'}_{\text{\text{cusp}}}}{8\beta_0^2}
\left[\frac{4\pi}{\alpha_s(\mu_h)}
\left(1-\frac{\alpha_s(\mu_h)}{\alpha_s(\mu_s)}
-\ln\frac{\alpha_s(\mu_s)}{\alpha_s(\mu_h)}\right)\right.
\nonumber\\
&&\hspace*{-1.5cm}\left.
+\left(\frac{\Gamma^{(1),r}_{\text{\text{cusp}}}
+\Gamma^{(1)r'}_{\text{\text{cusp}}}}{\Gamma^{(0),r}_{\text{\text{cusp}}}
+\Gamma^{(0)r'}_{\text{\text{cusp}}}}
-\frac{\beta_1}{\beta_0}\right)
\left(1-\frac{\alpha_s(\mu_s)}{\alpha_s(\mu_h)}
+\ln\frac{\alpha_s(\mu_s)}{\alpha_s(\mu_h)}\right)
+\frac{\beta_1}{2\beta_0}\ln^2\frac{\alpha_s(\mu_s)}{\alpha_s(\mu_h)}
\right],\quad
\\[0.2cm]
a_{\Gamma}(\mu_h,\mu) &=& 
\frac{\Gamma^{(0),r}_{\text{\text{cusp}}}
+\Gamma^{(0)r'}_{\text{\text{cusp}}}}{4\beta_0}
\ln\frac{\alpha_s(\mu)}{\alpha_s(\mu_h)}
\, , \\
a^{V}_i (\mu_h,\mu) &=&
\frac{\gamma_i^{(0),V}}{2\beta_0}
\ln\frac{\alpha_s(\mu)}{\alpha_s(\mu_h)} \, .
\end{eqnarray}
and an analogous expression for $a^{\psi,r}$.
Here we
have introduced the perturbative expansions of the beta function 
and the anomalous dimensions:
\begin{align}
\beta(\alpha_s) &= -2 \alpha_s \sum_{n=0} \beta_n \left(\frac{\alpha_s}{4
\pi}\right)^{n+1}\, ,\\
\gamma(\alpha_s) &=\sum_{n=0} \gamma^{(n)}
\left(\frac{\alpha_s}{4\pi}\right)^{n+1}\, ,
\end{align}
with
\begin{equation}
  \beta_0=\frac{11}{3}C_A-\frac{2}{3}n_f\;,\quad
 \beta_1=\frac{34}{3}C_A^2-\frac{10}{3}C_An_f-2C_F n_f\, .
\end{equation}
Explicit expressions for the anomalous dimensions required 
in the present work can be found, e.g. in \cite{Beneke:2009rj}.

\section{Fixed-order expansions}
\label{app:expand}

The expression for the resummed cross section~\eqref{eq:fact-resum} can be
expanded to a fixed order $\alpha_s^n$ in the strong coupling, providing an
approximation to the full $\mathcal{O}(\alpha_s^n)$ QCD calculation.  According
to the counting~\eqref{eq:syst}, the expansion of the NLL resummed cross
section to order $\alpha_s$ compared to the leading order cross section is
accurate up to terms of the order $\alpha_s\times\beta^0$.  This approximation
to the NLO cross section is obtained from~\eqref{eq:resum-NLL} by expanding the evolution function $U_i$
to $\mathcal{O}(\alpha_s)$, performing the convolution of the leading order
Coulomb- Green function for arbitrary $\eta$,
identifying $\eta=2a_{\Gamma}^{R_\alpha}(\mu_s,\mu_f)$ and expanding up to
$\mathcal{O}(\alpha_s)$. Inserting the explicit results for the one-loop
cusp and soft anomalous dimensions, the result reads
\begin{eqnarray}
f^{\text{NLL}(1)}_{pp'(i)} &\!\!=\!\!& 
- \frac{2 \pi^2 D_{R_\alpha}}{\beta} +
(C_r+C_{r'}) \bigg[
4\left(\ln^2\left(\frac{8 E}{\mu_f}\right)
-\ln^2\left(\frac{8 E}{\mu_s}\right)\right)\nonumber\\
&&+ \ln^2 \left(\frac{4 M^2}{\mu_h^2}\right)
-\ln^2 \left(\frac{4 M^2}{\mu_f^2}\right)
\bigg]  
- \,4 \,(C_{R_\alpha}+ 4\,(C_r+C_{r'})) \,
\ln\left(\frac{\mu_s}{\mu_f}\right) \nonumber\\
&&  +\,2\,(\gamma^{\phi,r(0)}+\gamma^{\phi,r'(0)}+2C_{R_\alpha}-2\beta_0)
 \ln \left(\frac{\mu_h}{\mu_f}\right) + {\cal O}(1)\,.
\label{eq:NLLexp}
\end{eqnarray}
We have introduced the expansion of the cross section in the strong coupling 
\begin{equation}
\hat \sigma_{pp'}^{(i)}(\beta,\mu_f) = \hat \sigma^{(0,i)}_{pp'} \Bigg\{ 1
+ \frac{\alpha_s(\mu_f)}{4\pi} f^{(1)}_{pp'(i)} 
+\left(\frac{\alpha_s(\mu_f)}{4\pi}\right)^2
f^{(2)}_{pp'(i)} 
+ {\cal O}(\alpha_s^3)
\Bigg\}\,.
\end{equation}
The term involving $2\beta_0$ in the last line of~\eqref{eq:NLLexp} appears, 
since we express $ \hat \sigma^{(0,i)}_{pp'} \propto \alpha_s^2$ 
in terms of the strong coupling at scale $\mu_f$, while in the factorization 
formula the hard function involves the scale $\mu_h$. 
The expansion~\eqref{eq:NLLexp} is used in~\eqref{eq:cross_matched}
 to match the NLL resummed prediction for
squark-antisquark production to the fixed-order NLO calculation.  At the
order considered here, one can further approximate
$E=M\beta^2+\mathcal{O}(\beta^4)$ to write the result in a more familiar form.

The NLL approximation to the NLO cross section~\eqref{eq:NLLexp} can
be improved to include in addition all terms of order
$\alpha_s\times\beta^0$ by inserting the Laplace transform of the
one-loop soft function~\eqref{eq:soft-one} as well as the hard
function in~\eqref{eq:fact-resum}.  Performing the derivatives with
respect to $\eta$ and expanding to $\mathcal{O}(\alpha_s)$ afterwards
we obtain for the NLO corrections to the cross section in the
colour-channel $i$~\cite{Beneke:2009ye}:
\begin{eqnarray}
f^{(1)}_{pp'(i)} &\!\!=\!\!& 
- \frac{2 \pi^2 D_{R_\alpha}}{\beta} +
4\,(C_r+C_{r'}) \bigg[\ln^2\left(\frac{8 E}{\mu_f}\right)
+8 -\frac{11 \pi^2}{24}\bigg]
\nonumber\\
&&  
- \,4 \,(C_{R_\alpha}+ 4\,(C_r+C_{r'})) \,
\ln\left(\frac{8 E}{\mu_f}\right) 
+ 12  C_{R_\alpha} +h^{(1)}_i (\mu_f)+ 
{\cal O}(\beta)\,.
\label{eq:NLOexp}
\end{eqnarray}
We note that  the dependence on the soft scale has canceled between the
one-loop soft function and the expansion of the evolution function. 
Similarly the evolution equation~\eqref{eq:evolution-hard-2} implies 
that the dependence on the hard scale cancels between
$h_i^{(1)}(\mu_h)\equiv H_i^{(1)}(\mu_h)/H_i^{(0)}$ 
and the evolution function. This can also be
checked explicitly for the case of top-pair production where the 
one-loop hard functions can be obtained from~\cite{Czakon:2008cx}. 
In contrast, the NLL result~\eqref{eq:NLLexp} contains residual
dependence on the soft and hard scale that is formally of higher order
if the scales are chosen of the order $\mu_s\sim E\sim M\beta^2$ and 
$\mu_h\sim 2M$. We observe that the $\ln 8$ constant terms in the 
NLO result~\eqref{eq:NLOexp} can be reproduced exactly with $\mu_s=8 E$.

Expanding the NNLL resummed cross section in the same way 
results in an approximation to the
NNLO cross sections $f_{pp}^{(2)}$ that is accurate up to terms
$\alpha_s^2\beta^0$.  
The results for the
production of heavy particles of arbitrary spin and colour and the
application to top-pair production have already been presented
in~\cite{Beneke:2009ye} and will not be repeated here. 


\providecommand{\href}[2]{#2}\begingroup\raggedright\endgroup

\end{document}